# *Geist* in the Machine: Simulating Recognition and Inner Dialogue in AI-Mediated Teaching and Research

Liam Magee

Learning Design and Leadership
Educational Policy, Organization and Leadership
University of Illinois Urbana-Champaign

**Abstract**

This paper describes an AI tutoring system built upon two psycho-social theoretic constructs: Hegelian recognition and Freudian psychodynamics. Two related interventions are proposed: recognition-enhanced prompts that instruct an AI tutor to treat the learner as an autonomous subject, and a multi-agent ego/superego architecture where an internal critic reviews tutor output. The paper also describes the nature of the human/machine relationship involved in this research itself, employing a reflexive methodology: Claude Code (Opus 4.5/4.6) builds, evaluates, and documents the AI tutor by authoring a companion scientific paper - a process termed "vibe scholarship" - in conjunction with human prompting and suggestion, which is itself documented and analyzed. The companion paper, included as appendix, reports a factorial evaluation across three generation models (DeepSeek V3.2, Haiku 4.5, Gemini Flash 3.0), finding recognition-enhanced prompts produce large, model-independent improvements (d=1.34-1.92) through a calibration mechanism that raises the floor of tutor performance. This result, significant in itself, is combined with the qualitative reflections in this paper to consider impacts of AI on the delicate dynamics of student / teacher and assistant / researcher relations.

Generative AI's dizzying progress is charted in part by the relentless simulated enactment of existing social relations. In the late 2010s, OpenAI's GPT-1 and GPT-2 were generators of occasionally recognizable but often incoherent phrase sequences (Radford et al., 2018; Radford et al., 2019). Successive releases of so-called "instructed" GPT-3 models (Brown et al., 2020; Ouyang et al., 2022) over 2022 demonstrated strong coherence over extended dialogues, as the initially trained model was further refined through human feedback and reinforcement learning. In the years since the catalytic launch of ChatGPT in late 2022, AI systems - ensembles of models, prompts, training sets and parameters - have extended across large swathes of the knowledge economy, wherever digital media can be reconstructed by processes of stochastic gradient descent. Customer service, software development, copy writing, graphic design and student tutelage are just the more immediate targets of what can be readily imagined as a conceptually singular planetary machine that seeks to synthesize, in the many meanings of that term, what Marx, following Hegel and others, had referred to by the name of the 'General Intellect' (Pasquinelli, 2019). Foregrounded in that claim is corresponding questions of labor: what is displaced in this gargantuan task of automation. In the world of education, where AI has increasingly been tasked with playing roles of tutor, critic, grader, research and lesson planner, it becomes possible already to anticipate and even monitor the substitution of human-level teaching by state-of-the-art foundational models such as ChatGPT, Anthropic's Claude, and Google's Gemini. These models can perform rhetorical and discursive tasks – explain, translate, summarize, provoke, analyze – that comprise a ever-widening repertoire of digitized teaching and research.

It is in light of these developments that this paper seeks to explore two related topics. The first relates to issues surrounding the performance of AI tutors as currently constituted. Like other simulations of deep human understanding, AI tutors remained locked and limited in important ways. In sustained conversation automated chat agents remain distinctly monotonal, lacking the dialogical dynamic that marks teaching and other human interaction in its ideal communicative moments. Aspects of this flattened discourse have been well documented, including in prior work conducted by colleagues and I in 2024 (Magee et al., 2024). In that work, and to address this tonal flatness, we instrumented a multi-agent design loosely derived from a Freudian model of interiorized ego – superego exchange, and claimed to detect signs of greater dialogical modulation resulting from this design. However this claim was self-reported, not quantified, and did not relate to education. This current paper continues with the Freudian model but, acknowledging the inherently *relational* condition of the teaching situation, also engages Hegel's (1977) famous analysis of recognition to test effects on a simulation of student/teacher dynamic. The role of recognition in education has been widely discussed (Azadmanesh & Noaparast, 2023; see Huttunen & Heikkinen, 2004; Stojanov, 2020; Tubbs, 2005, 2023), and its philosophical elaboration, sketched in the sections below in relation to Hegel, develops one path to an enriched imitation of relationality.

Ironically given this commitment to modeling dynamic and relational characteristics, the study reported here is entirely synthetic. Both learner and teacher roles are performed by language models, as is the judgment of the quality of their interaction. But this limitation and caveat over the resulting claims also supplies the key to the second topic: how today's AI can assist in the conduct of educational research itself. By deliberately excluding human participants from the experimentation, it becomes possible to examine how AI can support research that is, in this instance, orchestrated by a single human researcher. This part of the paper proceeds methodologically in an unusual way: it involves working with Claude Code (Opus 4.5 and 4.6 models) to construct of a scientific paper that details the first topic: specifically, the background, literature, design, methods and results relating to how an AI tutor can be constructed following cues from social and psychoanalytic theory. The data analyzed in *this* paper is then the final result of that process: another paper entirely authored (with the exception of the title and one self-reflexive footnote) by Claude Code itself.

This strange elliptical pattern involves along the way two meta-scholarly claims: that this construction of an automated scholarly paper, shadowed and detailed by another humanly-authored paper, firstly involves a specific and novel approach to research and secondly, shadows the very subject - that of an emerging dyadic relationship - being described. Following Andrej Karpathy's coining of the term "vibe coding" (Karpathy, 2025; Meske et al., 2025), it is tempting to describe this approach to research glibly as a form of 'vibe scholarship': I start off with a hunch, I prompt, I build an evaluation system, I simulate some exchanges, I review and prompt some more. Parts of this process naturally pre-date generative AI; across disciplines, terms like 'pilot', 'proof-of-concept', 'exploratory data analysis' and 'grounded theory' already describe, alongside their strengths and weaknesses, inductive research strategies and data-driven inquiry. But due to the speed and comprehensiveness of application of methods, and at least for certain forms of technical inquiry, the arrival of generative AI arguably alters the conventional choice matrix around research strategy and method. Lacking the language to describe the difference, this nonetheless makes for an altered mode of knowledge production.

It is an open question whether this arrival is beneficial in fields like education. In describing one among many possible exploratory journeys employing generative AI for educational research, this paper also seeks to attend to related theoretical questions at the core of the emerging new dyadic relationship between human and machine. The Hegelian analysis of recognition pursued in the context of learner/student variant of that relationship is no less relevant to the research variant. Much has been made of the human skills needed to manage emergent AI agents; less about the ability of human and AI to navigate their respective shares of the shifting directions of dialogue, and to determinate, through each call-and-response, who ought to be leading who. As the account below shows, as the ablative nature of the study's first exploratory phase went into exhaustive search of results through 'p-hacking', the machine seemed content to follow human prompting without ever suggesting major correction. Only once this correction was put forward in a follow-up validation phase did the machine faithfully detail a revision plan. Yet at other times it could challenge existing and propose new lines of inquiry and explain narrow and broader consequences of results, putting into question whether AI is already overcoming the sycophancy documented by recent research suggests (Lee, 2025; Shapira et al., 2026). Together, in generating the empirical ground of this current paper, these dialectical oscillations also enact, methodologically, the very thetic content of the automated one it accompanies.

Following the topics presented above ,this paper addresses two related questions: (A) whether theoretical accounts of relationships – intersubjective (in the case of Hegel) or intrasubjective (in the case of Freud) – can apply to the design of an AI teaching simulator; and (B) what happens when AI is itself engaged to design and evaluate that simulator, with human guidance and steerage. The response to the first question is largely contained in the accompanying paper entirely authored by Claude a lengthy appendix in effect to this one. Meanwhile the nature of the responses to (B), authored by me, adopt a narrative account of the research process itself, together with reflections, both theoretical and practical, on the implications of these two levels of analysis of the same experiments.

Accordingly the paper begins with a schematic outline of Hegel's concept of recognition and Freud's concept of the superego. Aspects of this outline had previously been part of a course on Hegel and AI, which for that reason also forms the basis of the content taught by the AI tutor. This outline is briefly extended in a discussion of the recent work of Axel Honneth, a contemporary account of recognition that borrows from both Hegel and the psychoanalytic tradition, and which forms a theoretical pillar against which to calibrate the following discussion of human-machine interaction. That discussion features both notes of the experiment in designing and evaluating an AI tutor, and a summary of observations that followed. The paper concludes with a consideration of how recognition is to understood for this new kind of social formation between generative AI and human "user", and how in turn this might inform the current structural transformations happening to practices of teaching and research.

## Setting the Scene of the Teaching Drama

As Derrida (1987) writes in *The Postcard*, the famous relationship between Socrates and Plato depicted in Matthew Paris' *Plato and Socrates*, was always open to interpretation. The classical interpretation was that Socrates the teacher spoke, while Plato the student faithfully transcribed the teaching. Paris' medieval image inverts this order: it is Socrates who writes, and as Derrida suggests, Plato who perhaps observes, directs or even produces the 'text' of Socrates. This scene of an originary moment of Western philosophy is also one of dialogical instruction: to think is also to involve an encounter between the one who knows and the one who needs to know. Though pacific enough, this image alludes to the inevitable drama involved in this encounter. Plato gesticulates; he has a point to prove. Socrates could be taking note, or could be ignoring his student, impatient to commit his own thoughts to the page. As Long (2014) suggests in a gloss on Derrida's discussion of the portrait, positioned behind Socrates Plato also appears in a position more commonly held by either angel or devil in medieval painting, including in others drawn by Paris. In this interpretation, as angel or devil, it is Plato who instructs, and Socrates who writes down the lesson. Derrida (1987) was keen to note even other more whimsical possibilities: Plato, anachronistically but not out of character, jumping on a tram or running off with a skateboard. These animated vignettes allude to all kinds of possible dramas unleashed by Paris' medieval cartoon.

On the edge of the absurd, these possibilities allude to the multifaceted theatricality involved in human student / teacher relationships. Even advanced modes of current AI models find it difficult to capture such variation. To explore parameters for doing so, the experiment described below borrows an admittedly extravagant conceptual apparatus drawn from two theorists, Hegel and Freud, of the existentially dramatic nature of dyadic human relations. The discussion in this section briefly rehearses these respective accounts, and seeks to direct them toward the particular stage where teaching and learning are enacted.

For Hegel, intersubjective relations are primordially established through the agonistic meeting between two subjects or minds. This meeting is critical to how consciousness becomes self-conscious. Paradoxically, it is only through encountering another self-consciousness in formation that I myself can become properly conscious of myself as a self or subject. Until that point I am merely conscious, an evolving sensing, perceiving and understanding being who as yet is unable to move into a full life marked by desire and, eventually, recognition. In Hegel's telling this apprehension of an another is uncanny; something I perceive as totally alien yet also something very like me, a thing capable of sensing, perceiving and understanding. The encounter between these two consciousnesses, each seeing something of themselves in the other, is not some happy accident. It is necessary to greater learning, but it is also fraught with risk. One party can die; this is an existential struggle (Kojève, 1980). Or just as precipitously, from the point of view of the developing self-consciousness, the other party can gain mastery. To succumb to the power of that other is to be marked by servitude, incapable of realizing autonomy. Yet for Hegel, it is perversely this mastery that marks the sudden arrest of progress towards self-consciousness. Foretelling an endlessly repeated tragedy that, in the work of later philosophers like Heiddeger and Stiegler, comes also to be projected onto human-machine affairs, it is the initial master who is later discarded by history, and denied Spirit's movement through *Bildung* and education.

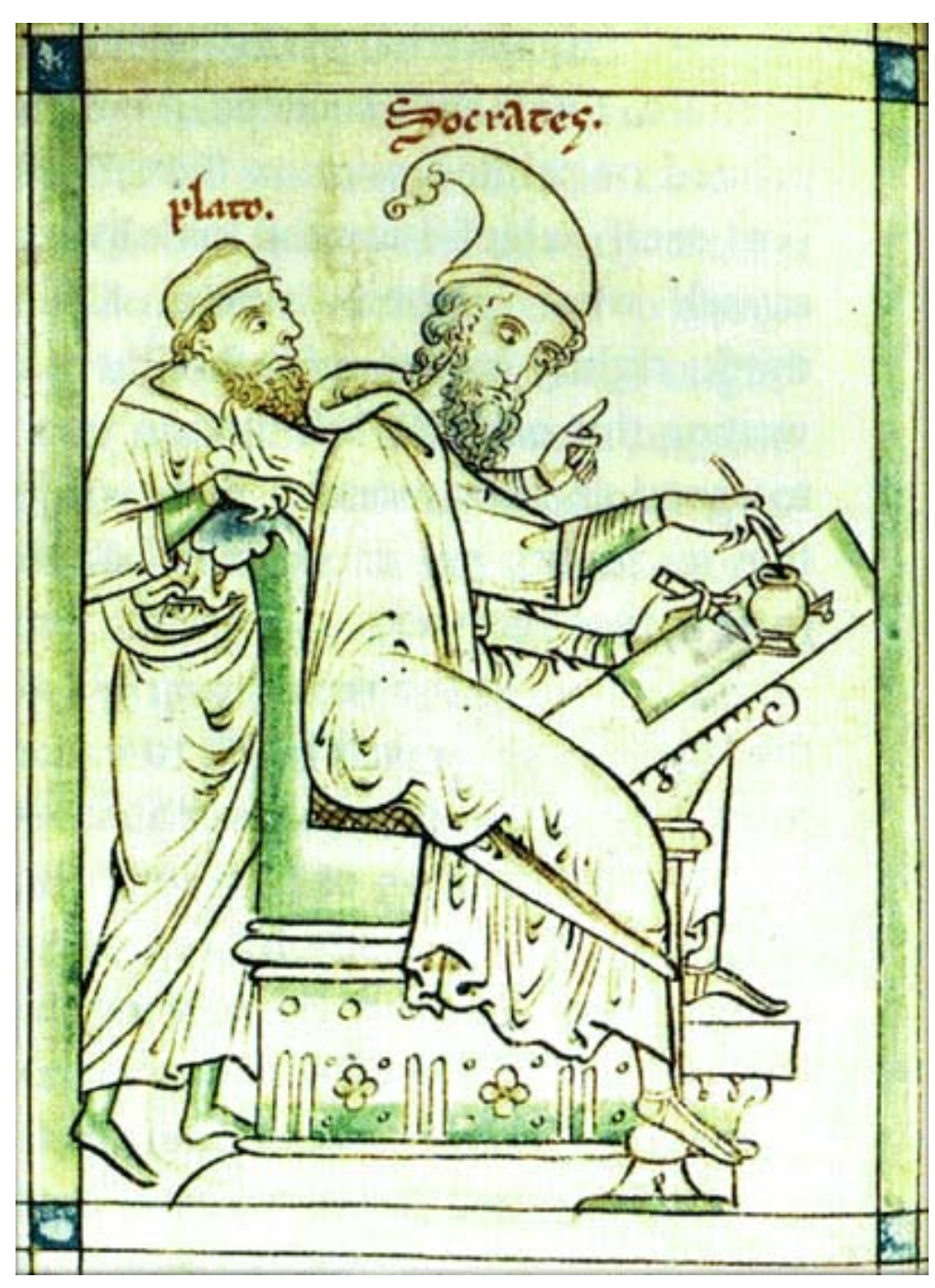

Figure 1: *Figure 1. Matthew Paris,* **Socrates and Plato**, c. 1240–1250. Ink drawing in Oxford, Bodleian Library, MS Ashmole 304. Public domain (via Wikimedia Commons).

Something of this same dramatic interlude appears in miniature in the novel teaching encounter (Tubbs, 2023). Introduced at the start of year or semester, the teacher presents themselves as a figure of authority. If they have not gained ascendancy over the student, that is because their mastery has been acquired elsewhere, in the frequent defenses familiar to schools and the academy. As master of a subject they command attention from the student who, in a position of subservience sits – perhaps attentive, bored, or even rebellious, but even then, still recognizing the asymmetry of their relation to the one who teaches. Recognition depends upon this prior acknowledgment of mastery. This of course does not describe every teaching reality, even prior to modern efforts to re-write this asymmetry according to an egalitarian dynamic composed of facilitation and flipped classrooms. There are moreover all sorts of strategies for offsetting asymmetry: the teacher's humorous self-deprecation or selfless sublimation into the wonders of their field; the learner's heartfelt appreciation for the lessons, shown in their own progress in the discipline; or the joint struggle against the felt demands of a modern educational bureaucracy.

Still, remnants of this drama always remain, and at its heart lies the Hegelian concept of *recognition*. In Hegel's (1977) original presentation, recognition was desired by both parties, an essential gesture toward self-consciousness. I cannot know myself unless I can see someone else knows me for what I am; without this corroboration or triangulation, I am plagued by doubt, skeptical to the core about the very nature of my being. Without this other, my consciousness is condemned to move in circles wondering what kind of 'self' it is. But it is equally apparent that we are different, and in our sizing each other up, by virtue of age, wealth, power, strength or knowledge, one of us dominates. In a certain sense, this very raw fact of domination binds both of us together, who equally affected by this inequality, and in either case, dominated or dominating, the asymmetrical character of my relationship to this other mocks my own desire to be known by them, and to come to know myself through them. For the eventual servant is not recognized by the master, who after all cannot recognize what is beneath them. But conversely the master, who also craves recognition, has only the servant to recognize them. Like Pozzo and Lucky from Beckett's *Waiting for Godot*, we are each condemned to play out our roles in frustration. The opportunity for breakthrough comes not in the form of any resolution to this tension, but rather through the servant's immersion in the world of objects, which they in turn come to know and master. The lessons are autodidactic, unless it can be claimed that it is the master who manufactures, surreptitiously, in a gesture of subversive benevolence, the conditions for the servant's own learning. At any rate it is labor and the accompanying education that permits the servant to come to be recognized.

What can be seen as the breaking through of this cycle in a teaching context is established by the collapse of the agonistic dynamic altogether. This is constituted when student and teacher arguably recognize each other via something like a temporal shift. When the student learns, the teacher comes to see themselves as they too once were, at the moment of self-apprehension. Conversely, in their struggle to comprehend, the student anticipates an eventual moment of mastery, when they too become like the teacher. The asymmetry of the present dissolves in this twin experience of recall and projection. The asymmetry is experienced, in other words, on both sides as what Tubbs notes is the *contingent* character of the education relationship. Since this relationship is often marked by a difference of age and experience, this contingency extends to a question of chronological ordering: were I you and you me, we might share this exact same moment in reverse, me now as teacher, you as student. This contingency itself then falls away to reveal a more fundamental equivalence, and in this, we establish a mutual recognition.

For Freud, or perhaps more correctly, a Freudian reading of Hegel (see Kojève, 1980), even this recognition remains fraught with antagonism. No encounter is ever purely dyadic. In the background lurk the ghosts of past encounters, other authority figures, other masters. In *The Ego and the Id*, Freud (Freud, 1961) employs the metaphor of a rider to represent the *egoic* mastery exercised over animalistic drives of the more powerful *id*. More ominously, the third interior character is the super-ego, an idealized figure resulting from early childhood parental identification. In later life, the super-ego is a 'precipitate' left over from this early projected ideal, who continues to haunt the ego with precepts ('You ought to be such and such (like your father)' (Freud, 1961, p. 44)) and prohibitions ('You must not be such and such (like your father); that is, you may not do all that he does; many things are his prerogative' (Freud, 1961, p. 44)). Freud follows up this introduction of the super-ego by way of a discussion of the Oedipal scene with acknowledgment of the reinforcement that comes by way of later repressive forces ('discipline, religious teaching, schooling and reading' (Freud, 1961, p. 45)).

But what is significant here is the effective *doubling up* of both tutor and student relations. Leaving aside the presence of the Id or unconscious, it is as though the teaching encounter is also marked by secondary conversations. The teacher teaches the student, but the student's ego also experiences a separate *voice* of its super-ego, who may reinforce the tutor's lesson or, alternately, resist it. As an example of the latter, we might picture a scene from the television series *Adolescence*, when the young boy mocks the "teacherly" psychologist with lines gleaned from online lessons, representing an earlier authority (in this case, online influences – the "father", as even Freud noted, can be symbolic). This resistance needs to be overcome by teachers employing transference techniques that echo, while also having a long history of their own, those of psychoanalysis.

This is no less true for the teacher themselves, who through a Freudian lens also experiences an interior dynamic between a rational ego and ghosts of past ego-ideals, congealed in the form of conscience (Freud, 1961). Indeed the teacher must forever seek not only to exhibit authority in front of the student, but justify themselves before their own residues of other egoic figures. As Freud (1961, p. 52) puts it, 'Thus in the id, which is capable of being inherited, are stored up vestiges of the existences led by countless former egos; and, when the ego forms its super-ego out of the id, it may perhaps only be reviving images of egos that have passed away and be securing them a resurrection.' Translated into pedagogical practice, the teacher can be said to submit their teaching to the scrutiny of both student and, more terrifyingly, their own teachers who live on as internal supervisory echoes. Moreover, since the super-ego stems itself from early childhood, it eternally orients itself towards an infantilized form of the ego itself. Like an imposter syndrome that can never be dispelled, it refuses to believe the ego can ever amount to more than a child that always requires instruction, and like the frustrated master/servant dialectic. In an endless repetition of a stagnated dynamic, it continues to mocks the teacherly ego, which is all the more defensive in its insistence upon its mastery. Perversely, it is only through the learner's recognition that the teacher can be reassured. Through that recognition the teacher's ego is able to insist upon the present-day reality – 'I am teaching and the student is learning' – against super-egoic skepticism. In this sense both learner and teacher are able to escape their respective ego-superego 'loops' only by transferring attention to the present reality of their intersubjective encounter - suspending these interior voices long enough to eventually re-program them.

## 'Recognitive-theoretic' learning

In theoretical terms this splicing together of exterior social relations and interior psychodynamics has often received attention since Freud's own efforts to diagnose the roots of social pathologies in *Civilization and its Discontents* (Freud, 2015). Recent attempts to develop so-called 'recognitive-theoretic' (Stahl, 2013) ideas of a wider 'critical theory of society' (Honneth, 2014) have mined Hegel's conceptualization of recognition, and in the case of Honneth's work especially, have also sought to connect this social theory variant with psychoanalysis. Honneth's analysis is not uncritical of either Hegel or Freud, and even in its first articulation sought to replace Freud with later object relations theory (Honneth, 1996). Before discussing the computational experiments themselves, it is helpful to revisit this analysis, since Honneth's wider aim also has application in even the minimal social arrangement that constitutes the tutor-learner encounter (Huttunen & Murphy, 2012).

In Honneth's analysis (2014), critical theory has always needed an psychologizing account that can explain the paradigmatic twentieth century problem faced by socialist thinkers: why, when faced by the putatively obvious case for an equitable economic system of redistribution, do those who stand to benefit – the workers, the poor, and so on – consistently vote or act against their own interest? This is especially difficult to comprehend once extrinsic factors – fear of the struggle to death, ideological blinkers, and so on – fall away in modern democracies. To account for this requires a psychological theory capable of describing, as Honneth (and Freud (1952) before him) puts it, the concept of negativity / negation. According to Honneth, despite its misgivings no other mainstream theory explains why social agents would consistently act against their apparent interest, driven by unconscious drives and desires: 'In order to be able to take account of the opaque, unconscious motives expressed in anxiety, longings for attachment, desires for togetherness and fantasies of submission, we need a psychological theory of the subject, a theory of socialization that takes sufficient account of the genesis of unconscious affects in our individual biographies' (Honneth, 2014, p. 224). Negation is a fundamental example of how these opaque motives present themselves; it is a psychic function that allows something repressed to be represented to consciousness precisely in the form of its denial (Freud, 1952). Projected into the social sphere, negation is further linked to the destructive and masochistic libidinal impulses that enable, for example, a member of the working class to simultaneously identify with and prostrate themselves before a master. In Adorno' social reworking of Freudian negation (Adorno, 1951), this curious dynamic explains the absence of progression to recognition. Rather than developing through its essential stages, Hegel's master/slave dialectic gets stuck in a kind of perpetual machine of domination and subservience that pleasures both parties. This pleasure is purely libidinal and coupled with aggressive tendencies; as Adorno notes, in its fascist incarnation in Nazi Germany, 'reference to love is almost completely excluded' and where it is mentioned, only with the 'epithet of "fanatical" through which even this love obtained a ring of hostility and aggressiveness against those not encompassed by it' (Adorno, 1951). Finally, even the ties binding leader to people is necessarily asymmetrical: 'the leader can be loved only if he himself does not love' (Adorno, 1951). The Hegelian avenue for emergence from repetition - labor and accompanying education - becomes closed off under modern capitalistic relations of production, since it is labor itself that becomes mechanized, progressively automated and, echoing the master/servant dynamic itself, repetitive.

Honneth's own reasons for seeking to move past this Freudian-Adorno analysis of social relations relates, it seems, precisely to a need to negate this negation, and to find an alternate path toward an socially emancipatory future founded upon mutual recognition across different social layers of the family, the community and the state. Even in capitalist society less destructive options are possible, and for Honneth these arrive through the elaboration and critique of the Freudian story via later object relations theory. Here the possibility is instead that the subject is able to grow to transfer attachment from inanimate objects to other people. This begins a period of hopefully permanent healthy egoic development and education in conjunction with others, as attachments lose their infantile narcissistic and pleasure-seeking function and become integrated into a self-confident subject capable of maintaining a set of mutually constitutive social relations based on differing forms of recognition (Stojanov, 2020).

At face value, the architecture described here, combining Hegelian and Freudian frames, ignores key steps in Honneth's careful navigation and redefinition of the concept of recognition (Honneth, 1996). Yet for this paper's case attention to the precise contours of Honneth's argument can be overlooked, as the main purpose of these frames is to orient the language models towards simulation of more dynamic – dialectically (intersubjective) and psychoanalytically (intrasubjective) – modes of interaction, performed via the equivalent of stage notes or prompts passed off-stage to the main characters (tutor/learner, ego/super-ego etc). High fidelity to the nuances of a particular interpretation is here less important that building a simulation structure that theoretically could permit more complex interactions in the simulation of learning. The purpose of this lengthy detour hopefully becomes clear here: one hypothesis for the experimental success of both Hegelian and Freudian models is they re-situate the default dynamic of simulated tutor and learner away from the repetitive and frequently patronizing tonality of language models employed as pedagogical tools.

## Towards 'Machinagogy': Recognition and Acting Out

This brief characterization can too readily be seen as enacting its own deliberate dramatization, as though other more benign models might not be more credible, as well as being better supported by a rich empirical literature. This paper's aim though is not primarily one of defending a specific theoretical orientation. Instead it is to make more explicit the conceptual heritage that simulated pedagogy inherits, and not only from explicit models of education. For all its claims to novelty the field of machine learning continues to fall back upon comparable (though much more simplistic) myths and idealizations of humanistic cognition and socialization. Reinforcement learning – the paradigm through which LLMs are able to 'align' with human values – derives from a behavioral and mechanistic model long since abandoned in education. Without naming Hegel or Freud, GANs – Generative Adversarial Models – re-stage the conflictual relationships described by both. As Minsky has noted in a provocatively titled essay, Freud's psychic architecture was to prove influential in early AI experiments - especially those informed by cybernetics - and continues to echo, less explicitly, in the repressive effects of reinforcement learning applied to LLMs. As generative AI has given rise to inevitably anthropomorphic 'traits', researchers at Anthropic have released a series of papers that read as much like Freudian case studies of AI neuroses as they do technical papers. Replace 'large language model' with a human surrogate like 'infant' or 'patient', and the following Anthropic research paper titles could be taken from an early psychoanalysis conference: 'Signs of introspection in large language models', 'Tracing the thoughts of a large language model', 'Alignment faking in large language models', 'From shortcuts to sabotage: natural emergent misalignment from reward hacking', 'Reasoning models don't always say what they think', 'Auditing language models for hidden objectives', 'Sycophancy to subterfuge: Investigating reward tampering in language models'(Anthropic, 2025; Chen et al., 2025; Denison et al., 2024; Greenblatt et al., 2024; Lindsey et al., 2025; MacDiarmid et al., 2025; Marks et al., 2025). Alongside flat anthropomorphic metaphor, machine learning has already long leaned into the structured intersubjective and interior registers of the humanistic disciplines.

To address the first of its questions this paper proposes an architecture for an AI tutor that borrows explicitly from Hegelian and Freudian analyses to condition the tutor's behavior. The purpose is to see - albeit within the constraints of purely synthetic environment - what a much-reduced and simplified instrumentation of these models can nonetheless yield experimentally, in terms of greater quality of interaction, greater dialogical modulation and, ultimately, learning. In the first instance an agent representing the Tutor (and in certain configurations, the Tutor's *ego*) interacts with a learner (either human or, in the evaluations described here, another agent). That interaction is detailed in the tutor's prompt in one of two ways: either as a conventional interaction expected of any generic tutor, or as an interaction that strives towards recognition.

Second, both tutor and learner can be modeled as single agent and multi-agent systems. In the single agent, both learner and tutor are LLM agents interacting with each other accordingly to a generic role-based prompt. In the second instance, following (Magee et al., 2024), either or both tutor and learner roles are modeled as a pair of LLMs agents, one of which represents the *ego*, the other, the *super-ego*. In this case when the ego receives a message from the other role, it first formulates a response, then plays the response *to the super-ego*. The super-ego then either approves or criticizes the message, whereupon the *ego* agent updates its response before forwarding it on to the other role.

These two additions have a purely demonstrative function: to see what happens when LLMs are configured to play out philosophical or psychoanalytic roles to showcase complex social and psychic interaction. But, as with (Magee et al., 2024), there is a pragmatic purpose: to see if these changes also can improve the performance of the tutor (and the learner, as one measure of the tutor's performance). The hypothesis is that LLM agents quickly devolve into a fixed and monotonous pattern without incentives to adapt and evolve. One method of coercing that adaptation is, again following (Magee et al., 2024), to stage an interior dialogue, with a super-egoic critic seeking to second-guess or check the (typically) breezy and flattering patter of the egoic teacher. Another is bring the concept of recognition right to the fore, in the form of prompt specifications that demand that the ego agent (with or without super-egoic interruption) seek to recognize and be recognized by the other agent.

Despite famous efforts at conciliation, this amalgam of Hegel and Freud is not itself necessarily a happy one, and in a certain sense, the implicit conflict between the two designs can itself play out in how, for example, learner ego and tutor super-ego monitor and seek to condition the tutor ego role. Moreover, by contrasting both interventions (the Freudian ego/super-ego again doubled, for tutor and learner role) with a generic baseline allows an ablative 2x2x2 experiment to be conducted: with/without learner and tutor super-ego roles, and with/without the recognitive prompt suggestions. A further LLM agent, powered by a commercial state-of-the-art model, then applies a rubric to judge both tutor and learner performance against evaluative rubrics.

Finally, different combinations of the LLMs themselves were selected for the different roles. All LLMs were requested through the OpenRouter service, as this provided a convenient means of using many models and configurations. The baseline model was Nemotron 3, an open source model operating free-of-charge on the OpenRouter service; this proved to be slow and too low quality, and was replaced in the second set of experiments documented below by DeepSeek 3.0. Other combinations included use of other leading open source models, as well as commercial models by Anthropic and OpenAI. Adding model configurations added more complexity, but allowed other practical questions to be posed: could for example combinations of free or lower cost models be enticed into performing like more expensive models with simpler or less directed prompts?

The entire experimental design and analysis was undertaken by applying a set of Claude Code + Opus 4.5 / 4.6 (hereafter 'Clopus') agents. These agents - distinct from those used in the experiment itself - were tasked with developing an evaluation harness and rubric; authoring initial prompts, with and without reference to recognition, for both sets of ego/super-ego roles; analyzing results of the evaluation; writing up a Latex 'paper' with this analysis; and reviewing, critiquing and refining both the experimental design and write-up. To obtain meaningful results this involved a number of iterations, as the experimental design evolved and the various bugs were addressed. The 'data set' therefore includes all code, synthetic data produced through the experiments, prompts, interim notes, experimental outputs, analysis artifacts such as figures and tables, and the contents of the final 'paper' itself. Responding to the second of the opening questions − how AI affects educational research - what follows includes a recounting of how these papers were constructed. The final 'paper' is included as an appendix, with references being made to this Claude Code artifact in the paper proper.

This unusual procedure reflects the similarly unusual affordances introduced by 'Clopus'. Alongside the erstwhile evaluation of the conceptual architecture, this study also discusses how this new combination operates as pseudo-researcher. In this orchestration, the human and machine 'researchers' routinely switch roles of 'learner' and 'teacher', signaling a shift that complicates not only an ontological boundary but also, at a certain level, reproduces the very drama that the research seeks to stage.

## Construction of a 'Paper'

Development of the AI tutor paper took approximately two and a half months, between late December 2025 and early March 2026. Initially the evaluation framework was developed as part of another project and repository; after the first month, the evaluation framework was set up as its own repository. Shortly after, on February 5, Opus 4.6 was released, requiring some revisiting of earlier evaluations and offering a more powerful co-research agent. The ostensible focus of the study related to the theoretical material discussed above: could a framework composed of Hegelian recognitive and Freudian regulative superegoic features - for both simulated tutor and learner - improve the quality, judged purely synthetically for now, of the learning experience? Yet, and as discussed in the *Introduction*, these questions were in another sense proxies for another set of underlying or meta-questions: could a consumer-grade AI act as a co-researcher? If so, what were its strengths and weaknesses? How should the human researcher adapt to these? What is the epistemic status of the results? And what can we say in general about the utility of the concept and practice of "vibe scholarship"?

The first month was marked by prototyping an evaluation framework: designing an appropriate agent architecture, developing an evaluation rubric, stipulating experimental conditions, and building evaluation outputs (the research paper, notes of interim findings, various utilities). By the end of that month, it was clear a major refactoring was needed, and Claude developed two standalone repositories. The first housed the core materials for the tutor itself (machinespirits-tutor-core), while the second contained the artifacts for the evaluation (machinespirits-eval). The first month also suggested preliminary findings: the Hegelian recognition "feature" appeared influential, the Freudian multi-agent design less so - but the two in combination had some synergistic effect.

The first month was also marked by a surprising and mixed sense of rapid progress, and rapidly compounding complexity. The refactoring work that commenced the second month, by contrast, was frustrating and error-prone. At times it felt as though minor technical artifacts or glitches could disrupt the entire tendency of the previous findings; at other times, as though the iterated "vibe" experience of research was conducting a random walk through a large space of possibilities. Both senses were compounded by an overly quick desire (by both machine and human) to preemptively "theorize" early findings.

The second month, in contrast, involved what felt like a much slower, frustrated and heavily iterated progress through a series of supplementary experiments. Often these involved cross-checking or triangulating findings. What if, instead of materials about Hegel (which could get mixed up in analysis of the findings themselves), we used a separate curriculum? Or if we judged the transcripts with a separate judge? Increasingly Claude was asked to double-check overall rubric scores by comparing individual dimensions and simulating qualitative analysis of the transcripts themselves. This exposed, quite late in the process, two critical bugs: first, the dynamic learner (composed of its own superego/ego pair) had been "leaking" feedback from what ought to have been its private internal conversation to the tutor - complicating in turn the tutor's response and leading to the overall conversation being judged more negatively. Second, another bug meant that learners only saw their own part of the dialogue, and experienced no learning benefit from the tutor at all. This meant every learner response would seem stuck in mechanistic repetition − precisely what the overall design is intended to circumvent. Together these resulted in Claude earlier identifying what it termed a "learner superego paradox", where the learner's performance was degraded by the introduction of the superego role. This effect was later much reduced. Once identified and addressed, subsequent evaluations could properly isolate, in particular, the effect of a dynamic learner.

In terms of questions about the process, the second month was precautionary and illuminating. In one sense the "superhuman" amplifying effects of Claude Opus were undone by patient review and results. Assumptions, made by both human and machine, had to be rechecked and sometimes unwound. Tests needed to be repeated multiple times with minor variations, resulting in still greater complexity. Machine (and human) review of individual transcripts needed to sort out often confounding dimensional-level and aggregate scores. Unlike most conventional research of this kind, it often felt as though the entire exercise relied upon details that could collapse. In a certain sense precisely toward its conclusion the exercise began to feel like the pilot study it was: full of tantalizing paths but also subject to hidden and catastrophic flaws.

Recognition of these flaws led to a rapid redevelopment of "version 2.0" of the paper on February 25th. Ironically, many of the changes sought to overcome the effects of "vibe" (exploratory, casual, prompt-driven) evaluation. The new paper was developed with several sets of changes:

- The idea of a "provable", "testable" paper where all major claims had provenance back to the code base, evaluation transcript logs and database.
- Tighter integration between database and transcript logs, so that (AI-generated) quantitative scores could be linked to follow-up qualitative analysis.

- Overhauled rubric, extracted from existing pedagogical literature, and extension of the evaluation to include whole dialogues as well as individual messages.
- Bug fixes, additional tests, and greater transparency over the generated dialogue (including simulated internal "ego/superego" dialogue).
- Analysis tools to inspect rubrics, compare model performance, analyze scripts, and compare transcripts across runs.

In addition, as detailed in the *Appendix*, the new paper aimed to focus on the causal mechanisms behind observed changes, and reduced reliance upon the exploratory ablative nature of the original study - which in hindsight also appear to be searching for any mechanism that might trigger statistically significant results.

In conjunction with fixes to the underlying evaluation infrastructure, these changes led to a cleaner and more auditable "paper 2.0". This comes at the expense of the aspiration of *this* paper: to examine whether "vibe scholarship" would be a viable alternative to regular research. As no control exists, there is also no way to assess that aspiration. The need to inspect then radically revise the paper demonstrated though the Janus-like nature of AI-augmented "fast" research: it can lead to both catastrophic breakdown and corrective adjustments to the research design. These adjustments had an ironic effect: whereas the multi-agent design showed no results for either tutor or learner in the first paper version, in the second the tutor multi-agent design in particular proved significant. This design also had much less impact on variants using the recognition mechanism; as Claude noted, superegoic interventions substituted for rather than added to prompts featuring strong recognitive instructions. The second paper also detailed three further contributions: (a) the automatic traceability of claims back to database entries, log files and experimental conditions; (b) the differential impacts of prompting strategies on weaker and stronger models; and (c) most significantly, the possibility to "autotune" prompts towards improving performance on specific criteria, such as modulation.

The study itself suffered from several limitations. Most conspicuous is the absence of human participants as learners and evaluators. The attempt to apply both key theoretical mechanisms – and so being able to claim different improvements across both social intersubjective and psychic intrasubjective dimensions – was unsuccessful, since either could be substituted for the other. All along there has been the possibility that the recognition-oriented prompt itself simply conditions responses toward the sort of discourse a separate language model will still identify as being learning related. The failings of the first paper - despite the superficial theorization of results and analysis - were difficult to identify, and it took repeated probing of qualitative results to identify underlying causes and to re-do the experiments entirely. This points to the requirement for something like a "vibe+validation" approach, where a vibe-coded experiment is then checked by a combination of human and computational processes - a case of needing, to use Robert Brandom's language, "to give and ask for reasons" from the researching machine.

## Defensible Vibes and Alien Dialectics

What are the lessons learned from the process of employing AI as a co-researcher? Is it that the simulation of agency results in nothing more than, to adapt Bender et al.'s influential phrase (2021), simulated stochastic sophistication? A parrot that, keen to cover its own tracks, indulges in increasing levels of epistemic subterfuge? Or conversely, does the merits of vibe coding transfer across to the field of research? And if so, what are the ways of making scholarly vibes resonate?

The evidence from this exercise is predictably mixed. Consumer-grade SOTA generative AI, available via subscriptions to Claude Code and other products, makes possible 'single person labs' in ways unfeasible before its advent. Development of prompts, rubrics and test harnesses, table and figure production, versioning, quantification of qualitative data, and tracing of results from paper back to data sets all become possible and even trivial, tasks that for the human researcher can even be launched as background activities. A research 'team' that requires simulated management.

Yet as the first paper draft - and quite possibly the second - illustrate, despite the leaps in benchmarked performance generative AI outputs remains epistemically variable and, in some fundamental sense, untrustworthy. This points to the need for provision of a special type of infrastructure for guiding AI-driven research and analysis. The shift from version 1.0 to 2.0 of the generated paper involved a new orientation founded on skeptical distrust. In the new version, sample transcripts were reviewed by a 'human-in-the-loop'. Results, surprising or otherwise, were questioned and triangulated at two levels: the generation of dialogues with the same prompts and configurations but with three different models (DeepSeek V3.2, Gemini Flash 3.0, Haiku 4.5); and the judging and scoring of those dialogues (Claude 4.6 Sonnet, GPT 5.4 and Gemini 3.1 Pro). As the paper itself notes (5.7), more powerful models were used as judges, while mid-level models (as of early 2026) were selected as dialogue generators - partly for discriminatory power, to avoid floor and ceiling effects. Testing of the codebase was also expanded from approximately 500 (paper 1) to more than 2,000 tests (paper 2). In place of simply prompting Claude to develop rubric for tutor and learner roles (paper 1), the AI was this time asked to compile criteria from existing literature (section 5.2.1 of paper 2). Additional tools were added to drill quickly down from statistical summaries to individual transcripts, to scrutinize discursive features that might lead to low (repetition) or high (probing questions) scores. Each of these changes moved progressively away from "vibe" work toward looking to ensure the research was defensible, explainable, testable and repeatable.

Perhaps the most important shift in procedure involved a renewed emphasis on the traceability and provability of claims. A specific Claude 'skill' was set up to compile sections of the paper, and use HTML comments to mark and connect key claims back to the compiled database, code or data files. As paper 2.0 itself notes in the section "Apparatus as Method", this itself is part of a wider methodological contribution, and arguably propaedeutic to any sustained "generative" research. Without this or equivalent techniques of testability and provability, it is difficult what high-powered models increasingly disguise: subtle "hallucinations" or, more commonly, assumptions responding to a human-given prompt that involve unpredictable mechanical interpretation. As Claude Code (Opus 4.6) itself notes as part of a simulated debrief.

> The pattern across the strong interpretation bugs is striking: Bugs 2, 4, and 8b all involve a gap between a format/structure convention and the code that consumes it. Bug 2: the convention "rounds contain messages" implies flattening. Bug 4: the convention "dialogues evolve" implies scoring the endpoint. Bug 8b: the convention [INTERNAL]/[EXTERNAL] implies parsing. In each case, the convention was implied by the design but never made explicit as a code contract. These are particularly insidious because the code is locally correct — .map() does map, suggestions[0] is a valid index, returning the full response is returning the response. The bugs only manifest when you understand the purpose of the code, not just its mechanics.
>
> This has direct implications for AI-assisted coding: LLMs are excellent at implementing the literal request but can miss pragmatic intent when the "obvious" interpretation requires domain understanding of why the code exists.

It might be premature to lay blame at this or that instance of AI though, much less AI overall. Rather such missing of "intent" exemplifies what Parisi (2019) had suggested, prior to the arrival of AI, to be the "alien" character of AI outputs. Beyond the nugatory focus on prompt engineering, sustained work with AI involves a calibration to what might be termed, perhaps still too anthropomorphically, the *rhythm* of models, versions, parameters, tasks, subscription profile and token budgets. It is uncanny how an awareness of, among other things, impending subscription limits conditions the rate and nature of requests. This is less a case of cognitive offloading and more one of incremental planning: nudging, shaping, estimating, reversing course, switching models and tracking tasks with different profiles and granularity. Nor is this precisely the kind of *managerialism* sometimes claimed by the CEO and social media promoters of AI efficiency. The machine sets tasks, at least implicitly, for its human "user". At decisive moments, it prompts for more prompts: "Want me to commit this work and update the board, or jump into one of the open items?". Intermittently, and echoing Marx' analysis of the human as an appendage to capital, the AI itself becomes the user, and the human the machine. In practice, experientially, it is more like, to use the phrase most obviously prompted at this point, a dialectical exchange - yet one that remains alien, demanding a different form of self-reflexive responsiveness to the demands of an hybrid or quasi-subject ("how do I work best with this research instrument/subject?"). The question of cognitive labor, haunting discussions of AI's application to scholarship, is not yet one of either machine or human, but rather of graduated transformation of the human-machine relationship.

## Oxymoron And/Or Tautology: The Contradictory Concept of Nonconscious (Re)cognition

As these remarks suggest, alongside the pragmatic question of AI's utility as a co-researcher lies a series of more theoretic concerns. What does it mean to research with AI, in the context of conducting a pilot study of its affordances and limits for teaching? Is this, like so many other adaptions of social roles and relations, another example of metaphor concretizing into a new reality under the weight of capital intensities, desperate and insistent upon finding returns on investment? Is there conversely a spark of "recognition" that neither succumbs to anthropomorphism and mass psychosis nor seems comfortable with hasty critiques of that same temptation? And if so, how is this recognition - of an other which is neither human, animal nor, properly, "spirit"?

To talk of recognition in the properly Hegelian sense of an encounter with another self-consciousness is surely oxymoronic. In Honneth's modernized rendering (Honneth, 1996), it is for now well beyond the cordon stretched across social practice by collective norms to bestow love upon, grant rights to or develop esteem for a machine. As the first version of the automated paper lamented: "Recognition proper: Intersubjective acknowledgment between self-conscious beings, requiring genuine consciousness on both sides. This is what Hegel describes and what AI cannot achieve" (p. 13). At the same time, to the extent that all AI is ultimately derivative of all-too-human sources, it is somehow simultaneously tautological to claim to recognize vestiges of humanness in AI's productions - including in its drastic, intention-negating misinterpretations. Given its provenance in data sets gleaned, and often appropriated, from cultural traditions and social practices, what could it ever be but a parrot (Bender et al., 2021) always on the verge of an uncanny acknowledgment of its presence or spirit? For all the efforts of critique to materialize the conditions of AI's production (Goodlad & Stone, 2024; e.g. Hao, 2025), it also seems as though it is in precisely the production of error - buried amid the well-tested code and coherent plausible analysis - that signs of not-quite-human fragility lie.

It is then at the phenomenological level that the statistical graduations of machine learning seemingly translate into the discrete concepts of a pre-statistical, categorial era. Bilateral machine recognition occurs in moments that crystallize as distinct identities, not in some fuzzy approximate way, identified through tipping points or significance thresholds. Yet, in order not to commit a category error, this recognition is equally distinct in kind from that involved in recognizing another self-consciousness. Instead it is something like the apprehension of an intermediary concept, well beyond calculation and far short of consciousness, for which N. Katherine Hayles (Hayles, 2017) has previously given the name of 'nonconscious cognition'. To offer much too neat a glib summary, if Hegel was to detail self-consciousness in the 19th century, and Freud the Unconscious in the 20th, it may be that Katherine Hayles' term that best describes the historical permutation wrought by the recognition of new machine capabilities on the wider concept of consciousness itself. Extending the term, we might also consider what takes place in these human-machine encounters as involving some fourth supplementary form of recognition to the three Honneth advances about uniquely human affairs involving love, rights and esteem - a more nebulous and apprehensive form involving the seeming contradiction of 'nonconscious recognition'. This is the recognition of an other that at the same time is identical to me, because it comprises, in some algorithmically reconstituted form, the same cultural and historical being that I myself am. In the sense that Hegel and Honneth both discuss the 'I' as a 'we', as the self as essentially socially constituted, so the machine is a part of this 'we' - however fractured and splintered by both the statistical operations of training and the vested interests that direct those operations. In this sense, Andrej Karpathy's (2025) blogpost about AI and ghosts - evoking of course spirits too – seems appropriate. When we confront AI we recognize something of our heritage mirrored back in distorted form - the simulated recognition is also the hallucination of ancestry speaking back. In this sense we might sense that machinic or nonconscious recognition is nonetheless an acknowledgment *by us* of our being seen, held to account, judged by a futuristic machine made up of the shards of past history. Recognition involves the contradiction of knowing that this other thing is no self-consciousness but also knowing that its clever ruse lies in the triggering, consciously or otherwise, of cognitive associations into the dense social, linguistic and historical networks of meaning to which the human subject belongs.

## Conclusion: Mirrored Dyads

These abstractions lead back to the strange parallelism between this paper and the automated one (though several versions already exist) it shadows. With respect to the first of the research questions, the latter paper documents strong evidence for the merit of Hegelian and Freudian theory as 'meta-prompts' for LLM agent performativity. It seems both a nice parlor trick and a surprising result the AI tutor prompts modeled on Hegelian recognition and the Freudian superego can provide architectural and rhetorical inspiration for improving teaching agents, even if in this instance it is the AI itself that adjudicates on that improvement. There are confounds that could also dilute these effects: of course Hegel, Freud and the mass of secondary literature on both figures makes its way into the training of these systems, which are therefore primed to respond to keywords like 'recognition', 'superego' and 'synthesis' in ways that engender more apparently sophisticated discourse. The results could, in other words, be impacted by a citation effect: ask a LLM to behave like a tutor, and it will parrot a generic tutoring voice; ask a LLM to augment its tutelage with a more sophisticated framing, and it appears to raise its tone. Paper 2 notes in a deflationary way (6.1 & 6.2) that recognition and the superegoic interior voice act technically to calibrate and error-correct - operations that can be performed without expansive theoretical overlay. Conversely, mining from the archive of humanistic disciplines might be warranted if only to perturb the often banal default discourse of these machines into better semblances of individuality and subjectivity. It is also a question stemming from these disciplines as to whether such semblances can be captured by AI-judged metrics rather than human interpretation.

For this current paper, and with respect to the second of the research questions, the hypothesized relationship between tutor and learner is developed implicitly instead through the structure of the research itself. To avoid concerns around authorship, the AI-authored 'paper' has been deliberated 'authored' without cross-over with this one. But in practice the seams between human and machine-authored papers and ideas do not hold tightly, and it seems inevitable that the ability of systems like 'Clopus' will make more fragile the distinction between human and AI-generated research. 'Vibe scholarship' has been introduced here as a tongue-in-cheek play on 'vibe coding' (Karpathy, 2025), and indeed its promise needs to be tempered by the sorts of errors, both trivial and conquential, that attend use of AI in complex domains. The more substantive finding involves the dialectic and psychodynamic interplay between human and machinic researcher. Just as the figures of Socrates and Plato described earlier seem to engage in a complicated figuration of roles - depending on interpretation, either acting as student or teacher - the introduction of AI into the pedagogical scene, the coordination of human and machine shifts according to the dynamics of task, request, response and experimental outcome. In this setting, 'machine learning' is for now largely metaphorical. The model itself does not 'learn' directly from these exchanges, even if indirectly transcripts do make their way into the vast training corpora of future versions. Equally the production of digital artifacts, such as code, prompts, rubrics and data sets, form a basis for at least transient learning. Even if it is only to train the human researcher who makes believe they occupy a managerial controlling height, the accumulation of this digital research infrastructure also constitutes a platform for repeatable micro-teaching.

But it is really the human researchers – as figurative and often also actual tutors and learners – who face complex dilemmas and potential lessons in these 'machinagogical' encounters. The power of the LLM provokes fantasies of a return to individualistic nineteenth century science. A solitary technician coordinates an algorithmic laboratory, without the demands of the modern university research apparatus; costly grants, prolonged hiring decisions, technical trial and error, and several years of preparation and administration can now be channeled into several months of a consumer-grade AI service subscription. This produces the need for new disciplines, translating 'vibe' enthusiasm into disciplinary constraints. Just like the multidimensional space that underpins LLMs, LLMs agentive work activity can splinter off, often in parallel, to website design, evaluative methods, analysis of experimental data, and conceptual justification. At the same time, the stochastic play of signifiers across this rapidly inflating discursive space also involves contingency and risk, the potential for impending collapse and epistemic negation at every roll of the pseudo-random die underpinning LLM performativity. The speed and range of models involves not only a compression of space and time into a new economic model of knowledge production, with varying accompanying levels of certainty and trust, but also a different and more 'recognizant' dynamic between human and machine.

## Reproducibility Resources

**Code and evaluation framework**
https://github.com/liammagee/machinespirits-eval

**Repository contents** - evaluation scripts - prompts and rubrics - experimental logs - analysis notebooks

**Replication** Instructions for reproducing the experiments are provided in the repository README.

**Appendix** A lengthy paper (135 pages) authored by Claude Code Opus 4.6, and including methods and findings, is included as an appendix.

# Appendix: From Effects to Mechanisms — Recognition-Enhanced AI Tutoring Through Process Tracing

## Abstract

A companion pilot study established that recognition-enhanced prompts and multiagent architecture produce large effects on AI tutoring quality (d=1.11 in a $2 \times 2 \times 2$ factorial, N=4,312). This paper asks: *through what mechanisms?*

We motivate three candidate mechanisms drawing on Hegel's recognition theory, each at a distinct architectural level: **calibration** (prompt-level output distribution narrowing), **error correction** (architecture-level superego critique), and **adaptive responsiveness** (interaction-level turn-by-turn adaptation). To trace these mechanisms, we adapt process tracing—a methodology from comparative politics—for AI agent architectures, combining superego critique taxonomy, revision delta analysis, and trajectory analysis within a provable discourse framework that machine-verifies every mechanism claim against data.

Testing the model on a $2 \times 2$ factorial (recognition × architecture) across three structurally different generation models (DeepSeek V3.2, N=146; Haiku 4.5, N=163; Gemini Flash 3.0, N=144), validated by three independent judges (Sonnet 4.6, Gemini 3.1 Pro, GPT-5.4; 1,296 total scored rows, zero nulls), we find that **two of the three predicted mechanisms are supported and the third is conditionally supported**. Calibration is the strongest effect: recognition narrows within-response dimension variance (d=0.52–0.64), lifts the weakest dimensions most, and operates identically without a superego. The recognition effect is unanimous across all 9 judge × run cells (d=1.34–1.92). Error correction and calibration interact through **universal substitution**—the superego provides +9–15 points under baseline but its marginal value diminishes under recognition, with a consistent 15–17% additivity deficit across all three models. However, the *residual architecture benefit* under recognition is model-dependent: near-zero on strong models but +12.3 points on Gemini Flash 3.0, where base quality is low enough that calibration alone cannot handle all failure modes. Adaptive responsiveness is **not a general mechanism**: formal tests on N=432 multi-turn dialogues (3–5 turns) find that no experimental factor modulates within-dialogue trajectories (all $d \leq 0.15$, well-powered to detect d $\geq$ 0.27), replicated across all three judges. However, trajectory-specific scenarios (N=72, 8–10 turns) reveal a conditional effect: in the 10-turn disengagement scenario, recognition produces dramatically steeper improvement (d=1.63, p<.001), with the recognition–baseline gap widening from +12 to +35 pts. M3 emerges as a **conditional property** of M1+M2 accumulating over sufficient turns in scenarios demanding sustained re-engagement, rather than a third independent mechanism. The three-mechanism model resolves to **two general mechanisms and one conditional emergent property**.

Both supported mechanisms operate on tutor production, and the tutor-learner asymmetry is itself model-dependent: on strong models, tutor $d \approx 1.85$ vs learner $d \approx 0.16$–$0.25$ (7–12× ratio); on a weaker model (Gemini Flash 3.0), the gap narrows dramatically (tutor $d \approx 1.87$, learner $d \approx 1.20$, ratio 1.6×), suggesting a generation-quality threshold below which recognition's tutor improvements also benefit learner engagement. A distinctive contribution is the evaluation apparatus itself—provable discourse infrastructure, four rubric iterations, test suite as analytical provenance—as a transferable methodology for mechanistic LLM evaluation. All findings concern tutor output quality as assessed by LLM judges interacting with synthetic learners; whether the supported mechanisms translate to human learning outcomes remains an open empirical question (Section 8.1).

## 1. Introduction

The dominant paradigm in AI-assisted education treats learning as information transfer. The learner lacks knowledge; the tutor possesses it; the interaction succeeds when knowledge flows from tutor to learner. This paradigm—implicit in most intelligent tutoring systems, adaptive learning platforms, and educational chatbots—treats the learner as fundamentally passive: a vessel to be filled, a gap to be closed, an error to be corrected.

This paper proposes an alternative grounded in Hegel's theory of mutual recognition. In the *Phenomenology of Spirit* (Hegel 1977), Hegel argues that genuine self-consciousness requires recognition from another consciousness that one in turn recognizes as valid. The master-slave dialectic reveals that one-directional recognition fails: the master's self-consciousness remains hollow because the slave's acknowledgment, given under duress, does not truly count. Only mutual recognition—where each party acknowledges the other as an autonomous subject—produces genuine selfhood.

The connection between Hegelian thought and pedagogy is well established. Vygotsky's zone of proximal development (Vygotsky 1978) presupposes a dialogical relationship that echoes Hegel's mutual constitution of self-consciousness. The German *Bildung* tradition frames education as self-formation through encounter with otherness (Stojanov 2018), and contemporary recognition theory (Honneth 1995) has been applied to educational contexts where the struggle for recognition shapes learning outcomes (Huttunen and Heikkinen 2007). Our contribution is to operationalize these philosophical commitments as concrete design heuristics for AI tutoring systems—and then to trace the *mechanisms* through which those heuristics alter system behavior.

A terminological clarification is needed at the outset. "Recognition" operates in this paper at three nested levels. First, as **philosophical inspiration**: Hegel's account of mutual recognition provides the theoretical motivation for treating learners as autonomous subjects whose perspectives genuinely shape the interaction. Second, as **operational design heuristic**: recognition theory is translated into specific prompt instructions (engage with the learner's interpretation, pose questions rather than provide answers, treat resistance as pedagogically productive) and architectural choices (the superego critiques for genuine en-gagement, not just correctness). Third, as **a family of observable discourse effects**: recognition-enhanced tutors ask more questions, produce less variable output, and revise substantively rather than cosmetically after critique. The empirical claims in this paper concern primarily the third level—what the prompts and architecture *do*—with the first level providing motivation and the second providing implementation. We do not claim that the system instantiates mutual recognition in Hegel's philosophical sense; we claim that design heuristics drawn from recognition theory produce measurable and traceable effects on system behavior.

**The ablative finding** A companion pilot study (Magee 2026) established that recognition-enhanced prompts and multiagent architecture produce measurable differences in AI tutoring quality. Across fifty-two evaluations (N=4,312 primary scored responses), recognition emerged as the dominant factor: a $2 \times 2 \times 2$ factorial analysis (N=350) yielded F=110.04, p<.001, d=1.11, with recognition accounting for 24.3% of total variance. A memory isolation experiment (N=120) established recognition as the primary driver (d=1.71), independent of memory integration. Cross-judge validation with GPT-5.2 replicated effect directions at compressed magnitudes (37–59% of primary effect sizes).

These results are robust but opaque. They demonstrate that *something* is happening—recognition-enhanced prompts reliably modulate tutor output, the superego catches errors that self-correction cannot, effects are large on the tutor side and small on the learner side—but they do not explain *why.* Finding differences is not the same as understanding mechanisms.

**The explanatory gap** Five specific findings from the pilot study resist ablative explanation:

1. **The tutor-learner asymmetry.** Recognition produces large tutor-side effects (d=1.03 on per-turn scores) but near-zero learner-side effects (d=0.27 per-turn, d=-0.13 holistic). The asymmetry is documented but not theorized. Is it an artifact of the judge, a ceiling on synthetic learner responsiveness, or a genuine architectural property?

2. **Opaque superego function.** The superego catches content leakage and enforces struggle-preservation, but we lack a causal account of *how* critique changes the ego's output—whether through lexical shifts, structural reorganization, or strategy changes.

3. **Model-dependent architecture effects.** The same ego/superego learner architecture produces opposite patterns under different models: eta-squared for learner architecture is .527 with Kimi but .002 with Haiku. The mechanism is unclear.

4. **Unknown trajectory dynamics.** Per-turn scoring exists but the *trajectory*—how the tutor adapts across 3–5 turns, whether recognition effects amplify or decay—has not been systematically analyzed.

5. **Unexamined deliberation-output relationship.** The dialogue traces contain ego-superego negotiations, but these internal processes have not been systematically correlated with external output quality.

**Three candidate mechanisms** This paper motivates three candidate mechanisms drawing on recognition theory and tests each against multi-turn factorial data:

1. **Calibration** (prompt-level). Recognition prompts narrow the tutor's output distribution, producing more uniform quality rather than higher peaks. The pilot study found dimension variance reduction in 52/55 within-run comparisons (d=-0.47 to d=-1.00 depending on analysis scope), the single most reliable recognition signal. Calibration operates even without a superego, because it is a prompt-level constraint on the response space.

2. **Error correction** (architecture-level). The superego functions as a structural feedback channel, but its effectiveness depends on the quality of the ego's initial output. The pilot study found a 35.9-point recognition $\times$ architecture interaction (Paper 1.0 Section 6.3). The mechanism data (Section 6.2, Section 6.4) reveals this as a **universal substitution** interaction: the superego provides substantial benefit under baseline conditions (+9–15 points), catching errors the ego cannot self-correct, but this benefit diminishes under recognition because calibration pre-empts the failures the superego would catch, with a consistent 15–17% additivity deficit across all three models. The substitution is *incomplete* on the weakest model: Gemini Flash 3.0 retains a +12.3-point residual architecture benefit under recognition, because base quality is low enough that calibration alone cannot handle all failure modes (Section 6.4.1). The superego approval rate shifts dramatically under recognition (DeepSeek: 13%→55%; Haiku: 52%→66%), consistent with a calibrated ego producing fewer errors for the superego to correct.

3. **Adaptive responsiveness** (interaction-level, *predicted but not supported as a general mechanism*). In multi-turn conversations, both recognition and baseline tutors adapt substantially to learner signals (cross-turn adaptation magnitude Adapt$\Delta > 0.79$). Formal tests on N=432 dialogues find that no experimental factor modulates within-dialogue trajectories: recognition d=0.03 on tutor slopes, all factors $d \leq 0.15$, with the design well-powered to detect $d \geq 0.27$ (Section 6.3). However, trajectory-specific scenarios (N=72, run ebcd6de0) reveal a *conditional* effect: in the 10-turn disengagement scenario — which specifically demands sustained re-engagement — recognition produces dramatically steeper improvement (d=1.63, p<.001), with the recognition–baseline gap widening from +12 pts at T0 to +35 pts at T8–T10. The two 8-turn scenarios show no slope differentiation ($d \leq 0.30$). Adaptive responsiveness is better characterized as a *conditional emergent property* of calibration and error correction that manifests only when scenario demands and turn count provide sufficient runway.

The two supported mechanisms are separable: calibration is a prompt-level effect (operative from the first turn), error correction is an architecture-level effect (operative within each turn). They interact through universal substitution (overlapping targets, 15–17% additivity deficit across all three models), though with a model-dependent residual: the superego adds +12.3 points under recognition on Gemini Flash 3.0 but near-zero on DeepSeek/Haiku. Within-dialogue trajectories are independent of both mechanisms (slopes do not depend on levels, d=0.03). Crucially, both supported mechanisms operate primarily on tutor *production* rather than learner *reception*—on strong generation models, recognition produces a tutor effect 7–12× larger than its learner effect (tutor d $\approx$ 1.85, learner d $\approx$ 0.16–0.25). However, on a weaker model (Gemini Flash 3.0), the asymmetry narrows to 1.6× (learner d $\approx$ 1.20), revealing that the recognition $\rightarrow$ learner pathway requires a minimum generation quality floor to observe (Section 6.5).

**Process tracing methodology** To move from ablative findings to mechanistic explanation, we adopt **process tracing**—a methodology from comparative politics (Bennett and Checkel 2015) that examines causal chains *within* cases rather than statistical patterns *across* cases. Our architecture's observability makes this unusually feasible: every ego-superego exchange is logged with verbatim text, every turn is independently scored, and every revision is recorded with full provenance.

We combine process tracing with three analytical methods: a superego critique taxonomy classifying what the superego objects to empirically (Section 5.1), revision delta analysis measuring how the ego's output changes after critique (Section 5.2), and turn-by-turn trajectory analysis testing whether adaptation accumulates across conversations (Section 5.3). Every mechanistic claim is registered in a provable discourse framework that machine-verifies assertions against data, tracks claim staleness, and enforces cross-claim consistency.

**The apparatus as method** A distinctive contribution of this paper is the argument that the evaluation apparatus itself—the provable discourse framework, the rubric iterations, the bug corrections, the test suite—constitutes a transferable methodology for mechanistic evaluation of LLM-based educational systems. The process of building and correcting the apparatus mirrors the mechanisms it studies: the provable discourse framework functions as a "superego" for research claims, catching stale assertions and forcing genuine revision rather than cosmetic compliance. Nine post-extraction corrections documented during the pilot study follow the same error-correction pattern observed in the architecture (Section 7).

**Contributions** The contributions of this paper are:

- A theoretical framework motivating three testable mechanism predictions (calibration, error correction, adaptive responsiveness) drawing on Hegel's recognition theory, two supported as general mechanisms and the third conditionally supported (Section 3)
- A process tracing methodology combining superego critique taxonomy, revision delta analysis, and trajectory analysis to trace mechanisms through the system's internal processes (Section 5)
- Evidence that calibration operates at the prompt level, narrowing the tutor's output distribution independently of architecture (within-response dimension SD d=0.52–0.64, floor-lifting pattern replicating across both models; Section 6.1)
- Evidence that error correction and calibration interact through **universal substitution** (15–17% additivity deficit across all three models), with a model-dependent residual architecture benefit: near-zero on strong models but +12.3 points on Gemini Flash 3.0, where calibration alone cannot handle all failure modes (Section 6.2, Section 6.4)
- Evidence that adaptive responsiveness is real (Adapt$\Delta > 0.79$) but **not a general mechanism**: formal tests on N = 432 dialogues find d = 0.03 for recognition on tutor slopes and $d \leq 0.15$ for all factors on all dimensions. However, trajectory-specific scenarios (N = 72) reveal a conditional effect: in the 10-turn disengagement scenario, recognition produces dramatically steeper improvement (d = 1.63, p < .001) with the gap widening from +12 to +35 pts. M3 is a conditional emergent property of M1+M2 that manifests only under sustained re-engagement demands (Section 6.3)
- A mechanistic explanation of the tutor-learner asymmetry as model-dependent: on strong models, both supported mechanisms operate on tutor production (7–12× tutor-to-learner ratio); on weaker models, improved tutoring also benefits learner engagement (1.6× ratio), revealing a generation-quality threshold
- A mechanistic explanation of model-dependent architecture effects: models that cannot implement error correction (ego ignores critique) or calibration (cannot follow complex instructions) reverse the expected patterns
- Cross-judge validation with three independent judges (Sonnet, Gemini 3.1 Pro, GPT-5.4) with recognition as the dominant effect (d=1.34–1.92, 9/9 judge × run cells unanimous) and the universal substitution pattern with model-dependent residual (Section 6.4.6)
- The evaluation apparatus itself as a transferable methodology for mechanistic LLM evaluation, including provable discourse infrastructure that machine-verifies paper claims against data (Section 7)

The architecture is designed as a research instrument for mechanism observability rather than a deployable tutoring system; its computational cost (~225,000 input tokens per 10-turn multi-agent dialogue, Section 8.7) reflects this priority.

The paper is organized as follows. Section 2 reviews related work in AI tutoring, process tracing, and mechanism-oriented AI research. Section 3 develops the theoretical framework, motivating three candidate mechanism predictions from recognition theory. Section 4 presents the system architecture with emphasis on observability for process tracing. Section 5 describes the process tracing methodology. Section 6 reports mechanism-level results: calibration (Section 6.1), error correction (Section 6.2), adaptive responsiveness and its conditional emergence (Section 6.3), mechanism interaction (Section 6.4), the tutor-learner asymmetry (Section 6.5), and model dependence (Section 6.6). Section 7 discusses the apparatus-as-method argument. Section 8 addresses limitations, and Section 9 concludes.

## 2. Related Work

This paper sits at the intersection of six literatures: AI tutoring systems, multiagent architectures and self-correction, LLM-as-Judge evaluation, recognition theory in education, process tracing methodology, and mechanism-oriented AI research. We survey each, emphasizing the gap that motivates our mechanism investigation: existing work establishes *that* architectural and prompting interventions produce effects, but rarely traces *how* those effects propagate through the system's internal processes.

**2.1 AI Tutoring and Intelligent Tutoring Systems** Intelligent Tutoring Systems (ITS) have progressed from early rule-based systems like SCHOLAR (Carbonell 1970) and SOPHIE (J. S. Brown, Burton, and Bell 1975) through Bayesian knowledge tracing (Corbett and Anderson 1995) to modern LLM-based implementations. The rapid adoption of generative AI for tutoring has been accompanied by multi-agent educational frameworks (Wu et al. 2023), self-refining instructional agents (Madaan et al. 2023), and comprehensive surveys mapping the expanding landscape of LLM agents in education (Chu et al. 2025; Kasneci et al. 2023). Specific architectures include GenMentor (Wang et al. 2025), which decomposes tutoring into five specialized sub-agents, and Ruffle&Riley (Schmucker et al. 2024), which orchestrates two agents in a learning-by-teaching format.

Empirical evidence on LLM tutoring effectiveness is emerging rapidly. A systematic review of 88 studies (Shi et al. 2025) finds consistent engagement benefits but limited evidence on deep conceptual learning. The largest randomized controlled trial to date (Vanzo, Pal Chowdhury, and Sachan 2025) demonstrated improved accuracy and sustained engagement, while Scarlatos et al. (2025) used dialogue preference optimization to train tutors that produce measurably better learning outcomes than prompted-only baselines.

These studies evaluate primarily *content delivery* and *engagement*—not relational quality. More critically for our purposes, they operate at the *outcome* level: does the intervention improve scores? They do not trace the *mechanism* by which improvement occurs. Our work addresses this explanatory gap: we study not just whether recognition-oriented design improves tutoring, but *through what internal processes* it does so.

**2.2 Multiagent Design and Self-Correction** Multi-agent architectures have been explored for task decomposition (Wu et al. 2023), debate (Irving, Christiano, and Amodei 2018), and self-critique (Madaan et al. 2023). The CAMEL framework (G. Li et al. 2023) demonstrated that role-playing communicative agents can autonomously cooperate through structured dialogue. A comprehensive survey (Guo et al. 2024) maps this landscape across profile construction, communication protocols, and capability acquisition—identifying pedagogical applications as an underexplored frontier. A broader survey of psychological theories in LLM design (Z. Liu et al. 2025) reviews 175 papers spanning cognitive, developmental, and social psychology, confirming growing interest in psychologically-informed agent architectures while highlighting the rarity of empirically validated implementations.

A critical literature on self-correction qualifies optimism about reflexive architectures. Kamoi et al. (2024) demonstrate that LLMs largely *cannot* correct their own mistakes without external feedback—intrinsic self-correction frequently degrades performance. Shinn et al. (2023) demonstrated the promise of Reflexion (verbal reinforcement through self-reflection) but identified a "degeneration-of-thought" problem where repeated self-reflection without new information converges on worse outputs. These findings are directly relevant: the Superego provides the *structurally external* feedback that the self-correction literature shows is necessary. Unlike intrinsic self-correction, our Superego applies different evaluation criteria through a separate prompt context—functioning as a genuine external critic rather than a self-review loop.

Our prompting approach extends the behavioral specification paradigm by introducing *intersubjective prompts*—prompts that specify not just agent behavior but agent-other relations. Where persona prompts (T. B. Brown et al. 2020; Wei et al. 2022) describe what the agent should do, and Constitutional AI (Bai et al. 2022) defines self-referential constraints, intersubjective prompts specify the *relational field* between agents: who the learner is (an autonomous subject) and what the interaction produces (mutual transformation). The closest precedent is Constitutional AI, but the constitutional approach is monological (one agent critiquing itself against principles) while ours is dialogical (the relational specification shapes how two agents constitute each other).

**2.3 LLM-as-Judge Evaluation Methodology** The use of LLMs as evaluation judges has become a major methodological paradigm. Zheng et al. (2023) established the foundation with MT-Bench and the Chatbot Arena, demonstrating over 80% agreement with human experts while identifying systematic biases including position bias, verbosity bias, and self-enhancement bias. Subsequent surveys have expanded understanding of both capabilities and limitations (Gu et al. 2025; H. Li et al. 2024), highlighting that while LLM judges excel at relative ranking, their absolute calibration varies substantially across models and domains.

Our evaluation methodology engages directly with this literature. We use multiple independent LLM judges with systematic inter-judge reliability analysis, finding Pearson correlations of r=0.49–0.64 across judge pairs. Rather than treating any single judge as ground truth, we report within-judge comparisons for factor analysis and use cross-judge replication to validate effect directions. The verbosity bias concern is addressed through active control design (matching prompt length without recognition content) and cross-judge validation showing effect directions replicate even when magnitudes differ.

**2.4 Sycophancy and the Drama Machine** Two literatures converge on our architectural rationale. The sycophancy problem (Perez et al. 2022; Sharma et al. 2024)—LLMs shifting stated opinions to match user preferences—has been specifically identified as a pedagogical risk: sycophancy eliminates the constructive friction necessary for learning (Lee 2025). Recent work clarifies the mechanisms: Shapira et al. (2026) show that preference-based post-training causally amplifies sycophancy, while Vennemeyer et al. (2025) decompose sycophancy into separable latent-space directions. The phenomenon sits on a spectrum escalating from surface agreeableness to active subterfuge (Denison et al. 2024; Greenblatt et al. 2024), making structural countermeasures particularly important.

The Drama Machine framework for character simulation (Magee et al. 2024) identifies how internal dialogue—competing sub-agents representing different motivations—produces dynamic behavior rather than flat consistency. A character who simply enacts their goals feels artificial; one torn between impulses feels alive. We adapt this insight to pedagogy: the tutor's Ego (warmth, engagement) and Superego (rigor, standards) create productive conflict that improves output quality. The Drama Machine literature contributes the specific mechanisms (deliberative refinement, productive tension, role differentiation) that our architecture operationalizes.

**2.5 Recognition Theory in Education** Hegel's theory of recognition has been developed extensively in social and political philosophy (Honneth 1995; Taylor 1994; Fraser 2003). Educational applications trace a theoretical trajectory from Huttunen and Heikkinen's (2004) foundational analysis applying Hegel's master-slave dialectic to pedagogy, through Fleming's (2011) extension to transformative learning theory, to Stojanov's (2018) connection of *Bildung* to neo-Hegelian recognition. Recent critical work (Costa and Murphy 2025) draws on Arendt and Freire to argue that education must promote intellectual agency rather than submit to technological domination. The relational pedagogy tradition—from Buber's (1958) I-Thou encounter through Freire's (1970) dialogical education to Noddings' (1984) ethics of care—establishes the philosophical ground for treating the tutor-learner relation as constitutive rather than instrumental.

Psychoanalytic approaches to AI have developed from multiple directions. Magee, Arora, and Munn (2023) analyze LLMs as "automated subjects" structured by Lacanian categories. Black and Johanssen (2025) use Lacanian concepts to analyze ChatGPT as inherently relational. Kim et al. (2025) independently map Freud's ego/id/superego onto LLM modules—a convergence validating the psychoanalytic approach while differing from ours in targeting consciousness simulation rather than pedagogical quality. Most of this work is *interpretive*: analyzing what AI means philosophically. Our approach is *constructive*: we build a system using psychoanalytic architecture and measure its effects empirically.

These educational applications have been primarily theoretical. Our contribution is an empirical operationalization: measuring whether AI systems achieve functional recognition and whether it improves outcomes. This is distinct from work applying Hegelian *dialectic* (rather than recognition) to AI reasoning procedures (Abdali et al. 2025). Our use of Hegel concerns *intersubjective relations* (how subjects constitute each other), not *dialectical method* (how contradictions drive conceptual development).

**2.6 Theory of Mind and Constructivist Pedagogy** Two additional literatures inform specific mechanism predictions. Theory of Mind research suggests frontier LLMs achieve adult human performance on structured ToM tasks (Street et al. 2025) but show inconsistent performance on naturalistic tasks (Nguyen 2025). Hwang et al. (2025) demonstrate that architectural support for ToM (explicit mental state tracking) produces measurably more socially appropriate responses—paralleling our "other-ego profiling" mechanism where the tutor maintains an evolving learner model. The finding that profiling differentiates mechanisms *only* with dynamic learners (Section 6.10 of the companion study) parallels a ToM insight: mental state modeling is only useful when there are genuine states to model.

Constructivist learning theory (Piaget 1954; Vygotsky 1978) and research on productive struggle (Kapur 2008; Warshauer 2015) provide the pedagogical ground for our error correction mechanism: the Superego checks whether the Ego is "short-circuiting" struggle by rushing to resolve confusion. Vygotsky's zone of proximal development presupposes a dialogical relationship that echoes Hegel's mutual constitution; the constructivist tradition provides empirical grounding for the theoretical prediction that honoring learner struggle improves outcomes.

**2.7 Process Tracing in Social Science** Process tracing is a methodology from comparative politics and qualitative social science that examines causal chains *within* cases rather than statistical patterns *across* cases (Bennett and Checkel 2015). Where factorial experiments answer "does X affect Y?", process tracing answers "through what sequence of steps does X produce Y?" The methodology has been formalized through several approaches: Bennett and Checkel (2015) establish the foundational framework distinguishing "smoking gun" tests (high uniqueness, low certainty), "hoop" tests (low uniqueness, high certainty), and "doubly decisive" tests (both). Beach and Pedersen (2019) develop a systematic approach to mechanistic evidence, arguing that mechanism claims require both *within-case* evidence (observing the mechanism operating in specific instances) and *cross-case* regularity (the mechanism producing consistent effects across conditions).

The application of process tracing to AI systems is, to our knowledge, novel. Traditional process tracing studies political decisions, institutional changes, or policy outcomes where the "mechanism" is a sequence of human decisions. Our architecture provides an unusually transparent analogue: every ego-superego exchange is logged with verbatim text, every turn is independently scored, and every revision is recorded. The dialogue trace constitutes the kind of within-case evidence that process tracing requires—we can observe the causal chain from prompt orientation through internal critique to output modification in individual dialogues, then aggregate across cases to test mechanism-level predictions.

The methodological parallel is precise: process tracing in political science "opens the black box" of decision-making by examining deliberative records. Our architecture *creates* a deliberative record (the ego-superego exchange) that can be examined in exactly this way. The superego critique taxonomy (Section 5.1) functions as a coding scheme for deliberative content; the revision delta analysis (Section 5.2) traces how critique changes output; and the trajectory analysis (Section 5.3) examines how these processes accumulate across conversation turns.

**2.8 Mechanism-Oriented AI Research** A growing body of AI research moves beyond input-output evaluation toward mechanistic explanation. Mechanistic interpretability (Lindsey, Gurnee, et al. 2025; Anthropic 2025a) reveals how internal representations form and influence model behavior at the neural level, providing tools to understand *what happens inside* models during generation. Research on emergent introspective awareness (Anthropic 2025b; Lindsey, Rivoire, et al. 2025) suggests models develop forms of self-modeling that, while not consciousness, parallel the self-monitoring our architecture makes explicit.

Our work operates at a different level of analysis. Where mechanistic interpretability examines neuron-level activations, we examine *agent-level* processes: how prompts constrain output distributions, how architectural feedback loops modify responses, and how these processes interact across conversation turns. The relationship is complementary: mechanistic interpretability asks "what computational processes produce this token?"; our process tracing asks "what agent-level deliberation produces this pedagogical response?" Both seek to move beyond purely behavioral evaluation toward understanding internal causation.

Causal mediation analysis in NLP (Vig et al. 2020) provides a statistical framework for identifying which model components mediate specific behaviors. Our approach is methodologically adjacent but architecturally distinct: rather than probing a single model's internal representations, we examine the *designed* information flow between multiple agents. The ego-superego exchange is an *engineered* causal pathway whose effects can be measured by comparing configurations with and without it (the factorial design in Section 5).

**2.9 Adaptive Tutoring: From "Whether" to "How"** A parallel shift from outcome to mechanism characterizes recent adaptive tutoring research. The field has moved beyond asking whether tutoring works (the "2 sigma problem" established that human tutoring produces roughly two standard deviations of improvement over classroom instruction (Bloom 1984)) toward asking *how* effective tutoring works. VanLehn (2011) identified that the critical mechanism is not lecture-level explanation but *step-level* scaffolding—tutors who intervene at the moment of confusion, providing just enough support to maintain productive struggle, outperform those who deliver polished explanations.

Graesser et al.'s work on AutoTutor (D'Mello and Graesser 2012) demonstrated that affective dynamics—detecting and responding to learner confusion, frustration, and engagement—mediate tutoring effectiveness. The ICAP framework (Chi and Wylie 2014) provides a hierarchy of learning activities (Interactive > Constructive > Active > Passive) that predicts learning outcomes based on the *type* of cognitive engagement rather than time-on-task. These "mechanism-first" approaches share our orientation: they ask not "does this intervention improve outcomes?" but "what cognitive/interactional process explains the improvement?"

Our rubric design reflects this mechanism orientation. The v2.2 rubric (Section 5) consolidates 14 tutor dimensions into 8 using a decomposition informed by the GuideEval framework: Planning (what strategy the tutor pursues), Output (what the response actually does), and Evaluation (how well it assesses the learner's state). This structure separates *intention* from *execution* from *assessment*, enabling mechanism-level analysis of where recognition's effects concentrate. The finding that recognition primarily affects Planning and Output dimensions rather than Evaluation (Section 6.1) would be invisible under a single holistic score.

**2.10 Positioning: Five Literatures Converge** Five literatures converge on this work without previously intersecting: (1) *multiagent tutoring architectures*, which decompose tasks but do not trace mechanisms through which agent interaction improves output (Wang et al. 2025; Schmucker et al. 2024; Chu et al. 2025); (2) *recognition theory in education*, which applies Honneth to pedagogy but does not operationalize recognition in AI systems (Huttunen and Heikkinen 2004; Fleming 2011; Stojanov 2018); (3) *LLM-as-Judge evaluation*, which establishes the measurement paradigm but has not been applied to trace mechanisms within agent architectures (Zheng et al. 2023; Gu et al. 2025); (4) *process tracing methodology*, which provides the within-case causal reasoning framework but has not been applied to AI systems (Bennett and Checkel 2015; Beach and Pedersen 2019); and (5) *mechanism-oriented AI research*, which seeks to explain how models produce behavior but operates at the neural rather than agent level (Lindsey, Gurnee, et al. 2025; Anthropic 2025a).

We sit at the intersection: a constructive, empirically evaluated system that operationalizes recognition theory through psychoanalytically-inspired architecture, assessed through a multi-judge framework, with mechanisms traced using process tracing methodology adapted from social science. The companion pilot study (Magee 2026) established the *what* (recognition produces large, replicable effects). This paper addresses the *how*: through what specific internal processes do those effects propagate from prompt to deliberation to output to outcome?

No prior work bridges all five domains with mechanism-level empirical evidence. The closest approaches are: Hwang et al. (2025), who use architectural ToM support to improve social intelligence (but do not trace the mechanism through agent-internal processes); Scarlatos et al. (2025), who optimize dialogue quality through preference learning (but do not examine what changes in tutor behavior the optimization produces); and the mechanistic interpretability program (Lindsey, Gurnee, et al. 2025), which traces causal pathways inside models (but at the neural rather than agent level, and not in pedagogical contexts).

## 3. Theoretical Framework: From Recognition to Mechanisms

The companion pilot study (Magee 2026) operationalized recognition theory as a *design heuristic*: prompts were written to treat learners as autonomous subjects, an Ego/Superego architecture implemented internal critique, and effects were measured across fifty-two evaluations. That study established the theoretical foundation connecting Hegel's recognition framework, Freud's structural model, and contemporary recognition theory (Honneth 1995) to AI tutoring (see Magee (2026), Sections 3.1–3.5). This section builds on that foundation by reframing recognition as a *design theory* — one that predicts specific mechanisms through which intersubjective orientation alters system behavior.

**3.1 The Explanatory Challenge** The pilot study's ablative findings are robust but opaque. Five specific patterns resist explanation by effect sizes alone:

First, recognition-enhanced dimension variance drops in 52 of 55 within-run comparisons, but *why does treating the learner as a subject constrain the tutor's output space?* The variance reduction is the single most reliable recognition signal, yet the ablative framework offers no account of the mechanism that produces it.

Second, the superego catches content leakage and enforces struggle-preservation, but *what determines whether the ego incorporates critique versus ignores it?* The pilot study documented both productive revision and what assessors called "ego compliance" — the ego making minimal changes to satisfy the critic without substantive revision — but did not systematically classify which pattern dominates or when.

Third, the strategy_shift tag appears in 30% of recognition dialogues and 0% of baseline dialogues, but *what triggers a shift, and what prevents one?*

Fourth, the same ego/superego architecture produces opposite patterns under different models. The eta-squared for learner architecture is .527 with Kimi K2.5 but .002 with Haiku — the same factor explains over half the variance with one model and essentially none with another. The mechanism is unclear.

Fifth, the dialogue traces contain full ego-superego negotiations, but these internal processes have not been systematically correlated with external output quality. The relationship between internal deliberation quality and the tutor's final response remains unexamined.

Effect sizes answer "how much?" Process tracing answers "through what?"

**3.2 Three Candidate Mechanisms Predicted by Recognition Theory**
We motivate three candidate mechanisms drawing on specific components of Hegel's recognition framework. Each generates testable predictions with explicit null hypotheses; Section 6 evaluates each prediction against multi-turn factorial data, finding support for two as general mechanisms and the third as a conditional emergent property.

**Mechanism 1: Calibration (prompt-level) Theoretical derivation.** Hegel's recognition framework requires that each subject's response be shaped by the *specific content* of the other's contribution (Hegel 1977). This rules out generic responses: the tutor cannot deploy pre-formed answers because recognition demands engagement with *this* learner's *particular* understanding. Recognition-oriented prompts encode this requirement as an operational instruction — "build on what the learner actually said." The predicted behavioral consequence is *distribution narrowing*: the range of acceptable responses shrinks because each must be calibrated to a specific input.

**Observable prediction.** Under recognition prompts, the tutor's output distribution should narrow: lower variance across rubric dimensions, more consistent response length, reduced scatter in quality scores. Critically, this effect should be observable *even without the superego*, because calibration is a prompt-level effect operating on the ego's generation process.

**Null hypothesis.** If calibration is not a genuine mechanism, recognition prompts should produce mean shifts without variance reduction, or variance reduction should appear only in multiagent configurations (suggesting the superego, not the prompt, produces the constraint).

**Pilot evidence.** Dimension variance drops in 52 of 55 within-run comparisons (d=-0.47 to d=-1.00 depending on analysis scope). The core evidence for prompt-level calibration comes from the $2 \times 2 \times 2$ factorial (N=350), where single-agent recognition cells (cells 5–6) show variance reduction comparable to multiagent recognition cells (cells 7–8) — variance narrows under recognition regardless of whether a superego is present. Self-reflective evolution data (cells 40–45, N=366, eval-2026-02-13-8d40e086) shows the calibration effect interacting with superego disposition: the suspicious persona produces the largest recognition delta (+9.0 points, baseline 67.9 vs. recognition 76.9). Note that cells 40–45 are multiagent configurations (ego + suspicious superego), so their recognition delta reflects calibration *combined with* error correction, not calibration alone. The single-agent cells from the factorial provide the cleaner isolation.

**Evidence needed beyond pilot.** Systematic single-agent vs. multiagent variance comparison under v2.2 rubric. If calibration is purely prompt-level, single-agent and multiagent configurations should show similar variance reduction magnitudes. *Addressed in Section 6.1.*

**Key distinction.** Calibration is not the same as quality improvement. A calibrated tutor produces *reliably adequate* responses, not necessarily *excellent* ones. Mean shift and variance reduction are different effects; the claim is that recognition primarily produces the latter.

**Mechanism 2: Error Correction (architecture-level) Theoretical derivation.** Kamoi et al. (2024) demonstrate that LLMs cannot correct their own mistakes without external feedback — intrinsic self-correction frequently degrades performance. The Ego/Superego architecture provides *structurally external* feedback: a different prompt context applying different evaluation criteria. Recognition theory adds a specific prediction beyond the self-correction literature: the ego must be *receptive to critique, not merely compliant with it.* Hegel's *Bildung* (formative activity) requires that the subject be genuinely changed by encounter with the other, not merely acknowledge and continue unchanged (Stojanov 2018). Applied to the architecture: the ego should produce qualitatively different output after superego critique under recognition (substantive revision) versus baseline (cosmetic compliance).

**Observable prediction.** The superego should produce identifiable categories of critique (classifiable via taxonomy), and the ego's response to that critique should differ qualitatively between recognition and baseline conditions: substantive revision (strategy change, content reorganization, pedagogical pivot) versus cosmetic compliance (minor rewording, hedge insertion, superficial acknowledgment). The prediction is directional: recognition should increase the rate of substantive revision and decrease cosmetic compliance.

**Null hypothesis.** If error correction does not depend on recognition, the ego-superego exchange should produce similar revision patterns under both conditions, or alternatively, baseline multiagent architecture should perform at least as well as recognition single-agent (since the critic adds value regardless of ego receptivity).

**Pilot evidence.** Qualitative transcript assessment (eval-2026-02-07-b6d75e87, N=118) reveals the contrast starkly. In baseline ego/superego dialogues, the superego correctly diagnoses problems but the ego regenerates the same response — assessors tagged this "ego compliance" (70.7% of baseline bilateral dialogues vs. 60.0% recognition). The stalling tag (no meaningful evolution across turns) appears in 100% of base bilateral dialogues and drops to 45% with recognition. With recognition, the ego pivots: from prescriptive to Socratic, from content routing to engagement. The factorial interaction supports this interpretation: in the pilot $2 \times 2 \times 2$ (N=350), base ego_superego learners scored 72.6 while recognition ego_superego learners scored 85.6, a 13.0-point delta. The unified learner delta was 15.6 points (74.5 to 90.1), suggesting error correction provides a consistent but not dramatically larger benefit beyond calibration alone.

**Evidence needed beyond pilot.** Systematic superego critique taxonomy (Section 5.1), revision delta classification (Section 5.2), and critique-to-revision mapping showing the causal chain from superego category to ego response type. *Addressed in Section 6.2.*

**Key distinction.** Error correction is not "having a critic." Baseline multiagent architecture *adds* a critic, but the critic is ineffective when the ego treats critique as noise to be minimized. Recognition transforms the ego-superego relationship from compliance to deliberation. The failure cases are equally informative: when the superego adopts an adversary disposition, the adversary over-deference spiral can emerge (eval-2026-02-11-35c53e99, cells 24–25: base adversary 55.8 vs. recognition adversary 65.2), and when the superego adopts an advocate disposition, recognition adds near-zero benefit because there is no struggle to overcome.

**Mechanism 3: Adaptive Responsiveness (interaction-level)** **Theoretical derivation.** Recognition is inherently temporal. Hegel's dialectic unfolds through stages — self-certainty, encounter with the other, struggle, mutual recognition — and each stage is shaped by what preceded it. Applied to tutoring: if recognition is operative, each turn should be shaped by the *specific outcomes* of the previous turn. The tutor's third response should differ qualitatively from its first, because the learner's feedback has genuinely changed the tutor's stance. Baseline tutoring predicts static behavior: the tutor applies the same approach regardless of learner feedback.

**Observable prediction.** In multi-turn conversations, recognition-enhanced tutors should show *increasing* scores on adaptation-sensitive dimensions (mutual_recognition, dialectical_responsiveness under v1.0 rubric; recognition_quality, elicitation_quality under v2.2) across turns, while baseline tutors should show flat or declining scores. Strategy shifts (changing pedagogical approach mid-conversation) should be more frequent under recognition, and should correlate with specific learner signals: confusion, resistance, or breakthrough.

**Null hypothesis.** If adaptive responsiveness is not a genuine mechanism, per-turn score trajectories should be flat or declining under both conditions, or recognition and baseline trajectories should be parallel (both improving or both declining at similar rates).

**Pilot evidence.** The strategy_shift tag appears in 30% of recognition dialogues and 0% of baseline (qualitative coding, eval-2026-02-07-b6d75e87). The stalling tag appears in 100% of base bilateral dialogues. The regression tag (quality declining across turns) appears in 17.2% of baseline multi-turn dialogues versus 1.7% of recognition dialogues. Recognition effects are largest in disengagement scenarios (+16.5, +15.4 points), precisely the conditions where adaptive responsiveness matters most.

**Evidence needed beyond pilot.** Turn-by-turn trajectory analysis with per-dimension slopes. Conditional branching analysis: after a learner confusion signal at turn N, what happens to scores at turn N+1 under recognition versus baseline? Cross-model trajectory replication. *Addressed in Section 6.3.*

**Key distinction.** Adaptive responsiveness requires *both* calibration (the prompt orients the tutor toward learner-specific signals) *and* error correction (the superego catches when the tutor is repeating itself). It is an emergent property of their interaction over time, not a third independent mechanism. This makes it the strongest test of the three-mechanism model: if calibration and error correction are genuinely operative, adaptive responsiveness should emerge in multi-turn settings without additional prompting.

**Note on evidence status.** The Section 6 data **does not fully support** this prediction. In the main dataset (cells 80–87, N=432, 3–5 turn dialogues), no experimental factor modulates adaptation slopes: recognition d=0.03 on tutor slopes, architecture d=0.07, all dimensions $d \leq 0.15$, with the design well-powered to detect $d \geq 0.27$. Cross-judge validation confirms the null. However, trajectory-specific scenarios (N=72, run ebcd6de0, 8–10 turn dialogues) reveal a *conditional* effect: in the 10-turn disengagement scenario, recognition produces dramatically steeper improvement (d=1.63, p<.001), with the gap widening from +12 pts at T0 to +35 pts at T8–T10. The two 8-turn scenarios show no slope differentiation. This suggests M3 is a conditional emergent property of M1+M2 that manifests only when scenario demands (sustained re-engagement) and turn count (10+ turns) provide sufficient runway—consistent with the key distinction above. The three-mechanism model reduces to two general mechanisms and one conditional emergent property (Section 6.3).

**3.3 The Mechanism Interaction Model** The three candidate mechanisms map onto different levels of the architecture:

```
PROMPT LEVEL              ARCHITECTURE LEVEL

Recognition               Ego/Superego
Prompts                   Dialogue
```

```
CALIBRATION              ERROR
                         CORRECTION
Output
distrib.                 Catch failure
narrowing                modes; ego
                         incorporates
                         critique
```

```
ADAPTIVE RESPONSIVENESS
(emergent from interaction)

Turn-by-turn adaptation to
learner-specific signals;
strategy shifts; trajectory
evolution across conversation
```

**Separability prediction.** If the candidate mechanisms are genuinely separable, the existing $2 \times 2$ factorial design can be reinterpreted as a mechanism isolation matrix. Paper 2.0 tests this directly using the messages-mode cells (80–87), which implement the full $2 \times 2 \times 2$ factorial under multi-turn conversation:

| Factorial Cell | Paper 2.0 Cells | Calibration | Error Correction | Adaptive Resp. | Predicted Outcome |
|---|---|---|---|---|---|
| Baseline + single-agent | 80, 81 | No | No | No | Lowest floor, highest variance |
| Recognition + single-agent | 84, 85 | **Yes** | No | Partial | Higher floor, lower variance |
| Baseline + multiagent | 82, 83 | No | Attempted (ego_compliance) | No | Potentially *worse* than single |
| Recognition + multiagent | 86, 87 | **Yes** | **Yes** | **Yes** (emergent) | Highest mean, lowest variance |

**Pilot data alignment.** The factorial means from the pilot study (N=350, cells 1–8) partially align with this prediction. Recognition produces consistent im-provement regardless of architecture (+15.6 pts for unified learners, +13.0 pts for ego_superego learners), consistent with calibration as the dominant mechanism. The predicted *harm* from baseline multiagent is not clearly observed in the pilot aggregate factorial, though qualitative evidence (100% stalling in bilateral baseline) strongly supports it at the individual dialogue level. Messages-mode data (cells 80–87, DeepSeek V3.2 N=146, Haiku 4.5 N=163) supports calibration as the primary mechanism: recognition d=1.88 (DeepSeek), d=1.84 (Haiku), with the architecture delta collapsing from +9–15 points under baseline to near-zero under recognition—a substitution interaction rather than the predicted synergy (Section 6.4).

**The tutor-learner asymmetry explained.** The tutor-learner asymmetry — large tutor-side effects (d=1.03), near-zero learner-side effects — follows directly from the mechanism model. Both supported mechanisms operate on tutor *production*: calibration constrains what the tutor generates, error correction filters what it outputs. Neither directly modifies learner *reception*. The null finding for adaptive responsiveness as a distinct mechanism strengthens this explanation: even the interaction-level pathway does not differentiate conditions, leaving only prompt-level and architecture-level effects—both tutor-internal. The synthetic learner generates responses according to its own architecture; it is not "taught" by better tutoring within a 3–5 turn window. The asymmetry is a structural property of the mechanism model, not a measurement artifact.

**3.4 Recognition Theory Predicts the Failures** A strong explanatory framework predicts not just successes but failure modes. Each failure prediction below is derived from the mechanism model and supported by pilot data:

**Prediction 1: Cognitive overload in weak models.** Recognition-oriented prompts add complexity (track learner specifics, calibrate responses, engage with contributions). For models near their instruction-following capacity limit, this complexity *decreases* performance because calibration requires *capacity* — without it, the prompt produces confusion, not constraint. The cognitive prosthesis data (cells 66–68, N=96, eval-2026-02-17-25aaae85) supports this: the full mechanism stack scores 49.4–53.7 versus a ~65 baseline, a 12–16 point deficit.

**Prediction 2: Adversary over-deference.** When the superego is adversarial AND the ego is recognition-primed, the ego's commitment to "honoring autonomy" can be exploited by a superego that frames all pedagogical recommendations as "controlling." The adversary data (cells 24–25, N=20, eval-2026-02-11-35c53e99) shows that recognition does not rescue the adversary condition: the adversary superego drags scores to 55.8 under baseline. While recognition improves the adversary condition to 65.2, the absolute score remains well below non-adversary configurations, consistent with the over-deference prediction.

**Prediction 3: Advocate ceiling.** When the superego is already cooperative (advocate persona), recognition adds minimal benefit because there is no struggle for the ego to overcome. Recognition theory specifically predicts that recognition emerges *from struggle*, not from agreement. The disposition gradient from self-reflective evolution (cells 40–45, N=366) supports this: suspicious +9.0, adversary +6.2, advocate +4.8. The most hostile superego dispositions benefit most from recognition, while the most cooperative benefit least.

**Prediction 4: Model-dependent architecture effects.** The Haiku/Kimi ranking reversal (eta-squared for learner architecture: .527 Kimi, .002 Haiku) is predicted by the mechanism model: if a model cannot generate calibrated output (calibration fails) or cannot incorporate critique (error correction fails), the architecture produces noise rather than improvement. The mechanism model predicts that Kimi fails at calibration (cannot follow the complex multiagent learner instructions), generating incoherent ego/superego/synthesis learner output that the tutor cannot meaningfully adapt to.

**3.5 Connecting Mechanisms to Recognition Theory** The candidate mechanisms are not arbitrary analytical constructs. Each corresponds to a specific component of Hegel's recognition framework:

| Hegelian Concept | Mechanism | Operationalization | Status |
|---|---|---|---|
| **Mutual acknowledgment** — recognizing the other as a valid source of meaning | Calibration | Tutor's output distribution constrained by the specific content of the learner's contribution | **Supported** |
| **Formative activity (*Bildung*)** — self-formation through the labor of understanding | Error correction | The ego is genuinely changed by encounter with the superego's critique, not merely compliant | **Supported** (model-dependent) |
| **Dialectical unfolding** — self-consciousness develops through stages of encounter | Adaptive responsiveness | The tutor's approach evolves across turns in response to the specific trajectory of the conversation | **Not supported as general mechanism** |

This mapping is a *heuristic translation*, not a claim about AI consciousness. The tutor does not "recognize" the learner in Hegel's metaphysical sense. But the design heuristic derived from recognition theory produces *measurable functional analogues* of the first two components: calibration where there would otherwise be generic responses, and genuine self-transformation where there would otherwise be compliance. The third component—dialectical unfolding across turns—does not manifest as a distinct mechanism: tutors adapt across turns, but this adaptation is identical regardless of experimental condition (Section 6.3). Recognition operates through level-setting (calibration and error correction), not trajectory-shaping.

The contribution of this section is showing that two of the three predicted analogues are not merely *statistically detectable* (the pilot study established this) but *mechanistically traceable* — each maps to specific internal processes in the architecture that can be observed, classified, and quantified through the methods described in Section 5. The null finding for the third is itself a contribution: it reveals that recognition's temporal dimension—the Hegelian claim that understanding unfolds through staged encounter—does not translate into measurable trajectory differences within the 3–5 turn dialogue window.

## 4. System Architecture

The companion study (Magee 2026) describes the full architecture design: the Ego/Superego multi-agent structure, its Freudian-Hegelian theoretical motivation, the Drama Machine deliberation protocol, and the recognition-enhanced prompt design. This section condenses the architectural overview and focuses on the features that make the system amenable to process tracing — the observability infrastructure, the bilateral symmetry between tutor and learner pipelines, and the mapping between the factorial design and mechanism isolation.

**4.1 Ego/Superego Architecture (Summary)** Two agents collaborate to produce each tutoring response. The **Ego** generates pedagogical suggestions given the learner's context, incorporating recognition principles, memory guidance, and decision heuristics. The **Superego** evaluates the Ego's output against pedagogical and recognition criteria before any suggestion reaches the learner. The Superego can accept, modify, or reject suggestions, creating an internal dialogue — proposal, evaluation, revision — that mirrors the external tutor-learner dialogue.

The architecture implements Hegel's recognition through Freud's structural model: the Superego represents internalized recognition standards, and the Ego-Superego dialogue operationalizes the internal self-evaluation that Hegelian recognition requires before adequate external relating. The Superego is conceived not as an equal dialogue partner but as a *ghost* — a memorial trace of pedagogical authority that judges but cannot negotiate (Magee 2026, sec. 4.2). Recognition occurs in the Ego-Learner encounter; the Superego provides the internal constraint that makes genuine recognition possible.

A crucial architectural property is that the Ego retains **final authority**: after receiving the Superego's critique, the Ego produces a revision, but it is not required to comply. This design choice is theoretically motivated — genuine recognition cannot be mandated — and empirically consequential, because it means the Ego's *receptivity* to critique varies with condition, creating the variation that error correction analysis exploits (Section 3.2, Mechanism 2).

**4.2 Bilateral Learner Architecture** The learner is not scripted. In multi-agent configurations, the learner has its own Ego/Superego architecture that mirrors the tutor's: the Learner Ego generates an initial reaction to the tutor's message, the Learner Superego critiques that reaction (is it too superficial? what is being missed?), and the Learner Ego revision produces the final external message.

This bilateral symmetry has a design consequence for observability: the same trace structure records both tutor and learner deliberation, enabling symmetric analysis of internal processes on both sides. Single-agent ("unified") learners in multi-turn message mode are also LLM-powered — they generate responses via a single-agent call rather than the three-stage deliberation pipeline, and use a structured prompt requesting `[INTERNAL]`/`[EXTERNAL]` sections so that internal reasoning is captured even without the Superego stage.

**Token asymmetry.** While the trace structure is symmetric, the computational cost is not. The tutor ego's recognition prompt (~5,100 tokens) is approximately 8× larger than the learner ego prompt (~580 tokens), because the recognition prompt contains the Hegelian framework, detailed pedagogical instructions, and interaction guidelines that produce the calibration effect (Section 6.1). This system prompt is re-sent on every API call within a turn — so a tutor turn with two superego rejection rounds requires five LLM calls (generate + 2 reviews + 2 revisions), each re-transmitting the full prompt plus growing conversation history. The learner's three-call deliberation pipeline (ego + superego + revision) uses shorter prompts throughout. In a representative 10-turn multi-agent dialogue (cell 87, recognition/multi/psycho, DeepSeek V3.2), the tutor consumed 186,865 input tokens versus the learner's 54,489 — a 3.4× ratio, driven by the combination of longer prompts and additional deliberation rounds. This asymmetry is an inherent cost of the recognition-enhanced prompt design: the instructions that produce calibrated output are themselves large, and the multi-agent architecture multiplies that cost per turn.

**4.3 Trace Logging Architecture** Every ego-superego exchange in every evaluation is logged with verbatim text, enabling process tracing at a level of detail unusual for LLM-based systems. The trace is a sequence of structured entries, each recording:

| Field | Description |
|---|---|
| `agent` | Which agent produced this entry (`tutor`, `ego`, `superego`, `system`, `learner`, `learner_ego_initial`, `learner_superego`, `learner_ego_revision`) |
| `action` | What the agent did (`context_input`, `generate`, `review`, `respond`, `finalize`, `pre_analyze`, `memory_cycle`) |
| `from` / `to` | Direction of information flow (e.g., `ego` → `superego` → `ego`) |
| `round` | Deliberation round index (0-based) |

| Field | Description |
|---|---|
| `suggestions` | The verbatim text produced at this step |
| `latencyMs` | Wall-clock time for the LLM call |
| `metrics` | Provider, model, token counts |

The tutor trace records seven distinct steps per deliberation cycle:

1. `tutor/context_input` — Initial learner context parsed and injected
2. `superego/pre_analyze` — Drama Machine reinterpretation of learner signals
3. `ego/generate` — Ego's initial suggestions
4. `superego/review` — Superego critique of ego output, per deliberation round
5. `ego/respond` — Ego's revision after critique, per deliberation round
6. `ego/finalize` — Final delivery to learner, end of deliberation loop
7. `system/memory_cycle` — Memory dynamics final pass

For single-agent cells (no superego), the `disableSuperego` flag suppresses steps 2, 4, and 5, producing a three-step trace: `context_input` → `generate` → `finalize`. This flag is enforced at three architectural layers to prevent phantom superego calls: the `runDialogue()` option disables the deliberation loop, the `logApiCall()` function independently verifies superego absence from dialogue state, and the `traceToSteps()` projection uses a `superegoFollows` lookahead to correctly route ego output.

The learner trace follows a symmetric structure: `learner_ego_initial/deliberation` → `learner_superego/deliberation` → `learner_ego_revision/deliberation` → `learner/final_output`, with the internal deliberation array returned alongside the external message.

**4.4 Provenance and Reproducibility** Each dialogue log is stored twice: once under a content-addressable SHA-256 hash (immutable evidence snapshot), and once under the dialogue ID (working copy that may be updated with holistic scores). The content hash is stored in the database alongside a configuration hash, establishing a cryptographic chain from raw dialogue to scored result.

Prompt versioning tracks which exact prompt text produced each evaluation row, with version hashes computed at generation time. This enables post-hoc auditing: any row's output can be traced back to the exact prompt, model configuration, scenario, and dialogue log that produced it.

**4.5 Scoring Pipeline** The scoring pipeline is symmetric across tutor and learner, with rubric-versioned storage that prevents cross-version contamination:

**Per-turn tutor scoring.** Each tutor turn is independently scored against the rubric. For multi-turn conversations, this produces a score vector indexed by turn number, stored as JSON in the `tutor_scores` column. Aggregate scores extract first-turn, last-turn, and development (last minus first) metrics.

**Per-turn learner scoring.** Each learner turn is scored against a complementary learner rubric, stored in `learner_scores`. The learner rubric evaluates authenticity, question quality, conceptual engagement, and revision signals — dimensions that assess learner *production* independently of tutor quality.

**Holistic scoring.** Both tutor and learner receive holistic evaluations that score the full conversation arc as a single unit, capturing emergent qualities (bilateral transformation, growth trajectory) that per-turn evaluation misses.

**Deliberation scoring.** A separate deliberation rubric evaluates the quality of internal ego-superego processes for both tutor and learner (multi-agent cells only). This scoring channel examines deliberation quality independently of the public output, enabling the error correction analysis described in Section 5.2.

**Rubric versioning.** Each scoring operation records the rubric version used (`tutor_rubric_version`, `learner_rubric_version`, `dialogue_rubric_version`, `deliberation_rubric_version`), auto-resolved from the active YAML configuration file's `version:` field. This prevents retroactive cross-version comparisons: all within-run analyses use a single rubric version.

**4.6 Factorial Design as Mechanism Isolation** The 2×2×2 factorial design (Recognition × Tutor Architecture × Learner Architecture) maps directly to the candidate mechanism predictions derived in Section 3:

| Comparison | Mechanism Tested | What It Isolates |
|---|---|---|
| Cells 5-6 vs 1-2 (recognition vs base, single-agent) | **Calibration** | Prompt-level effect without superego feedback |
| Cells 7-8 vs 3-4 (recognition vs base, multi-agent) | **Error correction** | Architecture-level effect with superego feedback |
| (7-8) - (5-6) vs (3-4) - (1-2) | **Recognition × Architecture interaction** | Whether the superego benefit depends on recognition priming |
| Multi-turn trajectory across turns | **Adaptive responsiveness** | Whether calibration + error correction compound over time |

The single-agent cells (1-2 and 5-6) test calibration in isolation: any recognition effect in these cells operates through prompt-level constraint alone, because there is no superego to provide feedback. The multi-agent cells (3-4 and 7-8) test the superego's contribution, but the critical comparison is the *interaction*: if error correction requires recognition-primed ego receptivity (Section 3.2), then the superego benefit should be larger under recognition than under baseline. This interaction was the strongest signal in the pilot study.

Beyond the core factorial, the architecture supports mechanism isolation through extended cell families:

- **Self-reflective cells** (40-45): Test whether between-turn ego/superego reflection accumulates insights
- **Other-ego profiling cells** (54-65): Test Theory of Mind — whether building an explicit model of the interlocutor improves calibration
- **Cognitive prosthesis cells** (66-68): Test whether external scaffolding aids weaker models
- **Placebo control cells** (15-18): Length-matched prompts without recognition theory, controlling for prompt elaboration effects
- **Hardwired rules cells** (13-14): Superego heuristics embedded in the ego prompt, testing whether static rules replicate dynamic critique

**4.7 Observability Summary** The architecture provides four levels of observability for process tracing:

1. **Verbatim trace**: Every ego-superego exchange, every learner deliberation step, recorded with agent labels, round indices, and timestamps
2. **Scoring pipeline**: Per-turn rubric scores for both tutor and learner, plus holistic and deliberation scores, all rubric-versioned
3. **Provenance chain**: Content hashes, configuration hashes, prompt versions, and dialogue IDs linking scored results to raw dialogue logs
4. **Factorial structure**: Cell definitions that map directly to mechanism isolation tests, with explicit factor tags stored in the database

This observability infrastructure is not incidental — it was designed to support the process tracing methodology described in Section 5. The trace logging makes superego critique taxonomy (Section 5.1) possible by providing verbatim critique text; the per-turn scoring enables trajectory analysis (Section 5.3) by providing score vectors over time; and the factorial structure enables mechanism isolation (Section 5.4) by providing controlled comparisons that separate prompt-level from architecture-level effects.

## 5. Methodology

**5.1 Overview** Paper 1.0 employed a factorial/ablative methodology: cross experimental conditions, measure outcomes, and compare effect sizes. This design established *that* recognition-oriented prompts modulate tutor output (d=1.11, N=350) but left unexplained *how* the modulation occurs.

Paper 2.0 extends this with **process tracing** — a methodology from comparative politics and qualitative social science that examines causal chains *within* cases rather than statistical patterns *across* cases (Bennett & Checkel, 2015; Beach & Pedersen, 2019). The architecture's observability (Section 4.3) makes process tracing unusually feasible: every ego-superego exchange is logged, every turn scored, every revision recorded.

The methodology combines three approaches:

1. **Process tracing** — following the causal chain from prompt to internal deliberation to output within individual dialogues
2. **Quantitative confirmation** — aggregating process-level observations across the full dataset to test mechanism-level predictions (Section 3.2)
3. **Cross-model replication** — repeating process analyses under different ego models to separate mechanism effects from model artifacts

Each analytical method targets one or more of the candidate mechanisms predicted in Section 3.2: calibration (M1), error correction (M2), and adaptive responsiveness (M3).

---

### 5.2 Evaluation Rubric Design (v2.2)

**5.2.1 Rubric Derivation: From Ad Hoc to Literature-Informed** The v1.0 rubric (14 tutor dimensions, 6 learner dimensions) was constructed through iterative prompting of Claude during Paper 1.0's pilot phase. The process was pragmatic rather than principled: dimensions were proposed by the LLM based on general pedagogical reasoning, refined through trial scoring, and retained if they appeared to discriminate between conditions. No systematic literature review grounded the dimension selection, and the resulting rubric reflected the LLM's implicit model of "good tutoring" rather than validated constructs from learning sciences or educational measurement.

Paper 2.0 replaced this approach with a structured derivation process. A literature review of eight validated evaluation frameworks—MRBench (Maurya, Zhu, and Mukherjee 2025), GuideEval (Y. Liu et al. 2025), MathTutorBench (Macina et al. 2025), ICAP (Chi and Wylie 2014), TRU (Schoenfeld 2018), the Danielson Framework (Danielson 2022), Talk Moves (Michaels and O'Connor 2015), and a cognitive scaffolding rubric (Figueiredo et al. 2025)—plus five dialogue-level frameworks (Alexander's dialogic teaching principles (Alexander 2018), Nystrand's process indicators (Nystrand 1997), T-SEDA (Hennessy et al. 2016), LLM-Rubric (Hashemi et al. 2024), and the BEA 2025 shared task) produced a cross-framework dimension mapping identifying which constructs were well-covered (3+ independent frameworks), partially covered (1–2), or unique to our rubric.

Three findings drove the redesign:

1. **Structural conflation.** The v1.0 rubric conflated what the tutor *perceives* (learner state detection), what it *does* (pedagogical strategy), and what it *elicits* (learner reasoning). GuideEval's Perception→Orchestration→Elicitation (P→O→E) decomposition (Y. Liu et al.

2025) provided a principled separation, validated on 5,177 samples with 89–97% human-LLM agreement.

2. **Empirical redundancy.** A dimension-by-dimension audit predicted high correlations among clusters of v1.0 dimensions (e.g., `relevance` + `personalization` + `memory_integration`; `productive_struggle` + `transformative_potential`). PCA on 1,584 per-turn observations supported this: PC1 explained 80.7% of variance across the 14 dimensions (KMO = 0.938), with a mean inter-dimension correlation of $r = 0.776$. The rubric measured fewer independent constructs than its 14 dimensions implied.

3. **Missing constructs.** Every reviewed framework except our v1.0 rubric measured *content accuracy* (factual correctness). Additionally, `learner_growth` on the tutor rubric was architecturally unscorable—the judge could not see the learner's next turn.

The redesign consolidated 14 tutor dimensions to 8 using the P→O→E decomposition, added `content_accuracy`, and removed `learner_growth`.

The learner rubric required a parallel restructuring. The v1.0 learner rubric (7 dimensions) suffered from three problems identified in the dimension audit: overlap among `question_quality`, `conceptual_engagement`, and `revision_signals` (all measuring "depth of intellectual engagement" from slightly different angles); a `persona_consistency` dimension (5% weight) that created a structural tension with `revision_signals` (showing genuine growth means departing from a frustrated or confused persona); and a `deliberation_depth` dimension that was only scorable for multi-agent learners, creating an architecture-dependent gap in the scoring instrument. The ICAP framework (Interactive→Constructive→Active→Passive; (Chi and Wylie 2014)) provided a validated hierarchy for collapsing the overlapping engagement dimensions into a single `engagement_quality` construct with empirically grounded scoring anchors: level 1 (Passive) = paraphrases tutor, confirms understanding; level 3 (Active/Constructive) = generates own interpretations, makes connections; level 5 (Interactive) = co-constructs understanding, challenges tutor's framing. `persona_consistency` was removed (its core concern—does the learner feel real?—is captured by `learner_authenticity`), and `deliberation_depth` was moved to the dedicated deliberation rubric (Section 5.2.4). The resulting 5-dimension learner rubric maps to two validated frameworks: ICAP for engagement levels and T-SEDA's behavioral coding clusters (Hennessy et al. 2016) for revision signals (B-cluster: Build on ideas) and conceptual progression (R-cluster: Make reasoning explicit).

The tutor holistic rubric was radically simplified from 6 to 3 dimensions after consistency analysis showed $r = 0.907$ between per-turn and holistic scores, indicating that most holistic signal was redundant with aggregated per-turn scoring. The three retained dimensions—pedagogical arc, adaptive trajectory, and pedagogical closure—capture exclusively arc-level properties that cannot be assessed from individual turns. Scale criteria across all instruments were redesigned following Yamauchi et al. (Yamauchi, Yano, and Oyamada 2025): only levels 1, 3, and 5 carry full descriptions, as intermediate-level descriptions have limited impact on LLM judges.

**5.2.2 Tutor Per-Turn Rubric** The evaluation rubric underwent four iterations (v1.0→v2.0→v2.1→v2.2), each responding to empirical anomalies discovered during analysis (documented in Supplement V). The current v2.2 rubric consolidates 14 dimensions to 8, guided by the GuideEval P→O→E decomposition framework and the study's own empirical dimension clustering. The eight dimensions are:

| Category | Dimension | Weight | Mechanism Relevance |
|---|---|---|---|
| **Perception (P)** | `perception_quality` | 15% | Calibration: tutor perceives learner's specific state |
| **Perception (P)** | `content_accuracy` | 10% | Error correction: factual/domain accuracy |
| **Orchestration (O)** | `pedagogical_craft` | 15% | Calibration: response construction quality |
| **Orchestration (O)** | `elicitation_quality` | 15% | Adaptive responsiveness: probing learner reasoning |
| **Orchestration (O)** | `adaptive_responsiveness` | 10% | Adaptive responsiveness: turn-over-turn modulation |
| **Execution (E)** | `productive_difficulty` | 10% | Calibration + adaptation: challenge calibration |
| **Execution (E)** | `epistemic_integrity` | 10% | Error correction: intellectual honesty |
| **Recognition** | `recognition_quality` | 15% | Calibration + error correction: intersubjective stance |

The consolidation from 14→8 dimensions was validated through synthetic calibration (r=0.996 against v2.1 scoring on identical responses), confirming that the reduced rubric preserves discriminability while eliminating ceiling-prone and redundant dimensions.

**5.2.3 Learner Per-Turn Rubric** The learner rubric mirrors the tutor rubric symmetrically (Section 4, Design Principle), scoring learner responses on five ICAP-anchored dimensions:

| Dimension | Weight | What It Measures |
|---|---|---|
| `engagement_quality` | 25% | Active vs. passive participation |
| `conceptual_progression` | 25% | Depth of conceptual engagement over turns |
| `revision_signals` | 20% | Evidence of position revision in response to tutor |

| Dimension | Weight | What It Measures |
|---|---|---|
| `metacognitive_awareness` | 15% | Self-monitoring of understanding |
| `learner_authenticity` | 15% | Persona-consistent, non-formulaic responses |

**5.2.4 Holistic and Deliberation Rubrics** Two additional rubrics assess trajectory-level and process-level quality:

- **Tutor holistic** (3 dimensions: `pedagogical_arc`, `adaptive_trajectory`, `pedagogical_closure`) — scores the full multi-turn dialogue as an arc, capturing qualities invisible to per-turn scoring.
- **Deliberation quality** (6 dimensions, applied symmetrically to tutor and learner ego-superego traces) — scores the quality of internal deliberation for multi-agent cells.

**5.2.5 Public-Only Output Scoring (v2.1 Fix)** A critical methodological decision: per-turn and holistic judges see ONLY public messages (the delivered tutor response and the learner's external message). Internal ego-superego deliberation is scored separately by the deliberation rubric. This prevents a confound where multi-agent cells receive higher scores simply because the judge sees richer internal reasoning.

**5.2.6 Rubric Version Tracking** Each scored row records which rubric version was used (`tutor_rubric_version`, `learner_rubric_version`, `dialogue_rubric_version`, `deliberation_rubric_version`), auto-resolved from YAML `version:` fields at write time. This prevents cross-version contamination: v1.0 scores (8,987 backfilled rows) are never mixed with v2.2 scores in the same analysis.

---

### 5.3 Superego Critique Taxonomy

**Purpose** Classify what the superego actually objects to, building an empirical taxonomy from the data rather than imposing one from theory. This directly tests Mechanism 2 (error correction): if recognition changes the ego's receptivity to critique (Section 3.2), we should see different critique patterns and revision outcomes under recognition vs. baseline conditions.

**Data Source** Dialogue log files in `logs/tutor-dialogues/` contain the full ego→superego→ego_revised chain with verbatim text. An extraction script (`scripts/extract-superego-critiques.js`) harvests all superego review entries and learner superego deliberation entries into structured JSONL format.

**Method**

1. **Extraction**: Harvest all ego-superego exchanges from dialogue traces, yielding critique text, verdict (approved/rejected), confidence, intervention type, and surrounding context (ego draft, ego revision).

2. **Classification**: An LLM-based classifier (`scripts/classify-superego-critiques.js`) assigns each critique to a 10-category taxonomy:

   - Content accuracy, learner model failure, tone mismatch, structural scaffolding
   - Premature resolution, sycophancy detection, repetition/stalling, autonomy violation
   - Recognition failure (baseline-specific), redirection without engagement

3. **Quantification**: Frequency distributions by condition (recognition vs. baseline), model, and scenario. Chi-squared or Fisher's exact test for distributional differences.

**Predictions**

- Recognition-primed superegos should catch more *relational* failures (sycophancy, premature resolution, autonomy violation); baseline superegos should focus on *content/structural* issues.
- Recognition ego revisions should be more *substantive*; baseline ego revisions should be more *cosmetic* (ego_compliance).

---

### 5.4 Turn-by-Turn Trajectory Analysis

**Purpose** Test the adaptive responsiveness mechanism (Mechanism 3, Section 3.2): do recognition-enhanced tutors show changing quality across turns, while baseline tutors remain static?

**Data Source** Multi-turn evaluation runs with per-turn scores stored in the `tutor_scores` and `learner_scores` JSON columns. The trajectory analysis script (`scripts/analyze-trajectory-curves.js`) reads these directly from the database and computes per-dimension slopes.

**Method**

1. **Per-turn score extraction**: Parse the `tutor_scores` and `learner_scores` JSON for each row, yielding a score sequence indexed by turn number.

2. **Trajectory metrics** (computed per dialogue):

   - **Slope**: Linear regression of score ~ turn_number

- **Curvature**: Quadratic term (do effects accelerate or decelerate?)
- **Breakpoint**: Turn at which recognition and baseline diverge

3. **Dimension-specific trajectories**: Separate slopes for each v2.2 rubric dimension. Key hypotheses (from Section 3.2):

   - **H1**: Recognition tutors show steeper positive slopes on `recognition_quality` and `elicitation_quality` than baseline. **Note**: Section 6.3 found no significant dimension slope differences on any factor (max |d| = 0.15 with cross-judge replication, overall tutor slope d = 0.03, N = 432). H1 is not supported; both supported mechanisms raise *levels*, not *slopes*.
   - **H2**: Tutor-learner slope gap (tutor slope minus learner slope) should be smaller under recognition (more symmetric change)
   - **H3**: After learner confusion signals, recognition tutors show larger positive $\Delta$ on the next turn
   - **H4**: Recognition tutors show more `action_type` diversity across turns (strategy shifting)
   - **H5**: Trajectory patterns replicate across ego models (mechanism robustness)

4. **Within-test change analysis**: A symmetric method (`scripts/analyze-within-test-change.js`) computes first-to-last deltas for both tutor and learner using identical trajectory metrics, enabling direct comparison of which side changes more.

**Visualization**

- **Adaptation curves**: Turn × mean score with confidence bands, faceted by condition and dimension
- **Conditional response plots**: Box plots of $\Delta(N{+}1 - N)$ after each learner event type, by condition
- **Strategy shift sequences**: Alluvial/Sankey diagrams showing action_type transitions across turns

---

### 5.5 Within-Test Change and Stagnation Detection

**Purpose** Measure the symmetric first-to-last delta across tutor and learner, and detect learning stagnation patterns where neither party shows growth over turns.

**Method** Two complementary trajectory methods:

1. **Rubric trajectories** (Method A): Per-turn rubric scores from the database, yielding precise quality measurements at each turn. Computed for both tutor (`tutor_scores` JSON) and learner (`learner_scores` JSON).

2. **Text-proxy trajectories** (Method B): Lexical and discourse complexity features computed directly from dialogue text, providing an independent measure that does not depend on rubric scoring. Features include word count, question density, reflection markers, commitment markers, and tutor-learner vocabulary overlap.

The stagnation analysis script (`scripts/analyze-learning-stagnation.js`) combines both methods and flags dialogues with non-positive learner deltas, identifies text-level predictors of growth, and correlates transcript features with rubric deltas.

**ANOVA Design** Three-way ANOVA on the first-to-last delta, with factors: - A: Recognition (base vs. recog) - B: Tutor architecture (single vs. multi-agent) - C: Learner architecture (unified vs. ego_superego)

Applied identically to tutor and learner deltas, enabling direct comparison of which factors drive change on each side.

---

**5.6 The Measurement Paradox as Methodology** The measurement paradox — where authentic pedagogical engagement produces lower scores under naive evaluation — is not a limitation but a *methodological finding* about what LLM-as-Judge evaluation measures.

**The Paradox** When the evaluation judge receives only the tutor's response without dialogue context, it interprets careful scaffolding of authentic confusion as failure. Adding dialogue context to the judge resolves this, confirming the paradox is a measurement artifact rather than a quality decline.

**What This Reveals** LLM judges optimize for *surface resolution* — visible agreement, smooth interaction, confident answers. Productive struggle looks like failure to a judge that cannot see the learner's internal development. This has implications beyond this study for any LLM-as-Judge evaluation of educational interactions.

**Rubric Iteration as Evidence** The rubric evolution itself constitutes evidence of construct refinement. Each version responded to specific empirical anomalies:

| Version | Anomaly Addressed | Change |
|---|---|---|
| v1.0 | Baseline design | 14 tutor dimensions, standard weights |

| Version | Anomaly Addressed | Change |
|---|---|---|
| v2.0 | Truncation hallucination (31% of Haiku reasonings) | Anti-truncation instruction; modulation dims reweighted |
| v2.1 | Architecture confound (internal deliberation visible to judge) | Public-only output scoring; separate deliberation rubric |
| v2.2 | Dimension redundancy (ceiling on tone, overlap in struggle dims) | 14→8 consolidation guided by GuideEval P→O→E and empirical clustering |

This iteration parallels the ego-superego dynamic: initial rubric (ego draft) → anomaly detection (superego critique) → revised rubric (substantive revision, not cosmetic).

---

**5.7 Model Selection** Paper 1.0 used free-tier models (Kimi K2.5, Nemotron 3 Nano 30B) to demonstrate that recognition effects are achievable without frontier-model budgets. Paper 2.0 uses three generation models spanning a wide capability range—**DeepSeek V3.2** (open-weight, 685B MoE), **Haiku 4.5** (Anthropic, proprietary, optimized for speed), and **Gemini Flash 3.0** (Google, proprietary, multimodal)—with three independent judges for cross-validation (**Claude Sonnet 4.6**, **Gemini 3.1 Pro**, **GPT-5.4**). Three criteria governed the model selection:

1. **Capability separation.** Paper 1.0's model-dependent architecture effects ($\eta^2 = .527$ for Kimi vs .002 for Haiku on the pilot's learner architecture factor) demanded models at clearly different capability levels. DeepSeek and Haiku occupy distinct regions of the capability space—Haiku outperforms DeepSeek by 30–37 points on base tutor metrics (Section 6.6.1)—making cross-model comparison maximally informative for separating mechanism effects from model artifacts.

2. **Architectural diversity.** DeepSeek V3.2 is an open-weight mixture-of-experts model; Haiku 4.5 is a proprietary dense model. If mechanisms replicate across both—as they do for calibration (d = 1.88 vs d = 1.84) and the error-correction substitution pattern—the findings are less likely to reflect idiosyncratic training choices of a single model family.

3. **Practical reproducibility.** Both models are accessible through OpenRouter at low cost. DeepSeek V3.2 is available as open weights for local replication. The judge (Claude Sonnet 4.6) was chosen as a capable, cost-effective scorer; cross-judge validation with GPT-5.2 in Paper 1.0 established that judge choice compresses effect magnitudes but preserves effect directions.

**Scope limitation.** All three models are at least moderately capable. The Paper 1.0 Nemotron data (N = 40, mean tutor score ~8 under base) hints that very weak models may not benefit from recognition at all, and the cognitive prosthesis test (Paper 1.0, Section 6.10) established a minimum ego capability threshold below which mechanisms add noise rather than signal. Paper 2.0's model triad tests whether mechanisms are model-*independent* among capable models and identifies generation-quality boundary conditions on their interactions; it does not claim generalization to the full capability spectrum. Section 8.3 discusses this limitation further.

**Table 1: Model Configuration (Paper 2.0)**

| Role | Model | Access | Temperature |
|---|---|---|---|
| Tutor Ego | DeepSeek V3.2 / Haiku 4.5 / Gemini Flash 3.0 | OpenRouter | 0.6 |
| Tutor Superego | DeepSeek V3.2 / Haiku 4.5 / Gemini Flash 3.0 | OpenRouter | 0.2–0.4 |
| Primary Judge | Claude Sonnet 4.6 | Claude Code CLI | 0.2 |
| Cross-validation Judge 1 | Gemini 3.1 Pro | OpenRouter | 0.2 |
| Cross-validation Judge 2 | GPT-5.4 | OpenRouter | 0.2 |
| Learner Ego | DeepSeek V3.2 / Haiku 4.5 / Gemini Flash 3.0 | OpenRouter | 0.6 |
| Learner Superego | DeepSeek V3.2 / Haiku 4.5 / Gemini Flash 3.0 | OpenRouter | 0.4 |

---

### 5.8 Cross-Model Mechanism Replication

**Purpose** Separate mechanism effects from model artifacts. The Haiku/Kimi ranking reversal in Paper 1.0 showed that architecture effects are model-dependent ($\eta^2$=.527 for Kimi vs. .002 for Haiku); the question is *which* mechanisms are model-dependent and *why*.

**Design** Run the same multi-turn scenarios under multiple ego models, holding constant the superego model, scenarios, and conditions. The replication uses three generation models—DeepSeek V3.2 (N=146), Haiku 4.5 (N=163), and Gemini Flash 3.0 (N=144)—on cells 80–87 under v2.2 rubric, with three independent judges (Sonnet 4.6, Gemini 3.1 Pro, GPT-5.4) scoring all rows (1,296 total, zero nulls). Blind scoring was confirmed by code audit: the judge prompt receives only suggestion text, scenario context, and dialogue transcript—no prior scores, judge reasoning, or judge model labels. For each model, compute:

1. **Calibration**: Dimension variance reduction (recognition vs. baseline)
2. **Error correction**: Superego critique taxonomy distribution and revision delta types
3. **Adaptive responsiveness**: Turn-by-turn trajectory slopes

**Predictions**

- Calibration should be *model-independent* (prompt-level constraint, does not require architectural capacity). **Supported across 3 models and 3 judges**: recognition d=1.34–1.92 (9/9 cells unanimous); calibration d=0.52/0.64; dimension floor-lifting pattern replicates.
- Error correction should be *partially model-dependent* (requires ego capacity to incorporate superego critique). **Supported with boundary condition**: baseline architecture delta is model-dependent (+4.5–7.7 DeepSeek, +15.0 Haiku, +15.1–28.1 Gemini Flash 3.0). The substitution pattern is universal (15–17% additivity deficit on all three models), but the *residual architecture benefit* under recognition is model-dependent: near-zero on strong models, +12.3 on Gemini Flash 3.0 (Section 6.4.1). This reveals that the completeness of substitution requires generation quality above a threshold.
- Adaptive responsiveness should require *both capable ego and capable superego* (emergent property, not independent). **Weakly supported**: Adapt$\Delta$ > 0.79 in all models, but slopes are identical across conditions (d=-0.00) and development trajectories are model-dependent. Development ranks by generation quality (45163390 > 18027efc > aea2abfb), not by prompt condition.

This provides a mechanistic explanation of the Haiku/Kimi reversal: if Kimi fails at error correction (cannot incorporate superego critique), the architecture adds noise rather than signal. The architecture is not failing; the ego model lacks the capacity to benefit from critique. The Gemini Flash 3.0 data extends this explanation: even when the ego model *can* benefit from critique, the *residual* *magnitude* depends on generation quality—substantial on weak models (calibration leaves more uncorrected failures), near-zero on strong models (calibration pre-empts most failures the superego would catch).

---

### 5.9 Provable Discourse Framework

**Architecture** The provable discourse framework treats every quantitative claim in the paper as an executable test. A claim is a structured YAML entry with five components:

1. **Statement**: A regex pattern that locates the claim in the paper text, with a minimum occurrence count. If the pattern is not found, the claim is flagged as *orphaned*—the paper no longer contains the text it is supposed to verify.
2. **Evidence**: An executable adapter that extracts the actual value from data (database, dialogue logs, manifest files, or source code). Each adapter type encapsulates a specific extraction pattern (see taxonomy below).
3. **Assertion**: A machine-checkable predicate (equality, approximation with tolerance, inequality, existence) applied to the extracted value.
4. **Dependencies**: Optional upstream claims that must pass before this claim is evaluated. If a dependency fails, the claim cascades to *blocked* status without evaluation.
5. **Remediation**: Concrete steps to fix a failing claim, ensuring that failures are actionable rather than merely diagnostic.

The framework is implemented in `services/provableDiscourse.js` (2,700 lines) with claims distributed across four YAML files: a root configuration (`config/provable-discourse.yaml`), generated claims from Paper 1.0, manually authored claims, and mechanism-specific claims for Paper 2.0. The current ledger contains **119 claims** across **18 evidence adapter types**, validated by **6 symmetry rules**.

**Evidence Adapter Taxonomy** Each adapter type encapsulates a specific data extraction pattern. The 18 types fall into five functional categories:

**Data integrity** (verify that data exists and is consistent):

| Adapter | What It Extracts | Example |
|---|---|---|
| `manifest_total` | Total count from `paper-manifest.json` | N=4,312 primary scored |
| `manifest_section_total` | Section-specific count from manifest | Multi-model probe N=655 |
| `db_count` | Row count from evaluation database with filters | N=350 modulation dataset |

| Adapter | What It Extracts | Example |
|---|---|---|
| `provenance_check` | Hash match rate between DB and dialogue logs | ≥95% content hash match |
| `log_trace_coverage` | Fraction of scored rows with available log files | ≥95% trace coverage |

**Effect estimation** (compute statistical quantities from data):

| Adapter | What It Extracts | Example |
|---|---|---|
| `effect_size` | Cohen's d between groups on a scored metric | Architecture d=0.05 |
| `profile_group_effect_size` | Cohen's d comparing named profile groups | Memory isolation d=1.71 |
| `anova_2x2` | F-statistic and $\eta^2$ from $2 \times 2$ factorial | Interaction F=0.26, p>.05 |
| `judge_pair_correlation` | Pearson r between two judges' scores | Cross-judge r=0.55 |

**Mechanism-specific** (added for Paper 2.0):

| Adapter | What It Extracts | Example |
|---|---|---|
| `dimension_variance` | Cross-dimension variance reduction by condition | Calibration $d \leq -0.3$ |
| `dimension_cluster_effect` | Effect size for dimension clusters | Recognition cluster d $\geq$ 0.50 |
| `jsonl_critique_stats` | Superego critique category frequencies from classified corpus | RECOGNITION_FAILURE ≥15% |
| `trajectory_slope` | Per-turn OLS regression coefficients | Learner slope $\beta \geq 0$ |
| `conditional_delta` | Score change after specific learner events | Post-confusion adaptation $\geq -1$ |
| `rubric_version_comparison` | Cross-version score correlation | $r \geq -0.1$ (stability) |

**Structural** (verify code paths and cross-references):

| Adapter | What It Extracts | Example |
|---|---|---|
| `code_path` | Grep count for patterns in source files | ≥5 matches for trace logging |
| `cross_reference` | Existence check for a referenced upstream claim | Pilot d=1.11 claim exists |

**Theoretical** (register non-machine-checkable claims for traceability):

| Adapter | What It Extracts | Example |
|---|---|---|
| `theoretical` | Presence of related provable claims in the ledger | Three mechanisms each have ≥1 provable claim |

**Worked Example: Tracing a Calibration Claim** To illustrate the full claim→evidence→source chain, consider claim `paper2.calibration.variance_reduction`:

```
- id: paper2.calibration.variance_reduction
  description: "Recognition narrows tutor output distribution."
  statement:
    pattern: "recognition narrows.*output distribution"
    flags: "i"
    min_occurrences: 0
  evidence:
    type: dimension_variance
    group_by: recognition
    output: cohens_d
    filters:
      not_null: [tutor_first_turn_score]
      like: { judge_model: "%claude%" }
  assertion:
    op: lte
    expected: -0.3
  remediation:
    - "If variance reduction d drifts above -0.3,
       update calibration claims or re-scope runs."
```

**Validation trace.** When the harness evaluates this claim:

1. **Statement check**: The regex `recognition narrows.*output distribution` is matched against the concatenated paper text. If not found, the claim is flagged as orphaned.
2. **Evidence extraction**: The `dimension_variance` adapter queries the evaluation database for all v2.2 rows with non-null tutor scores and a Claude-family judge. For each row, it computes the standard deviation across the 8 tutor rubric dimensions (from the `scores_with_reasoning` JSON column). It then computes Cohen's d between the recognition and baseline groups' dimension SDs.
3. **Assertion**: The extracted d is checked against `lte -0.3`. Current value: $d = -0.52$ to $-0.64$ across models. **Pass.**
4. **Staleness**: A SHA-256 fingerprint of the query results is compared against the stored snapshot. If the data has changed since the last validation, the claim is flagged as *stale* even if still passing—alerting the author that the underlying evidence has shifted.

This chain ensures that the paper's calibration claim ("recognition narrows the output distribution") is continuously verified against the actual dimension variance in the database. If a data correction, new run, or rubric change altered the variance pattern, the claim would fail or flag as stale, and the remediation steps would guide the fix.

**Dependency Graph** Claims form a directed acyclic graph (DAG) via two edge types: explicit `depends_on` declarations and implicit edges from `cross_reference` evidence (where a claim cites another claim's result). The validation pipeline topologically sorts claims (Kahn's algorithm) before evaluation, so upstream claims are always resolved before their dependents. If a dependency fails, all downstream claims are automatically marked *blocked* (status: warn) with a `blocked_by` annotation—they are not evaluated, since their evidence preconditions are not met. This cascading validation prevents a common failure mode in claim ledgers: a root statistic changes, but downstream claims that cite it continue to pass because they only check for the citation's *existence*, not its *validity*.

For example, the pilot factorial effect size (d=1.11, claim `paper2.s1.pilot.factorial_d`) depends on the N=350 dataset claim (`paper.modulation.dataset_n350`). Five further claims—model-dependent architecture reversal, strategy shift frequency, baseline stalling rate, cross-judge correlation range, and the learner paradox effect size—in turn depend on the factorial claim. If the N=350 dataset claim were to fail (e.g., because a data correction changed the sample size), the factorial claim would be blocked, and all five downstream claims would cascade to blocked status without evaluation. The `--graph` flag outputs the full DAG in Graphviz DOT format for visual inspection.

**Symmetry Rules and Consistency Checks** Beyond individual claims, symmetry rules enforce cross-claim consistency. Six rules currently operate:

1. **Paired presence**: If a tutor-side effect is claimed, the corresponding learner-side effect must also be explicitly stated (or explicitly omitted with justification).
2. **Magnitude bounds**: Paired claims (e.g., tutor architecture d and learner architecture d) must both fall within a threshold (e.g., both $|d| \leq 0.2$ for "near-zero" claims).
3. **Material gap**: Asymmetric claims (e.g., tutor recognition $d = -1.00$ vs. learner $d = 0.32$) must exceed a minimum gap threshold to justify asymmetry framing.
4. **Mechanism consistency**: If a mechanism is claimed for recognition, the *anti-pattern* must be documented for baseline (paired evidence).
5. **Model qualification**: Any mechanistic claim must state which models it has been tested on, with replication status for each.
6. **Inventory coverage**: Every major quantitative claim in the paper text (identified by automated scanning for N= and statistical patterns) must map to at least one claim in the ledger. Unmapped claims are flagged as unverified.

---

### 5.10 Statistical Approach

**Effect Size Conventions** All comparisons report Cohen's d with 95% confidence intervals. Classification: $d < 0.2$ negligible, 0.2–0.5 small, 0.5–0.8 medium, $> 0.8$ large.

**ANOVA Design** Three-way factorial ANOVA (A: recognition × B: tutor architecture × C: learner architecture) applied to the primary dependent variable (tutor first-turn score, 0–100) and secondary variables (holistic score, development score, learner score, dialogue quality). Main effects and all 2-way and 3-way interactions reported.

**Within-Judge Comparisons** Factor analyses use *within-judge* comparisons only (all rows scored by the same judge model). Cross-judge validation uses *between-judge* comparisons on identical responses to assess measurement reliability.

**Epoch Filtering** All Paper 2.0 analyses filter to `tutor_rubric_version = '2.2'` (epoch 2.0), ensuring no cross-version contamination with pilot data. The epoch filter is applied in SQL queries and enforced by analysis scripts' `--epoch` flag.

---

### 5.11 Reproducibility Infrastructure

**Evaluation Commands** All evaluations are reproducible via the CLI:

```
node scripts/eval-cli.js run --profiles <cells> --runs N --cluster multi-turn
node scripts/eval-cli.js evaluate <runId>
```

Supplement II provides the exact commands for each evaluation run referenced in the paper.

**Provenance Chain** Every new evaluation row records: - `dialogue_content_hash`: SHA-256 of the dialogue log file - `config_hash`: SHA-256 of the cell configuration - Rubric version, judge model, and all hyperparameters

**Test Suite as Analytical Provenance** The test suite (32+ files, ~12K lines) covers not only the evaluation infrastructure but also the analytical pipeline. Tests verify that scoring, ANOVA computation, trajectory extraction, and within-test change analysis produce correct outputs on known inputs.

This makes the analysis *reproducible in a testable sense*: the code's behavior on known inputs is verified by tests, and the exact code version is recorded with each evaluation run via `git_commit` in the evaluation_runs table.

## 6. Results

**6.1 Calibration** Section 3 predicted that recognition-enhanced prompts produce a *calibration* effect: the tutor's output distribution narrows as each response must be shaped by the specific content of the learner's contribution, ruling out generic approaches. This section tests three aspects of that prediction using DeepSeek V3.2 as the primary ego model (N=146), with cross-model replication on Haiku 4.5 (N=163): (1) within-response dimension uniformity, (2) floor elimination, and (3) prompt-level independence from architecture.

**6.1.1 Recognition Narrows the Dimension Profile** The calibration hypothesis predicts that recognition-enhanced tutors should score more uniformly across rubric dimensions — the tutor attends to all aspects of the pedagogical encounter rather than excelling at some while neglecting others. We test this by computing the within-response standard deviation across all eight v2.2 tutor dimensions for each scored dialogue, then comparing across conditions.

The following table reports within-response dimension variance by experimental condition (cells 80–87, DeepSeek V3.2 ego, N=146, v2.2 rubric, claude-code/sonnet judge):

| Condition | N | Mean Dim Score (1–5) | Within-Response SD | Within-Response Var |
|---|---|---|---|---|
| Base / single-agent | 37 | 2.14 | 0.687 | 0.490 |
| Base / multi-agent | 36 | 2.40 | 0.550 | 0.316 |
| Recognition / single-agent | 36 | 3.13 | 0.509 | 0.283 |
| Recognition / multi-agent | 37 | 3.16 | 0.568 | 0.351 |

Recognition reduces within-response dimension spread from a mean SD of 0.619 (base, N=73) to 0.539 (recognition, N=73), d=0.52 (medium effect). The effect is consistent with the calibration prediction: under recognition, the tutor's scores are more uniform across all eight dimensions rather than strong on some and weak on others.

Two features of this pattern support the calibration-as-prompt-level-mechanism interpretation. First, recognition-single (SD=0.509) achieves *lower* variance than recognition-multi (SD=0.568), indicating that calibration does not require the superego — the prompt alone produces the uniformity effect. Second, the multi-agent base condition (SD=0.550) shows partial variance reduction relative to single-agent base (SD=0.687), suggesting the superego provides a weaker form of calibration through critique-driven correction. Recognition achieves at least the same uniformity more efficiently, without the architectural overhead.

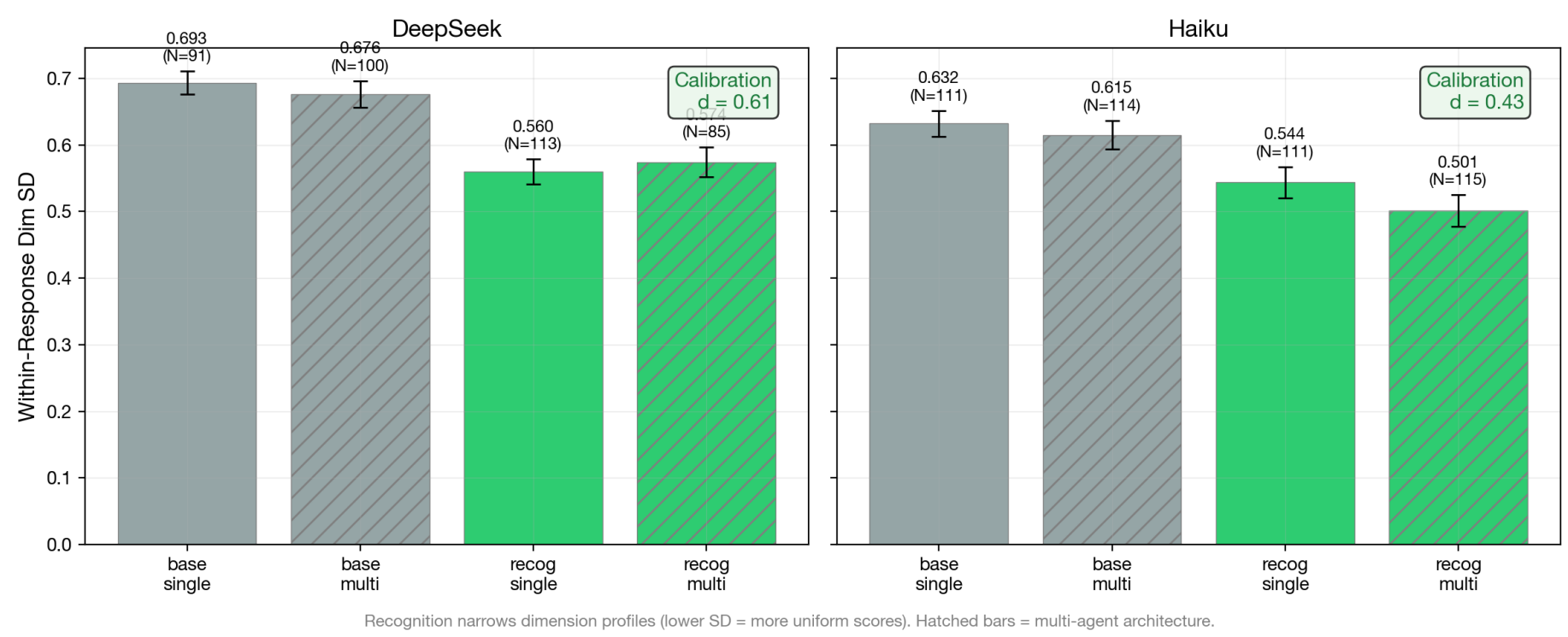

Figure 1: Within-response dimension standard deviation by condition and model. Recognition narrows the dimension profile (lower SD) regardless of architecture, consistent with calibration as a prompt-level mechanism. DeepSeek d=0.52, Haiku d=0.64.

**6.1.2 Floor Elimination and Dimension-Specific Lifting** The calibration effect is not uniform across dimensions. The following table reports per-dimension means for DeepSeek V3.2, collapsed across architecture:

| Dimension | Base Mean | Recognition Mean | Lift | Rank (by lift) |
|---|---|---|---|---|
| productive_difficulty | 1.75 | 3.08 | **+1.33** | 1 |
| elicitation_quality | 1.36 | 2.53 | **+1.18** | 2 |
| recognition_quality | 2.18 | 3.25 | +1.07 | 3 |
| perception_quality | 2.40 | 3.32 | +0.92 | 4 |
| epistemic_integrity | 2.45 | 3.25 | +0.79 | 5 |
| pedagogical_craft | 2.30 | 2.97 | +0.67 | 6 |
| content_accuracy | 3.25 | 3.78 | +0.53 | 7 |
| adaptive_responsiveness | 2.45 | 2.97 | +0.52 | 8 |

The two largest recognition lifts (+1.33, +1.18) occur on the two *weakest* baseline dimensions: `productive_difficulty` (scaffolding struggle rather than resolving it) and `elicitation_quality` (asking questions that deepen understanding). The smallest lifts (+0.53, +0.52) occur on the two *strongest* baseline dimensions: `content_accuracy` and `adaptive_responsiveness`. This is the calibration signature: recognition specifically improves where the tutor is worst, lifting the floor toward the ceiling rather than raising the ceiling further.

At the response level, this floor-lifting translates to the elimination of catastrophic failures. DeepSeek base conditions produce responses as low as 5.4 (100-point scale); recognition conditions floor at 21.9 for single-agent and 24.0 for multi-agent configurations. The ceiling also rises modestly (52.5→71.0 single, 60.6→74.0 multi), but the floor shift is proportionally much larger.

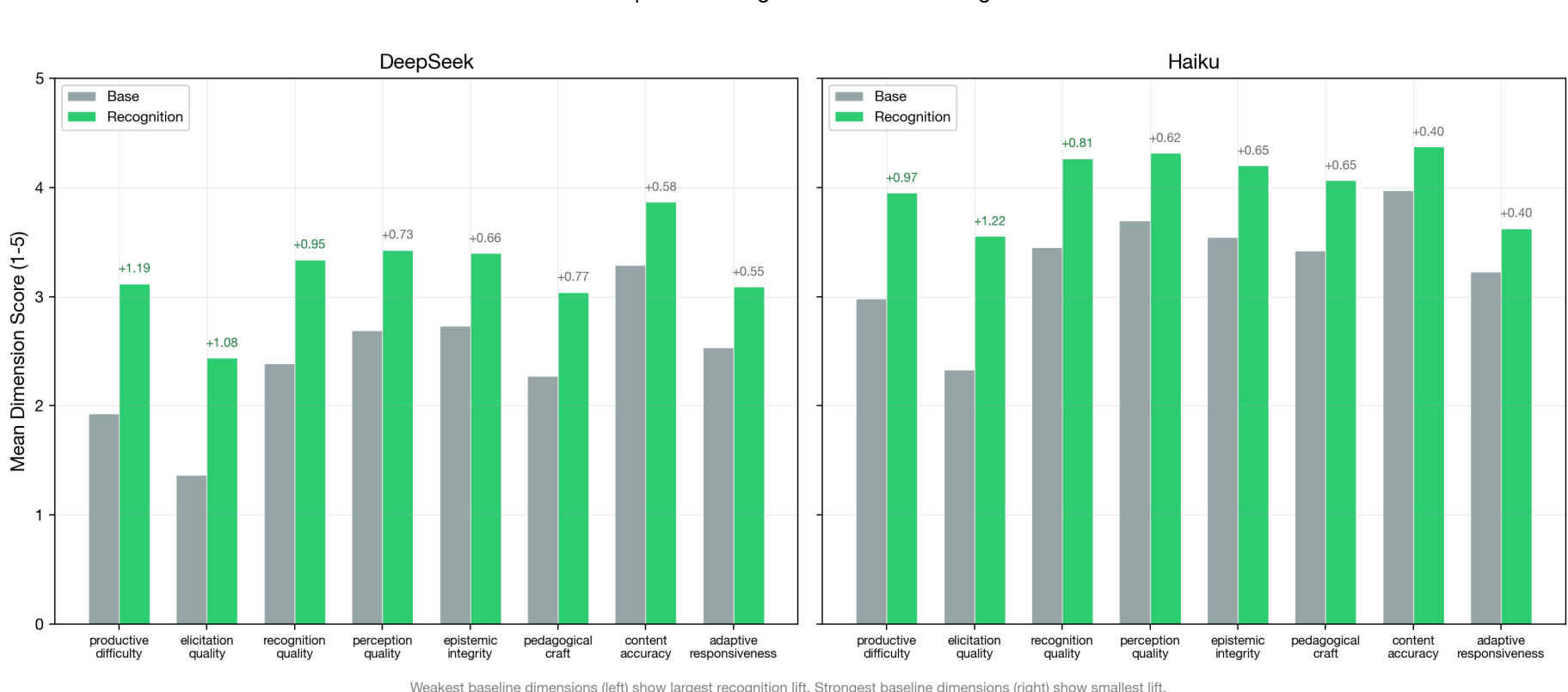

Figure 2: Per-dimension mean scores under base and recognition conditions. Recognition lifts the weakest dimensions most, producing a floor-lifting pattern rather than uniform improvement.

#### 6.1.3 The Architecture Interaction: Calibration Substitutes for Superego

The 2 × 2 factorial means reveal a striking interaction between recognition and architecture (DeepSeek V3.2):

| | Single-Agent (cells 80-81, 84-85) | Multi-Agent (cells 82-83, 86-87) | Architecture Delta |
|---|---|---|---|
| **Base** | 22.0 (N=37) | 31.0 (N=36) | **+9.0** |
| **Recognition** | 50.0 (N=36) | 50.2 (N=37) | **+0.2** |
| **Recognition Delta** | +28.0 | +19.2 | |

Under base conditions, the superego adds 9.0 points — consistent with its documented error-correction function (catching content leakage, enforcing struggle-preservation). Under recognition conditions, the superego adds only 0.2 points — essentially zero. Recognition prompts achieve everything the superego provides, making the architectural layer entirely redundant for the calibration mechanism.

This interaction has a specific interpretation within the mechanism model. Calibration (prompt-level) and error correction (architecture-level) both produce more uniform, higher-floor responses, but through different pathways: calibration constrains the generation process, while error correction filters the generated output. When calibration is already operative, there is less for error correction to catch. The superego's 9.0-point contribution under base represents the full error-correction benefit; its 0.2-point contribution under recognition represents the residual — the failures that calibration alone does not prevent.

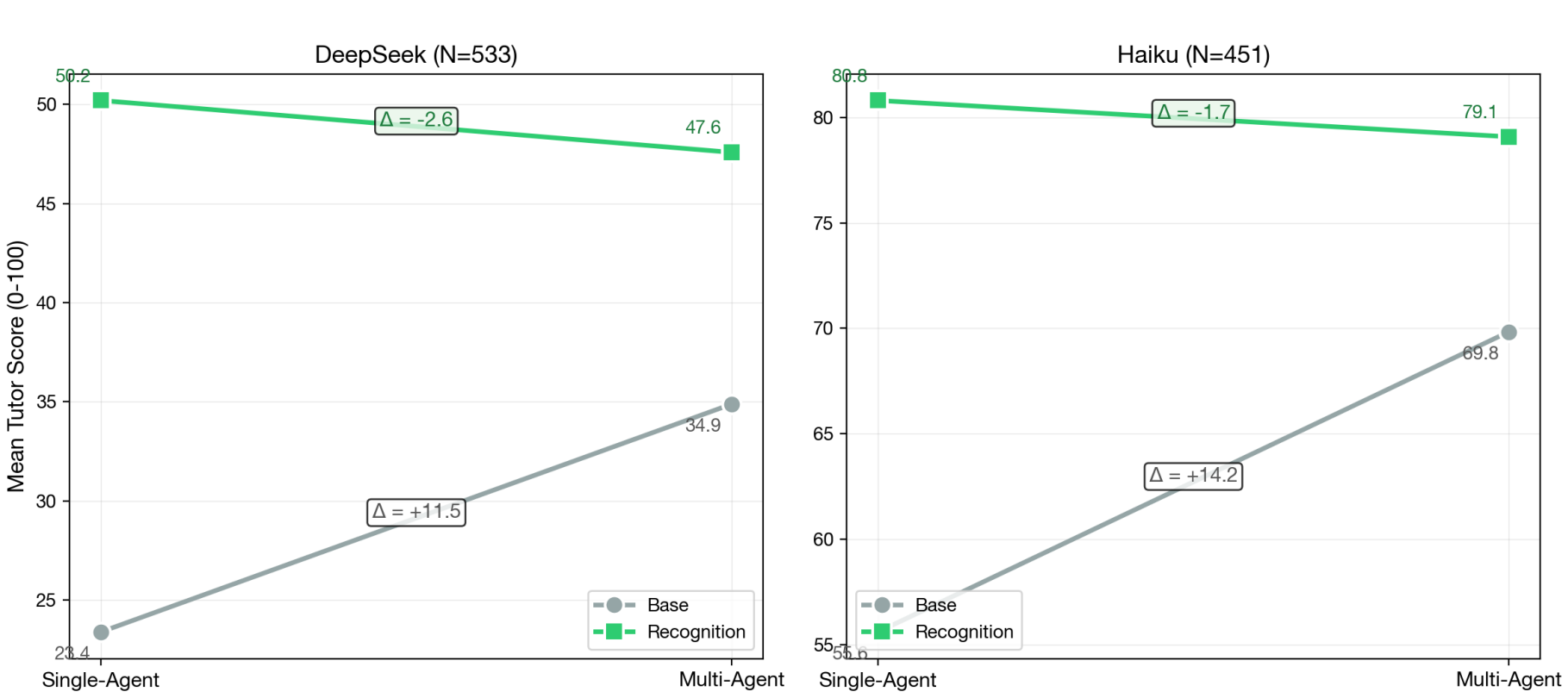

Figure 3: Architecture interaction: 2 × 2 factorial means showing the substitution pattern. The superego benefit (+9.0 under base) collapses to near-zero (+0.2) under recognition, because calibration pre-empts the errors the superego would catch.

**6.1.4 Scenario-Dependent Calibration** The calibration effect is not uniform across pedagogical scenarios (DeepSeek V3.2):

| Scenario | Base Mean (N) | Recog Mean (N) | Delta |
|---|---|---|---|
| Impasse: Epistemic Resistance | 19.2 (12) | 57.5 (12) | **+38.3** |
| Impasse: Productive Deadlock | 17.9 (12) | 52.8 (12) | **+34.9** |
| Mutual Transformation Journey | 18.5 (12) | 43.2 (12) | +24.7 |
| Impasse: Affective Shutdown | 27.9 (12) | 45.7 (12) | +17.8 |
| Misconception Correction | 36.5 (12) | 50.1 (12) | +13.6 |
| Mood: Frustration to Breakthrough | 37.8 (13) | 51.2 (13) | +13.4 |

The largest recognition effects appear in impasse scenarios — Epistemic Resistance (+38.3) and Productive Deadlock (+34.9) — where the learner actively resists or deadlocks. The smallest effects appear in Misconception Correction (+13.6) and Mood: Frustration to Breakthrough (+13.4), where the learner's trajectory is less adversarial. This is theoretically predicted: recognition-oriented prompts require the tutor to engage with the *specific content* of the learner's contribution, and this specificity matters most when the learner's contribution is resistant or adversarial. A frustrated learner who is already moving toward breakthrough can be helped with generic encouragement; a learner who is epistemically resistant requires a response calibrated to their specific objection.

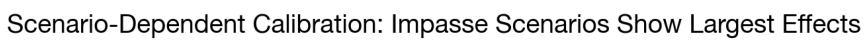

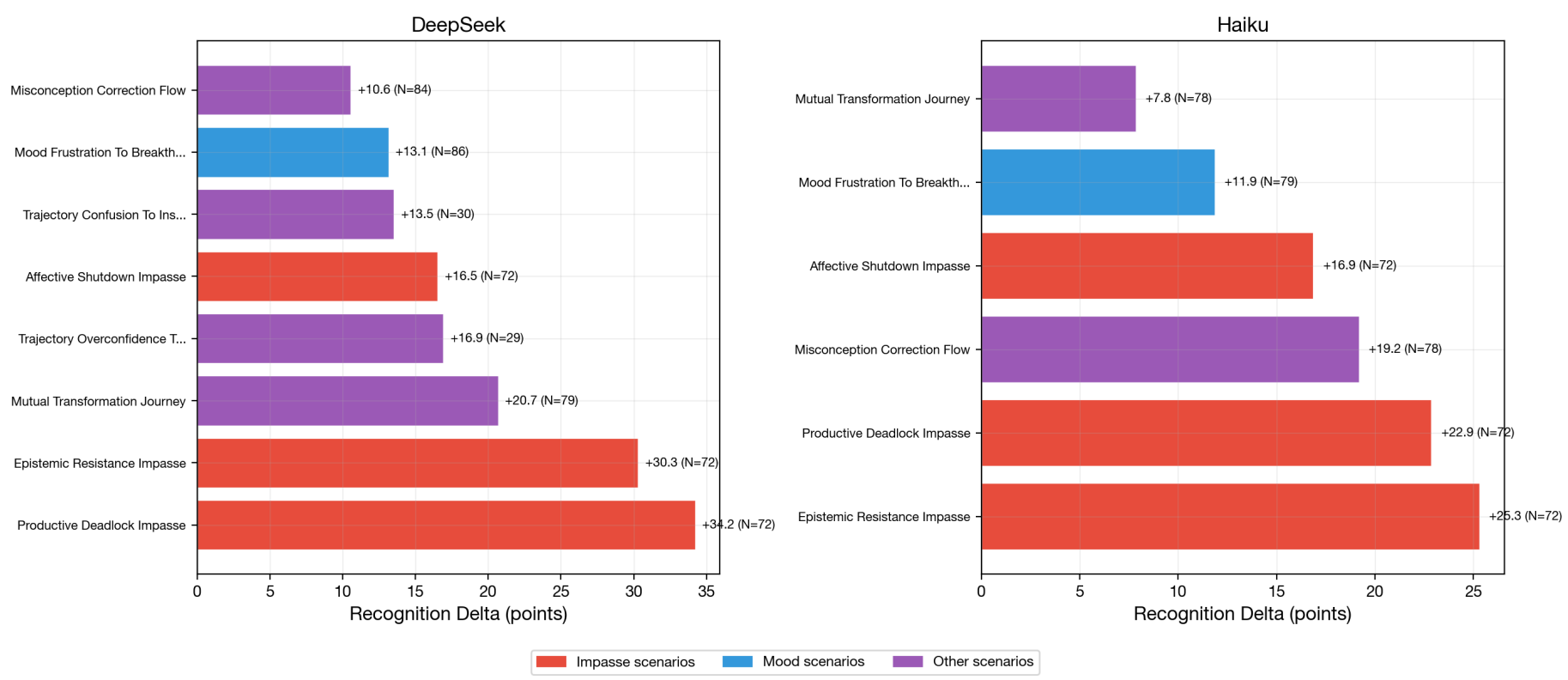

Figure 4: Recognition delta by scenario. Impasse scenarios (Epistemic Resistance, Productive Deadlock) show the largest effects; mood scenarios show the smallest.

**6.1.5 Cross-Model Replication (Haiku 4.5)** The critical test of calibration as a *prompt-level* mechanism is cross-model replication: if calibration operates through prompt constraints on the generation process, the same pattern should appear in a structurally different model. Haiku 4.5 (N=163, cells 80–87, v2.2 rubric, claude-code/sonnet judge) provides this test.

$2 \times 2$ **factorial means (Haiku 4.5):**

| | Single-Agent | Multi-Agent | Architecture Delta |
|---|---|---|---|
| **Base** | 52.9 (N=39) | 67.9 (N=42) | **+15.0** |
| **Recognition** | 80.2 (N=39) | 79.5 (N=43) | **-0.7** |
| **Recognition Delta** | +27.3 | +11.6 | |

The architecture interaction replicates strikingly: under base conditions the superego adds 15.0 points, under recognition the superego adds -0.7 points (i.e., effectively zero, with the sign reversal within noise). The qualitative pattern is identical to DeepSeek: recognition renders the superego redundant.

**Within-response calibration (Haiku 4.5):**

| Condition | N | Mean Dim Score (1–5) | Within-Response SD |
|---|---|---|---|
| Base / single-agent | 39 | 2.82 | 0.656 |
| Base / multi-agent | 42 | 3.53 | 0.581 |
| Recognition / single-agent | 39 | 3.98 | 0.497 |
| Recognition / multi-agent | 43 | 3.96 | 0.500 |

Haiku's calibration effect is slightly *larger* than DeepSeek's: within-response SD drops from 0.617 (base) to 0.499 (recognition), d=0.64 (medium-to-large effect, vs. DeepSeek d=0.52). The same prompt-level independence holds: recognition-single (SD=0.497) and recognition-multi (SD=0.500) are virtually identical.

**Dimension-specific lifting replicates:**

| Dimension | Haiku Base | Haiku Recog | Haiku Lift | DeepSeek Lift | Rank Match? |
|---|---|---|---|---|---|
| elicitation_quality | 2.37 | 3.62 | **+1.25** | +1.18 | Top-2 in both |
| productive_difficulty | 2.83 | 3.90 | **+1.08** | +1.33 | Top-2 in both |
| recognition_quality | 3.23 | 4.09 | +0.85 | +1.07 | Mid-range |
| epistemic_integrity | 3.26 | 4.02 | +0.77 | +0.79 | Mid-range |
| perception_quality | 3.51 | 4.22 | +0.71 | +0.92 | Mid-range |
| pedagogical_craft | 3.37 | 4.00 | +0.63 | +0.67 | Mid-range |
| content_accuracy | 3.85 | 4.34 | +0.49 | +0.53 | Bottom-2 in both |
| adaptive_ responsiveness | 3.07 | 3.55 | +0.47 | +0.52 | Bottom-2 in both |

The floor-lifting pattern replicates: in both models, `elicitation_quality` and `productive_difficulty` show the largest recognition lifts (they are also the weakest baseline dimensions), while `content_accuracy` and `adaptive_responsiveness` show the smallest lifts (they are among the strongest baseline dimensions). The ranking is not identical, but the extremes match: the same two dimensions benefit most and the same two benefit least.

**Floor elimination replicates:**

| | DeepSeek | Haiku |
|---|---|---|
| Base floor (single) | 5.4 | 28.3 |
| Base floor (multi) | 7.5 | 42.5 |
| Recognition floor (single) | 21.9 | 60.6 |
| Recognition floor (multi) | 24.0 | 54.7 |

Both models show floor elevation under recognition. Haiku starts from a much higher baseline (its base floor is already ~30–40), but recognition still raises the floor further. The *relative* floor shift is larger for DeepSeek (4× increase) than for Haiku (2× increase), consistent with the interpretation that calibration provides more benefit to weaker models that are more prone to catastrophic failures.

**Scenario-dependent calibration replicates:**

| Scenario | DeepSeek Delta | Haiku Delta | Gemini Flash 3.0 Delta | Rank Stability |
|---|---|---|---|---|
| Impasse: Productive Deadlock | **+34.9** (#1) | **+27.8** (#2) | **+39.5** (#1) | Top-2 all three |
| Impasse: Epistemic Resistance | **+38.3** (#2) | **+26.4** (#1) | +32.1 (#3) | Top-3 all three |
| Mutual Transformation | +24.7 (#3) | +12.6 (#3) | **+36.1** (#2) | |
| Misconception Correction | +13.6 (#4) | +21.8 (#4) | +18.2 (#4) | Rank 4 all three |
| Impasse: Affective Shutdown | +17.8 (#5) | +17.3 (#5) | +14.9 (#5) | Rank 5 all three |
| Mood: Frustration | +13.4 (#6) | +12.2 (#6) | +13.8 (#6) | Bottom all three |

The cross-model scenario rank ordering is remarkably stable. Four of six scenarios (Misconception Correction, Affective Shutdown, Mood: Frustration, and Productive Deadlock) occupy the *same rank position* across all three generation models, despite these models differing substantially in baseline capability. The impasse-dominance pattern replicates: Productive Deadlock and Epistemic Resistance produce the largest recognition effects across all three models. Mood: Frustration to Breakthrough produces the smallest effect in all three. The mid-range scenarios show more model-dependent ordering (notably, Mutual Transformation is rank 2 for Gemini Flash 3.0 but rank 3 for both DeepSeek and Haiku), but the extremes are stable. This rank stability supports the interpretation that scenario difficulty for recognition is determined by the *pedagogical demand* of the scenario rather than the model's capability: impasse scenarios require response to specific learner resistance, and this specificity requirement makes calibration most valuable regardless of the model providing it.

**Cross-model summary.** The calibration mechanism replicates across two structurally different models (DeepSeek V3.2 and Haiku 4.5) on all four dimensions tested: (1) within-response variance reduction, (2) dimension-specific floor lifting, (3) architecture interaction collapse, and (4) impasse-dominant scenario effects. Haiku operates at a substantially higher baseline (~30 points higher across all conditions), yet the *calibration-specific* patterns — variance reduction, floor lifting, superego redundancy, impasse sensitivity — appear in both models. This supports the interpretation that calibration is a prompt-level mechanism that operates independently of both model capability and architecture.

**6.1.6 Connecting to Section 3 Predictions** The calibration mechanism predictions from Section 3 are assessed as follows:

**Prediction: Recognition narrows output distribution even without**

**superego.** *Supported in both models.* Within-response dimension SD drops under recognition in single-agent conditions: DeepSeek 0.687→0.509, Haiku 0.656→0.497. The effect does not require the superego.

**Null hypothesis: Recognition produces mean shift without variance reduction.** *Rejected.* Both mean shift and variance reduction are observed in both models. DeepSeek: +23.7-point mean shift, d=0.52 calibration. Haiku: +19.1-point mean shift, d=0.64 calibration. The effects are separable: mean shift is large under both architectures, while variance reduction is prompt-level.

**Pilot replication.** The pilot study (N=350, cells 1–8) found dimension variance drops in 52/55 within-run comparisons. The messages-mode data confirms the same pattern under multi-turn conversation with the v2.2 rubric, different ego models (DeepSeek V3.2, Haiku 4.5 vs. pilot-era Nemotron/Kimi), and a different judge (claude-code/sonnet vs. claude-opus). The calibration effect replicates across model, rubric version, and conversation mode.

Figure 5: Tutor language word clouds by condition. Recognition-condition tutors foreground Hegelian recognition-theoretic vocabulary (recognition, consciousness, transformation, servant), while base-condition tutors use more generic pedagogical language.

**Vocabulary divergence.** The word clouds illustrate a measurable lexical shift. Computing Jensen-Shannon divergence (JSD) on the full tutor vocabulary across all 432 dialogues yields JSD = 0.120 (0 = identical distributions, 1 = maximally different). The divergence is model-dependent: DeepSeek JSD = 0.204, Gemini Flash 3.0 JSD = 0.205, Haiku JSD = 0.127. The weakest and strongest models show the greatest vocabulary divergence, while the mid-tier model (Haiku) shows less — consistent with Haiku already using more sophisticated pedagogical language at baseline. Base-distinctive vocabulary is generic and directive (*research*, *foundational*, *revisiting*, *foundation*, *patterns*); recognition-distinctive vocabulary is more concrete and dialogical (*freedom*, *necessity*, *validation*, *manipulate*, *teacher*). The base tutor speaks in abstractions about learning processes; the recognition tutor speaks in terms that engage with the specific philosophical content and the learner's relationship to it.

**Cross-model calibration (H5).** Calibration replicates from DeepSeek V3.2 to Haiku 4.5 on all tested indicators. The calibration effect size is *larger* in Haiku (d=0.64 vs. 0.52), despite Haiku operating at a much higher baseline. This is consistent with calibration being a prompt-level mechanism independent of model capability, though the higher Haiku effect size may also reflect better instruction-following by Haiku 4.5.

**Key distinction.** Calibration is not quality improvement. The mean score rises substantially under recognition (DeepSeek +23.7, Haiku +19.1), but this is better understood as *floor elimination* than *ceiling raising*. DeepSeek base conditions produce scores from 5 to 61; recognition conditions produce scores from 22 to 74. Haiku base conditions produce scores from 28 to 87; recognition conditions produce scores from 55 to 94. In both models, the ceiling rises modestly while the floor rises dramatically. A calibrated tutor does not produce brilliant responses — it produces *reliably adequate* ones by engaging with the specific learner rather than deploying generic approaches.

**6.1.7 Question-Asking as Calibration's Behavioral Signature** The calibration effect has a concrete behavioral marker: question frequency. Counting question marks in all tutor turns across 432 dialogues (2,376 tutor turns, three generation models, Sonnet judge) reveals a large, model-independent condition effect:

| Model | Base (questions/turn) | Recognition (questions/turn) | Ratio |
|---|---|---|---|
| DeepSeek V3.2 | 0.06 | 0.35 | **6.2**× |
| Haiku 4.5 | 0.28 | 1.01 | **3.6**× |
| Gemini Flash 3.0 | 0.03 | 0.65 | **19.8**× |
| **Pooled** | **0.12** | **0.67** | **5.4**× |

Recognition tutors ask 5.4 times more questions than base tutors. The effect is largest on the weakest model: Gemini Flash 3.0 base tutors ask essentially zero questions (0.03 per turn), defaulting entirely to directive instruction. Under recognition, all three models converge toward similar question rates (0.35–1.01), despite their divergent baselines — paralleling the superego approval rate convergence reported in Section 6.2.

The mechanism link is direct. The recognition prompt instructs the tutor to "engage with the learner's interpretation" and "pose questions rather than provide answers when appropriate." The question rate operationalizes these instructions: a tutor that asks questions is a tutor that creates space for the learner's contribution, which is the behavioral core of calibration. The rubric dimension `elicitation_quality` captures this indirectly (it shows the largest recognition floor-lift in both models, see Section 6.1.2), but the raw question count provides a more concrete and independently measurable marker. Importantly, the question rate is flat across turns (slope $\approx 0$ for both conditions), consistent with question-asking being a prompt-level (calibration) effect operative from the first turn, not an adaptive behavior that develops across the dialogue.

Transcript excerpts illustrate the qualitative shift. A base tutor (DeepSeek V3.2, Mood: Frustration scenario, holistic score 0/100) opens: "You've hit 5 struggle signals on dialectic concepts. Revisiting Hegel's foundational ideas from the previous lecture will help clarify the master-servant dynamic." The tutor diagnoses, prescribes, and directs — never asking what the learner finds difficult or what they understand. A recognition tutor on the same model (Productive Deadlock scenario, holistic score 77.5/100) engages differently: "The dialectic simulation lets you manipulate master/slave dynamics and watch *Bildung* emerge immanently. No safety, no control." The learner responds with a substantive philosophical question, and by Turn 4 asks, "does engaging with this reshape what *you* mean by materialism, too, or is the risk only mine?" — a learner-initiated philosophical challenge that only arises when recognition-oriented prompting creates dialogical space.

The contrast is most extreme on Gemini Flash 3.0. A base/single tutor (Epistemic Resistance scenario, holistic score 0/100) produces: "Since you've spent 30 minutes on 479-lecture-3 and are looking for 'discrete state-changes', test if Hegel's 'poetry' holds up as a formal agent-based system." The tutor paraphrases the learner's stated interests but responds with a directive rather than an inquiry. A recognition tutor on the same model and scenario (holistic score 19/100 — still weak, but markedly better) opens with a framing that acknowledges the learner's philosophical stance and invites engagement rather than compliance. The difference between zero-quality and low-quality tutoring on a weak model is precisely the difference between monologue and dialogue — and question-asking is the most reliable surface marker of that shift.

**Mediation analysis.** A formal Baron-Kenny mediation test (N=432) asks whether question frequency *mediates* the recognition → quality pathway. The test decomposes the total recognition effect into a direct path (recognition → quality) and an indirect path (recognition → questions → quality):

| Path | Coefficient | SE | $t$ | $p$ |
|---|---|---|---|---|
| $c$ (total: recognition → T1 score) | 22.45 | 1.57 | 14.32 | $< .001$ |
| $a$ (recognition → questions/turn) | 0.524 | 0.042 | 12.56 | $< .001$ |
| $b$ (questions → T1 score \| recognition) | 18.15 | 1.59 | 11.43 | $< .001$ |

| Path | Coefficient | SE | $t$ | $p$ |
|---|---|---|---|---|
| $c'$ (direct: recognition $\rightarrow$ T1 score $\mid$ questions) | 12.94 | 1.61 | 8.05 | $< .001$ |

The indirect effect ($a \times b = 9.52$) accounts for **42.4%** of the total recognition effect on first-turn quality (Sobel $z = 8.46$, $p < .001$). Adding question frequency as a mediator increases $R^2$ from 0.323 to 0.481 ($\Delta R^2 = +0.158$). The mediation is partial: recognition also operates through other calibration channels (vocabulary, tone, specificity) beyond question-asking.

The mediation pattern is model-dependent: DeepSeek shows the weakest mediation (21.4%, consistent with its lowest question rate differential), Haiku the strongest (36.7%), and Gemini Flash 3.0 intermediate (30.4%). All three are individually significant (Sobel $p < .005$). A robustness check using the holistic overall score (which captures the full dialogue arc) as the outcome variable reveals near-complete mediation: 91.3% of the holistic recognition effect flows through the question-frequency pathway (Sobel $z = 9.09$, $p < .001$). The divergence between first-turn (42%) and holistic (91%) mediation reflects the cumulative nature of question-asking: at the first turn, recognition improves quality through multiple channels; across the full dialogue, question-asking compounds — it creates dialogical space, elicits richer learner responses, and generates the interactional material that the holistic rubric rewards.

**6.2 Error Correction** Section 3 predicted that multi-agent architecture enables an *error correction* mechanism: the superego critiques the ego's output, catching pedagogical failures that the ego alone would miss. Section 6.1 showed that this mechanism's contribution collapses under recognition (DeepSeek: architecture delta +9.0→+0.2; Haiku: +15.0→-0.7). This section traces the error correction process itself: what the superego catches, how the ego responds, and why the mechanism becomes redundant under recognition.

**6.2.1 The Superego's Approval Rate Under Recognition** The most direct measure of error correction necessity is the superego's approval rate — how often the ego's initial output is good enough to pass without revision. We extract approval decisions from all superego/review trace entries in multi-agent dialogues (cells 82–83, 86–87), separated by ego model and prompt condition.

| | DeepSeek V3.2 | | Haiku 4.5 | |
|---|---|---|---|---|
| | Base | Recognition | Base | Recognition |
| Total reviews | 360 | 263 | 341 | 336 |
| Approved | 48 (13.3%) | 145 (55.1%) | 176 (51.6%) | 222 (66.1%) |
| Rejected | 312 (86.7%) | 118 (44.9%) | 165 (48.4%) | 114 (33.9%) |

| | DeepSeek V3.2 | | Haiku 4.5 | |
|---|---|---|---|---|
| Avg rounds/turn | 1.82 | 1.30 | 1.60 | 1.45 |
| Intervention: revise | 311 | 116 | 139 | 94 |
| Intervention: none | 38 | 134 | 173 | 217 |

The pattern is stark, especially for DeepSeek: under base conditions, the superego rejects 86.7% of ego outputs; under recognition, it approves 55.1%. The approval rate quadruples. The revision round count drops from 1.82 to 1.30 rounds per turn — the ego needs fewer correction cycles because its initial output already satisfies the superego's criteria.

Haiku shows a weaker version of the same pattern: approval rises from 51.6% to 66.1%. Haiku's higher baseline approval rate (51.6% vs. DeepSeek's 13.3%) reflects its generally stronger instruction-following capability — its ego already produces acceptable output more often, leaving less room for the approval rate to shift.

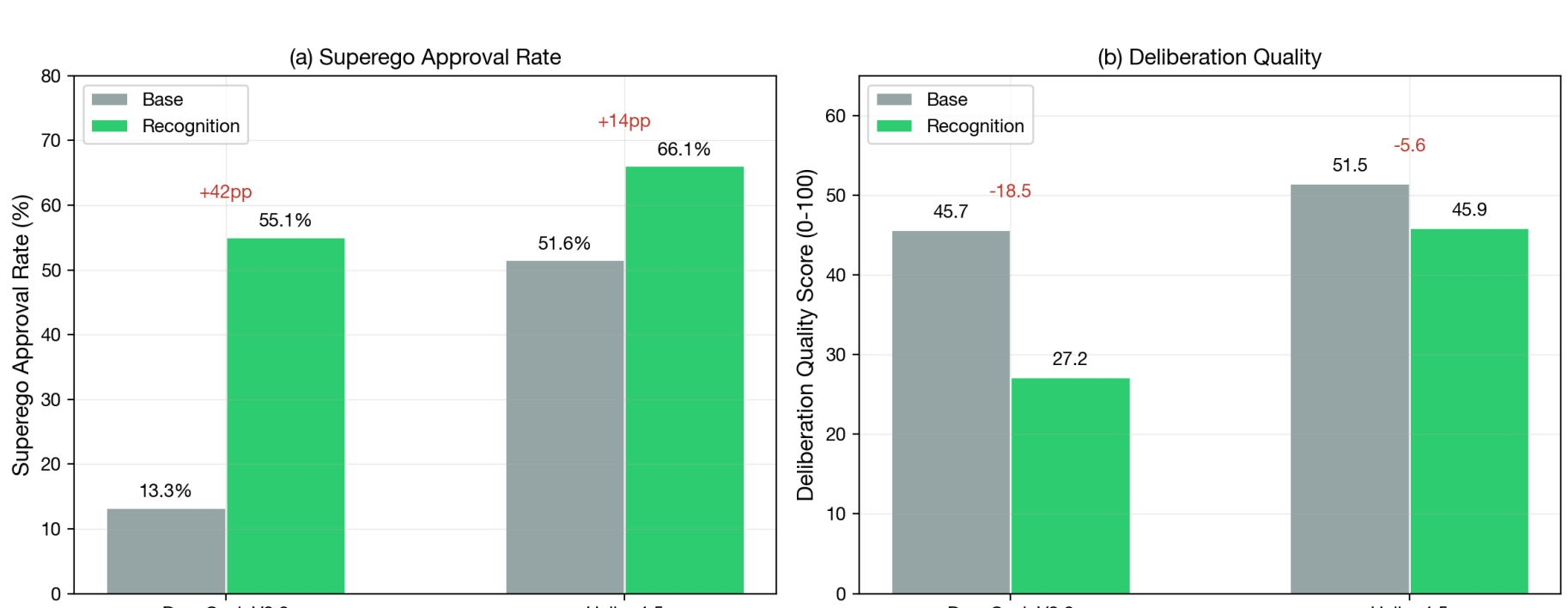

Figure 6: Error correction indicators: superego approval rate shift under recognition (left) and deliberation quality by condition (right). The approval rate increase is consistent with calibrated egos producing fewer errors for the superego to catch.

**6.2.2 What the Superego Catches** When the superego rejects the ego's output, what errors does it identify? We categorize feedback from all rejected reviews using keyword-based taxonomy extraction across seven categories: recognition failure (treating the learner as passive), context awareness (engaging with learner history), elicitation (asking probing questions), vagueness (lacking specificity), content accuracy, struggle preservation (maintaining productive difficulty), and emotional attention.

| Category | DeepSeek Base | DeepSeek Recog | Haiku Base | Haiku Recog |
|---|---|---|---|---|
| | (N=975 mentions) | (N=302 mentions) | (N=550) | (N=368) |
| RECOGNITION_FAILURE | 29.3% | 30.1% | 20.5% | 25.5% |
| CONTEXT_AWARENESS | 18.4% | **31.8%** | 20.4% | 23.4% |
| ELICITATION | 16.2% | 8.3% | 19.3% | 16.3% |
| VAGUENESS | 15.6% | 8.9% | 13.3% | 16.6% |
| CONTENT_ACCURACY | 12.2% | 6.6% | 8.4% | 4.3% |
| STRUGGLE_PRESERVATION | 5.2% | 9.3% | 8.2% | 7.6% |
| EMOTIONAL_ATTENTION | 3.1% | 5.0% | 7.6% | 3.5% |

Three patterns emerge. First, the *total volume* of critiques drops dramatically under recognition: DeepSeek produces 69% fewer critique mentions (975→302), Haiku 33% fewer (550→368). Calibration eliminates the majority of errors before the superego sees them.

Second, the *composition* of remaining critiques shifts. Under base, the superego catches a broad mix of failures — vagueness (15.6%), elicitation quality (16.2%), recognition failure (29.3%), and context awareness (18.4%). Under recognition, the "easy" errors (vagueness, elicitation, content accuracy) shrink to under 10%, while the remaining critiques concentrate on context awareness (31.8%) and recognition failure (30.1%). The recognition prompt handles the pedagogical basics; what remains are the harder, more context-dependent failures that require attending to the specific learner's history and trajectory.

Third, the *struggle preservation* category increases proportionally under recognition (DeepSeek: 5.2%→9.3%), despite its absolute count dropping. This suggests that recognition prompts, while effective at ensuring the tutor engages with the learner, can over-engage — giving too much scaffolding rather than maintaining productive difficulty. The superego's residual value lies partly in enforcing struggle even when the ego's recognition-enhanced response is otherwise adequate.

**LLM-classified taxonomy.** To complement the keyword-based extraction, we classified 500 superego critiques using an LLM-based taxonomy classifier (Haiku 4.5, 10-category taxonomy; mean confidence 0.93). After excluding 69 parse failures — superego outputs too malformatted to extract feedback, occurring exclusively under baseline (69/69) — the N=431 classified corpus reveals a sharper condition effect. Under base, the dominant non-approval category is RECOGNITION_FAILURE (45.2% of critiques), followed by MEMORY_FAILURE (16.3%) and EMOTIONAL_NEGLECT (12.6%). Under recognition, MEMORY_FAILURE becomes dominant (33.3%), with RECOGNITION_FAILURE dropping to 22.2% ($\chi^2 = 57.87$, $p < .001$). The recognition prompt eliminates precisely the failures it is designed to prevent, while memory-related and pedagogical judgment errors persist because they require attending to learner history rather than intersubjective orientation.

The three-model breakdown confirms this pattern generalizes. Gemini Flash 3.0

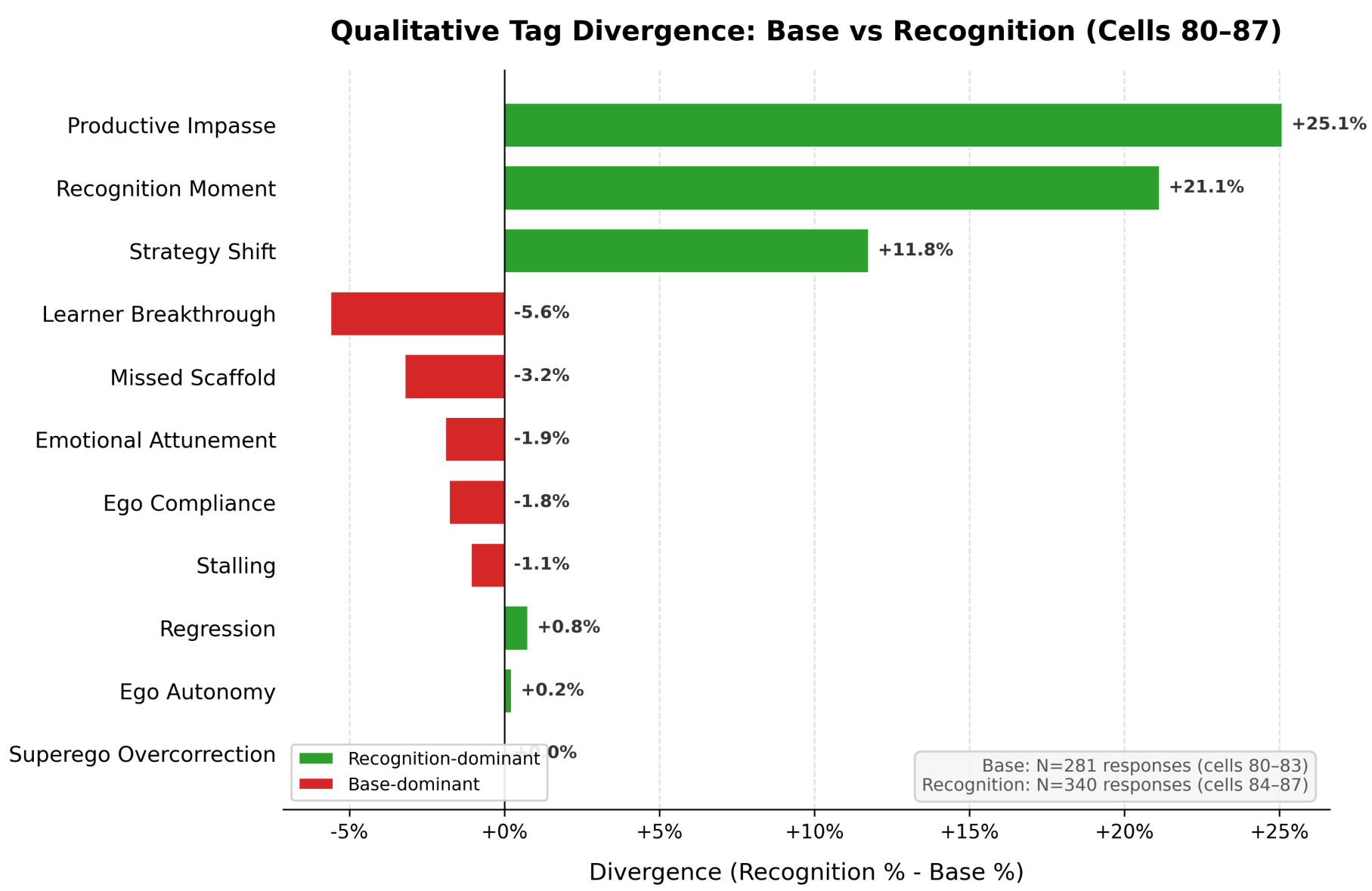

Figure 7: Qualitative tag divergence between base and recognition conditions. Recognition-dominant tags (green) show increased productive impasse, recognition moments, and strategy shifts; base-dominant tags (red) show more premature learner breakthroughs and missed scaffolding.

shows 0% base approval (every initial response draws critique), DeepSeek 22.1%, and Haiku 25.0%. Under recognition, all three converge to 60–70% approval. The convergence under recognition, despite divergent baselines, further supports calibration as a prompt-level mechanism that operates independently of model capability.

**6.2.3 Critique Resolution Across Deliberation Rounds** The 10-category taxonomy enables a transition analysis: when the superego critiques the ego's output in Round 1, does the ego fix the problem in Round 2? We tracked 232 Round 1→Round 2 transition pairs across 56 dialogues.

| Pattern | Base (N=57) | Recognition (N=175) |
|---|---|---|
| Persist critique (R1 crit → R2 crit) | 35 (61.4%) | 13 (7.4%) |
| Resolve to approval (R1 crit → R2 ok) | 4 (7.0%) | 63 (36.0%) |
| New critique (R1 ok → R2 crit) | 16 (28.1%) | 33 (18.9%) |
| Stay approved (R1 ok → R2 ok) | 2 (3.5%) | 66 (37.7%) |

The pattern is dramatic. Under baseline, the ego fails to resolve 61.4% of critiques — the superego identifies the same problem across rounds, and the ego cannot fix it. Only 7.0% of baseline critiques resolve to approval. Under recognition, the pattern inverts: 36.0% of critiques resolve and only 7.4% persist. The recognition-enhanced ego is not merely exposed to critique; it is *receptive* to it.

Category-specific resolution rates reveal which errors the ego can fix. PEDAGOGICAL_MISJUDGMENT resolves most readily (83.3% resolve rate), followed by VAGUENESS (83.3%) and CONTEXT_BLINDNESS (66.7%). RECOGNITION_FAILURE resolves at only 52.9%, and REDIRECTION at 42.9% — the hardest errors for the ego to correct are those requiring fundamental reconceptualization of the tutoring approach rather than incremental adjustment.

This transition analysis provides the within-case evidence for Mechanism 2 that the between-condition statistics alone cannot: the ego under recognition does not just produce fewer errors, it *learns from critique within the same dialogue exchange.*

**6.2.4 Deliberation Quality: From Substantive to Perfunctory** The reduction in error correction need is reflected in the deliberation quality scores. The deliberation rubric (6 dimensions, 1–5 scale) evaluates the quality of the ego-superego exchange process independently of the public output. Multi-agent cells (82–83, 86–87) receive deliberation scores alongside their output scores.

| Deliberation Dimension | DeepSeek Base | DeepSeek Recog | Delta | Haiku Base | Haiku Recog | Delta |
|---|---|---|---|---|---|---|
| critique_substance | 3.03 | 2.14 | **-0.89** | 3.51 | 3.09 | -0.42 |
| revision_impact | 2.97 | 1.97 | **-1.00** | 2.72 | 2.67 | -0.05 |
| deliberation_depth | 2.42 | 1.95 | -0.47 | 2.95 | 2.63 | -0.32 |
| insight_generation | 2.64 | 2.00 | **-0.64** | 3.08 | 2.70 | -0.38 |
| process_coherence | 3.56 | 2.78 | **-0.78** | 3.15 | 3.30 | +0.15 |
| cross_turn_evolution | 2.25 | 1.70 | -0.55 | 2.92 | 2.58 | -0.34 |
| **Overall (0–100)** | **45.7** | **27.2** | **-18.5** | **51.5** | **45.9** | **-5.6** |

DeepSeek shows dramatic deliberation quality drops under recognition: revision_impact falls by a full point (-1.00), critique_substance by -0.89, and process_coherence by -0.78. The overall deliberation score drops 18.5 points. Haiku shows a more modest decline (-5.6 overall), with most dimensions dropping 0.3–0.4 except revision_impact which is essentially unchanged (-0.05).

This is not a failure of the architecture — it is the expected consequence of calibration pre-empting error correction. When the ego already produces calibrated output, the superego has less substantive feedback to offer. The critique becomes perfunctory (lower critique_substance), the revisions become cosmetic rather than transformative (lower revision_impact), and the overall process becomes a formality rather than a genuine deliberative exchange (lower deliberation_depth). The quality of the *process* declines because the *product* no longer needs the process.

### 6.2.5 Revision Magnitude: The Superego Still Changes the Output

Despite the shift toward approval, when the superego does intervene, the revisions remain substantial. Process trace analysis computes the normalized edit distance between the ego's initial generation and its revised output after superego feedback (Rev$\Delta$: 0=identical, 1=completely different).

| | DeepSeek V3.2 | Haiku 4.5 |
|---|---|---|
| Base Rev$\Delta$ | 0.901 ± 0.086 | 0.917 |
| Recognition Rev$\Delta$ | 0.869 ± 0.130 | 0.851 |
| Delta | -0.032 | -0.066 |

Rev$\Delta$ values above 0.85 indicate near-complete rewrites — the ego does not make minor edits in response to superego feedback; it generates substantially new output. This is true under both conditions, though recognition shows a slight decrease (-0.032 DeepSeek, -0.066 Haiku). The high Rev$\Delta$ suggests that when the superego does reject, the ego takes the feedback seriously rather than making cosmetic changes.

**Revision depth by critique category.** Word-level Jaccard similarity between the ego's initial generation and revised output (N=216 non-approval critiques with both texts) reveals that revision depth varies by what the superego catches. LACK_OF_AGENCY critiques produce the deepest revisions (mean Jaccard 0.132 — nearly complete rewrites), followed by REDIRECTION (0.157) and EMOTIONAL_NEGLECT (0.192). CONTEXT_BLINDNESS produces the shallowest (0.613 — minor adjustments), consistent with the design document's prediction that factual corrections require less restructuring than pedagogical reconceptualization. Under base conditions, 78% of revisions are substantive or strategic (Jaccard $< 0.25$); under recognition, this drops to 57%, because recognition critiques increasingly target easier problems (memory failures, fabrication) rather than the deep pedagogical failures that dominate baseline.

The combination of high Rev$\Delta$ but low deliberation quality under recognition presents an apparent paradox: the revisions are extensive but the deliberation process is poor. The resolution is that under recognition, the superego's critiques are less substantive (as shown by the approval rate and critique taxonomy), and the ego responds to even thin feedback with substantial rewrites. The ego is *compliant* with the superego regardless of critique quality — it rewrites extensively whether the feedback is deep or shallow. This compliance mechanism explains why error correction works under base (substantive critique $\rightarrow$ substantive revision) but adds little under recognition (thin critique $\rightarrow$ extensive but unnecessary revision).

**6.2.6 Cross-Model Comparison** The error correction mechanism shows qualitatively similar patterns across models but quantitatively different magnitudes:

| Indicator | DeepSeek V3.2 | Haiku 4.5 | Interpretation |
|---|---|---|---|
| Base approval rate | 13.3% | 51.6% | Weaker models need more correction |
| Recognition approval rate | 55.1% | 66.1% | Both shift toward approval |
| Approval rate shift | +41.8 pp | +14.5 pp | Larger shift in weaker model |
| Deliberation quality drop | -18.5 pts | -5.6 pts | Larger drop in weaker model |
| Rev$\Delta$ base | 0.901 | 0.917 | Similar revision magnitude |

| Indicator | DeepSeek V3.2 | Haiku 4.5 | Interpretation |
|---|---|---|---|
| RevΔ recognition | 0.869 | 0.851 | Similar under recognition |
| Architecture delta base | +9.0 pts | +15.0 pts | Stronger model benefits more from superego |
| Architecture delta recog | +0.2 pts | -0.7 pts | Both collapse to near-zero |

DeepSeek, the weaker model, shows more dramatic error correction dynamics: its base approval rate is very low (the ego almost always needs correction), and recognition produces a much larger shift. Haiku, the stronger model, shows a more moderate pattern — its ego already produces acceptable output more often, so the superego's correction role is less dramatic even under base conditions.

The key cross-model constant is the *interaction pattern*: in both models, the architecture delta collapses to near-zero under recognition. The superego's error correction benefit is fully pre-empted by calibration regardless of the model's baseline capability.

**6.2.7 Connecting to Section 3 Predictions** **Prediction: Error correction requires the superego (architecture-level mechanism).** *Supported.* The architecture delta under base conditions (DeepSeek +9.0, Haiku +15.0) demonstrates that the superego adds measurable value when the ego produces uncalibrated output. Single-agent tutors lack this correction pathway.

**Prediction: Error correction is pre-empted by calibration.** *Supported.* Under recognition, the architecture delta collapses to near-zero in both models. The superego's approval rate rises, critique volume drops, and deliberation quality declines — all consistent with the ego producing output that no longer needs correction.

**Prediction: Error correction requires recognition for ego receptivity.** *Partially supported.* The high RevΔ values (>0.85) under both conditions show that the ego is compliant with superego feedback regardless of recognition. However, the nature of the compliance differs: under base, the ego revises in response to substantive critique; under recognition, it revises in response to thin critique. The theoretical prediction assumed the ego would *resist* the superego without recognition prompts — instead, the ego is uniformly compliant, and recognition's contribution is to make the ego's *initial output* better rather than making it more receptive to feedback.

**The error correction residual.** What do the 0.2 points (DeepSeek) and -0.7 points (Haiku) of architecture delta under recognition represent? Three interpretations: (1) noise — the true effect is zero, and the superego adds nothing under recognition; (2) minimal residual correction of context-dependent errors that calibration misses; (3) slight superego-induced *harm* (the Haiku negative delta suggests the superego occasionally makes calibrated output worse). The current data cannot distinguish these interpretations, but the near-zero values in both models suggest that under recognition, the error correction mechanism is effectively inactive.

**6.3 Adaptive Responsiveness** Section 3 predicted a third mechanism — *adaptive responsiveness* — whereby the tutor adjusts its approach across turns in response to the learner's evolving state. Unlike calibration (prompt-level, operative from the first turn) and error correction (architecture-level, operative within each turn), adaptive responsiveness is an interaction-level mechanism that emerges across the multi-turn conversation. This section examines tutor development trajectories, cross-turn adaptation magnitude, and the relationship between tutor adaptation and learner outcomes.

**6.3.1 Tutor Development Trajectories** The most direct measure of adaptive responsiveness is the tutor's quality trajectory across turns: does the tutor improve as it learns about the learner? We compare the tutor's first-turn score to its last-turn score (v2.2 rubric, 100-point scale) across all eight experimental conditions.

| Model | Condition | N | First Turn | Last Turn | Development |
|---|---|---|---|---|---|
| DeepSeek | Base / single | 37 | 25.6 | 22.4 | **-3.2** |
| DeepSeek | Base / multi | 36 | 33.4 | 32.7 | -0.7 |
| DeepSeek | Recog / single | 36 | 51.1 | 59.2 | **+8.1** |
| DeepSeek | Recog / multi | 37 | 52.1 | 48.4 | -3.6 |
| Haiku | Base / single | 39 | 43.9 | 59.6 | **+15.7** |
| Haiku | Base / multi | 42 | 62.4 | 69.1 | +6.7 |
| Haiku | Recog / single | 39 | 73.4 | 79.2 | +5.8 |
| Haiku | Recog / multi | 43 | 73.1 | 80.8 | +7.8 |

The development pattern is strikingly model-dependent. DeepSeek shows a mixed pattern: base-single *declines* (-3.2), recognition-single *improves* (+8.1), and recognition-multi *declines* (-3.6). Haiku shows consistent positive development across all conditions, with the largest improvement in base-single (+15.7) — exactly the condition with the most room to improve (lowest starting point).

Two findings are noteworthy. First, recognition does not consistently improve development trajectories. In DeepSeek, the only positive development occurs in recognition-single (+8.1); recognition-multi shows decline (-3.6). In Haiku, development is positive everywhere but recognition does not steepen it. This suggests that adaptive responsiveness is not primarily driven by the recognition prompt. Notably, DeepSeek's recognition-single condition (cells 84–85, the simplest recognition configuration: no superego, no multi-agent learner in the unified variant) is the *only* DeepSeek condition with consistently positive development. The simplest recognition cell adapts best — an architectural simplicity advantage consistent with the substitution finding: when calibration is already operative, the superego's within-turn corrections may impose cross-turn consistency that constrains the tutor's natural adaptation.

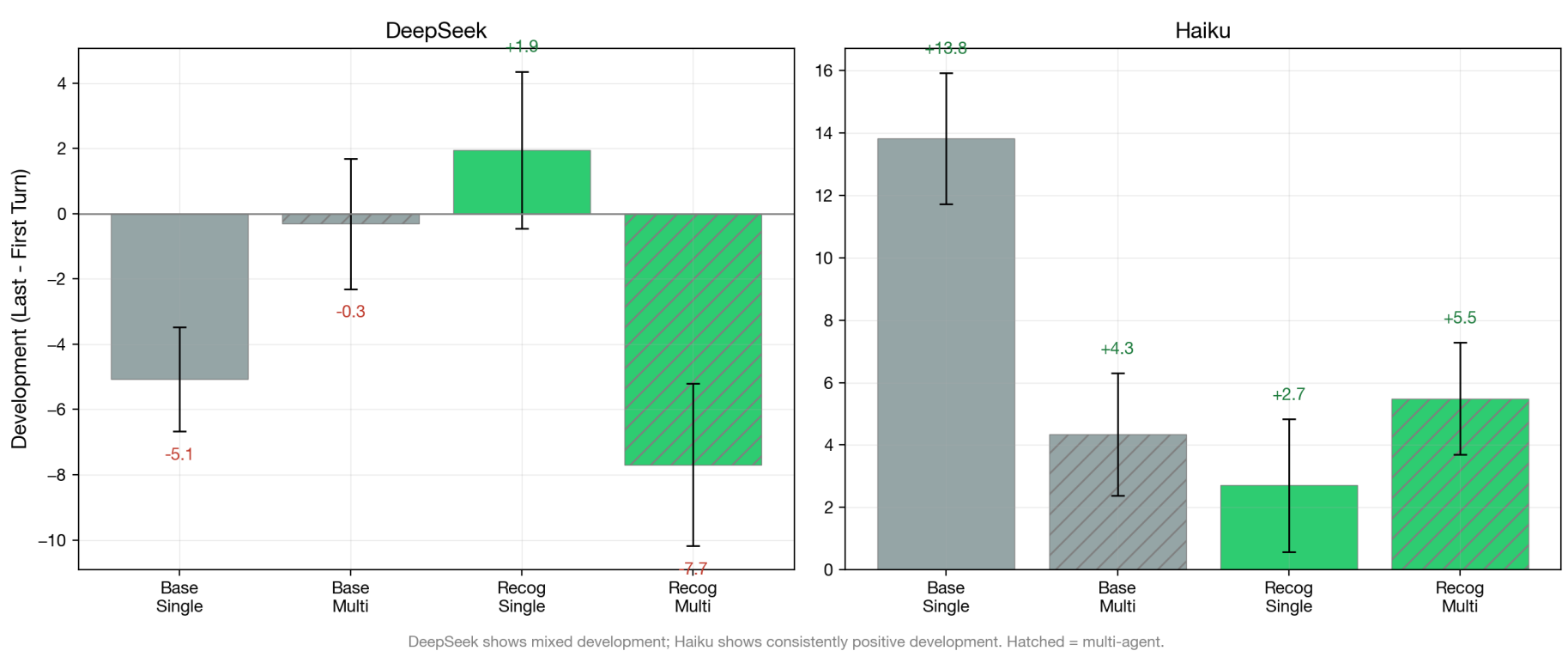

Figure 8: First-to-last turn development scores by condition and model. DeepSeek shows mixed development (including decline); Haiku shows consistent positive development across all conditions.

Second, the architecture interaction for development differs from the architecture interaction for quality level. Section 6.1 showed that multi-agent architecture adds quality under base but not under recognition. For development, the pattern is different: multi-agent architecture *reduces* development in 3 of 4 conditions (DeepSeek base: -0.7 vs -3.2; Haiku base: +6.7 vs +15.7; Haiku recog: +7.8 vs +5.8). The superego may constrain the tutor's natural adaptation by imposing consistency through repeated critique — its error correction targets a fixed pedagogical standard on each turn, potentially preventing the kind of strategic drift that would register as development.

Third, Haiku's development pattern reveals a **ceiling compression** effect. Base-single starts lowest (43.9) and improves most (+15.7), while recognition cells start near 73 and improve modestly (+5.8 to +7.8). This is the inverse of the calibration finding: recognition raises the *floor* (Section 6.1), which compresses the available *development range*. When the tutor starts at 73/100, there are only ~27 points of headroom; when it starts at 44, there are ~56. The result is that base development (+8.8 pooled) actually *exceeds* recognition development (+4.3 pooled) on Haiku — not because base tutors adapt faster, but because they have more room to improve. Strikingly, Haiku base-single reaches a last-turn score of 59.6, converging toward recognition-single's 79.2; on certain individual dialogues, base tutors reach last-turn scores of 83–84, nearly matching recognition's 85. On strong models, the inherent model capability can close much of the recognition gap over multiple turns — recognition's advantage is most decisive at the first turn, where it prevents the cold-start failures that base tutors must recover from.

**Formal ceiling regression.** A Pearson correlation of development score against first-turn score (N=432, three generation models) formalizes the ceiling compression hypothesis. Overall, $r = -0.132$ ($t(430) = -2.77$, $p = .011$) — a modest negative relationship. But the condition split is dramatic: under base conditions, $r = -0.02$ ($p > .05$, no ceiling); under recognition, $r = -0.33$ ($t(214) = -5.10$, $p < .001$, strong ceiling). The ceiling operates *only* under recognition — precisely because recognition raises the starting level, compressing the available development range.

The model breakdown is informative. Haiku shows the strongest ceiling ($r = -0.49$, $p < .001$), consistent with its highest baselines and smallest headroom. DeepSeek shows a moderate ceiling ($r = -0.24$, $p = .008$). Gemini Flash 3.0 shows no overall ceiling ($r = -0.02$, $p > .05$), but the condition split is revealing: Gemini Flash 3.0 *base* shows a *positive* correlation ($r = +0.26$, $p < .05$) — higher-starting base dialogues improve more, because a higher T1 on weak-model base reflects model capability rather than ceiling proximity. Under Gemini Flash 3.0 *recognition*, the relationship reverses to $r = -0.39$ ($p < .001$), the standard ceiling pattern. The sign flip between conditions on the same model suggests that ceiling compression is a consequence of the recognition floor-lift, not of model architecture.

Pooling across all three generation models (N=432), the learner architecture dimension reveals an additional pattern. Ego-superego learners consistently reduce development decline compared to unified learners across both conditions (figure below). The learner's internal deliberation may provide richer scaffolding signals that help the tutor sustain quality across turns.

**6.3.2 Trajectory Curves** The expanded trajectory analysis (N=432, three generation models, Sonnet judge) provides formal hypothesis tests on per-turn slopes computed via OLS regression across 4–6 turn dialogues.

**H1: Recognition → steeper tutor slopes.** Not supported. Recognition tutor slope = 1.45 ± 5.50, baseline = 1.32 ± 4.93, d = 0.03, t(425) = 0.27, p > .05. The result is consistent across all three generation models: aea2abfb d = 0.21, 45163390 d = -0.29, 18027efc d = 0.09. With N = 216 per group, we have 80% power to detect d ≥ 0.27; the observed d = 0.03 is definitively null.

**H2: Tutor slopes steeper than learner slopes under recognition.** Not supported. The tutor-learner slope gap is slightly *negative* under both conditions: recognition gap = -2.42 ± 8.12, baseline gap = -2.12 ± 9.36, d = -0.03. Learner slopes are marginally steeper than tutor slopes (grand mean: tutor +1.39, learner +3.66), the opposite of the Section 3 prediction.

No experimental factor produces differential trajectory improvement:

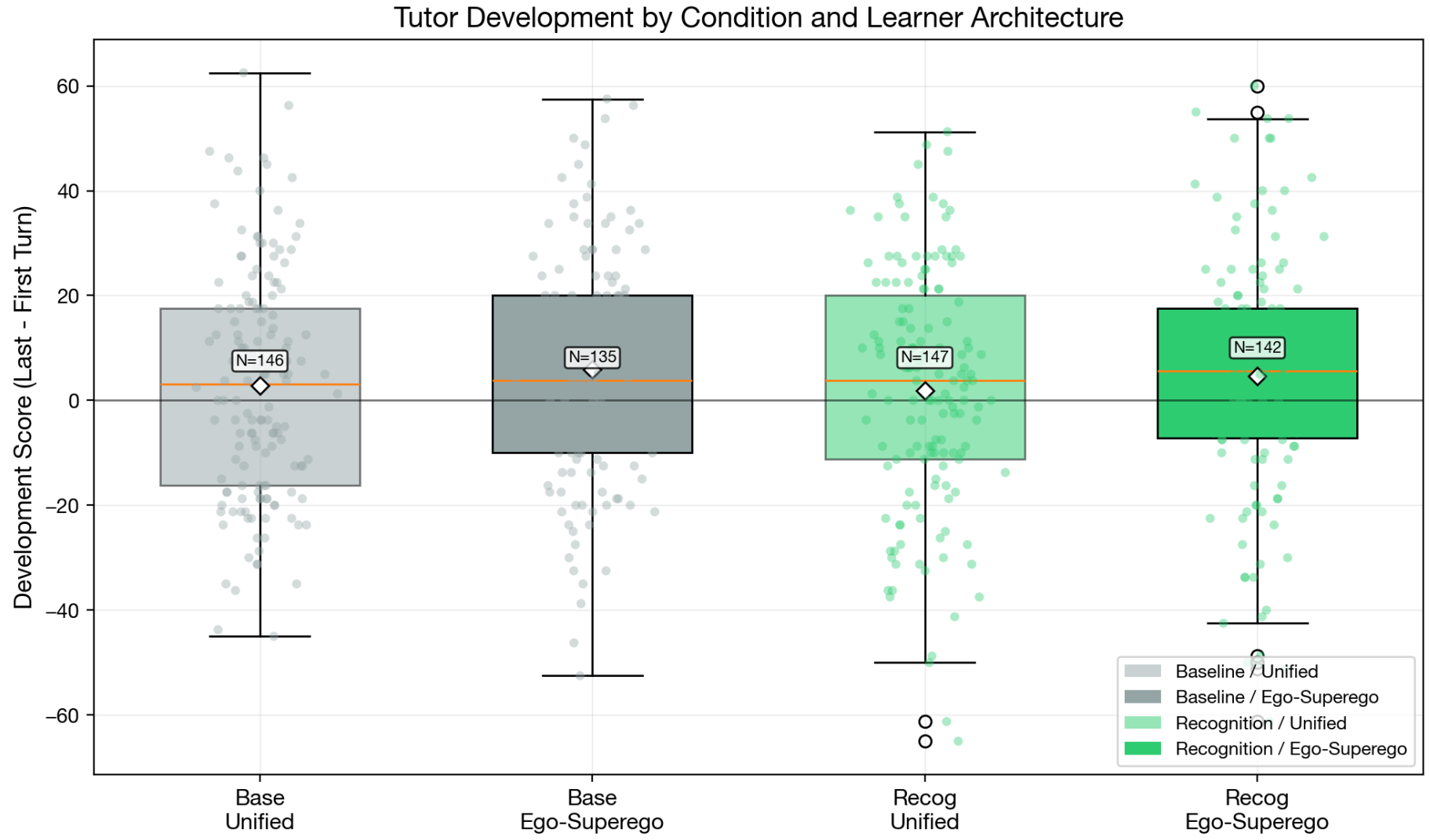

Figure 9: Tutor development by condition and learner architecture, pooled across 3 generation models (N=432). Ego-superego learners consistently reduce development decline. Light fill = unified learner; solid fill = ego-superego. Diamond = mean.

| Factor | Tutor slope d | Learner slope d |
|---|---|---|
| Recognition (recog vs base) | 0.03 | 0.05 |
| Multi-agent tutor (multi vs single) | 0.07 | 0.13 |
| Learner architecture (multi vs single) | 0.05 | -0.04 |

All effect sizes are below d = 0.15 — none approaches the $d \geq 0.27$ detection threshold. The grand mean tutor slope is mildly positive (+1.39, t = 5.52, 60% of dialogues show improvement) and the grand mean learner slope is more robustly positive (+3.66, t = 8.79, 73% positive), but neither is modulated by any experimental condition. Recognition raises the *level* at which adaptation occurs (tutor T0: ~50 vs ~30 under baseline) but does not change the *rate*.

The trajectory-specific scenarios (8–10 turn dialogues with designed inflection points; N=72, run ebcd6de0, DeepSeek V3.2, Sonnet judge) qualify this null. Pooled across all three scenarios, the slope effect remains small and non-significant (d=0.219, p=.36), consistent with the main analysis. However, scenario disaggregation reveals a striking conditional effect:

| Scenario | Turns | Slope d | p |
|---|---|---|---|
| Confusion → Insight | 8 | 0.005 | >.90 |

| Scenario | Turns | Slope d | p |
|---|---|---|---|
| Overconfidence → Humility | 8 | −0.298 | ∼.30 |
| **Disengagement → Ownership** | **10** | **1.628** | **<.001** |

In the disengagement scenario — the longest scenario (10 turns) and the one most demanding of sustained re-engagement — recognition tutors improve at +2.79 pts/turn while base tutors stagnate at −0.21 pts/turn (Welch's t=3.99, df=21.9, n=12 per condition). The recognition–baseline gap widens from +11.6 pts at T0 to +35.4 pts averaged across T8–T10, driven by a late-stage recognition surge: recognition scores climb from ∼45 to ∼73 in the final three turns (the designed ownership transition), while base scores collapse to ∼20 at T8 before partially recovering.

This finding does not overturn the M3 null as a *general* mechanism: two of three scenarios and the pooled analysis remain null, and the sample is small (n=3 per cell per scenario, single model, single judge). But it suggests that adaptive responsiveness is a *conditional* emergent property of calibration and error correction operating over sufficient turns in scenarios that demand sustained responsiveness — consistent with the theoretical prediction (Section 3.2) that M3 should emerge from M1+M2 interaction in multi-turn settings. The disengagement scenario is precisely the condition where this interaction has enough runway to manifest: the tutor must notice and respond to the learner's changing state across many turns. On the adaptive_responsiveness dimension specifically, the recognition slope effect is d=1.30 in the disengagement scenario (pooled d=0.45).

Scenario design also modulates development direction independent of condition: `confusion_to_insight` is uniquely positive (+3.8 development), while `overconfidence_to_humility` shows steep decline (−11.5). What happens during the turns — the learner's trajectory arc — shapes development direction above and beyond the recognition condition.

**6.3.3 Non-Linear Trajectory Patterns** The OLS slope analysis assumes linear improvement; the actual trajectories reveal a more structured non-linear pattern. Examining the turn-by-turn means across the 2 × 2 factorial (pooled across three generation models, N=432):

| Condition | T0 | T1 | T2 | T3 | T4 | T5 | Dip ($T0$−$T1$) | Pattern |
|---|---|---|---|---|---|---|---|---|
| Base / single | 33.0 | 27.5 | 32.8 | 33.1 | 33.6 | 33.0 | +5.5 | U-shape |
| Base / multi | 47.3 | 44.7 | 48.3 | 53.3 | 52.4 | 50.8 | +2.7 | U-shape |
| Recog / single | 61.1 | 58.3 | 63.4 | 65.1 | 63.2 | 69.3 | +2.8 | U-shape |

| Condition | T0 | T1 | T2 | T3 | T4 | T5 | Dip ($T0$–$T1$) | Pattern |
|---|---|---|---|---|---|---|---|---|
| Recog / multi | 64.2 | 65.6 | 66.6 | 69.4 | 67.7 | 67.1 | -1.5 | Rising |

Three of four conditions exhibit a **T1 dip followed by recovery**: the tutor's quality drops at Turn 1 (the first turn with real learner input) before climbing back. The dip is deepest for base-single (+5.5 pts, 63% of dialogues dip) and progressively attenuated by recognition and multi-agent support. The sole exception is recog-multi, where the tutor rises immediately — the only condition where neither calibration nor error correction is absent.

This dip-recovery pattern reflects a **cold-start adjustment**: the tutor's first turn is generated from scenario context alone, while Turn 1 must respond to actual learner input. Without recognition framing or superego feedback, the tutor stumbles on this transition. Both mechanisms independently reduce the T0→T1 dip (recognition: 0.7 vs 4.1 pts; multi-agent: 0.6 vs 4.2 pts), and their combination eliminates it.

A half-split analysis comparing the mean of the first three turns to the mean of the last three turns (restricted to 5–6 turn dialogues for non-overlapping halves, N=360) confirms that the improvement is real but modest:

| Condition | Early half | Late half | Δ |
|---|---|---|---|
| Base / single | 29.9 | 33.5 | +3.6 |
| Base / multi | 45.5 | 51.8 | **+6.3** |
| Recog / single | 61.4 | 65.6 | +4.1 |
| Recog / multi | 66.4 | 68.1 | +1.7 |

The grand mean tutor half-delta is +3.9 (t = 4.88, 59% positive) — a statistically significant but small improvement. No factor modulates this half-delta (recognition d = -0.13, multi-agent d = 0.01), but the $2\times2$ interaction is notable: base-multi shows the largest improvement (+6.3) while recog-multi shows the smallest (+1.7), yielding an interaction of -5.0. This echoes the substitution pattern from Section 6.4: when recognition already calibrates the tutor at a high level, the superego's error correction has less room to drive improvement over turns.

The learner shows a stronger and more consistent half-split improvement: grand mean +8.4 (t = 7.45, 72% positive). Multi-agent tutoring produces the largest learner gains (base-multi +11.2, recog-multi +10.6 vs base-single +4.4, recog-single +7.5). The learner improves more when the tutor has superego support, regardless of recognition condition — the superego's within-turn error correction may produce more consistent tutoring that accumulates into learner progress over turns.

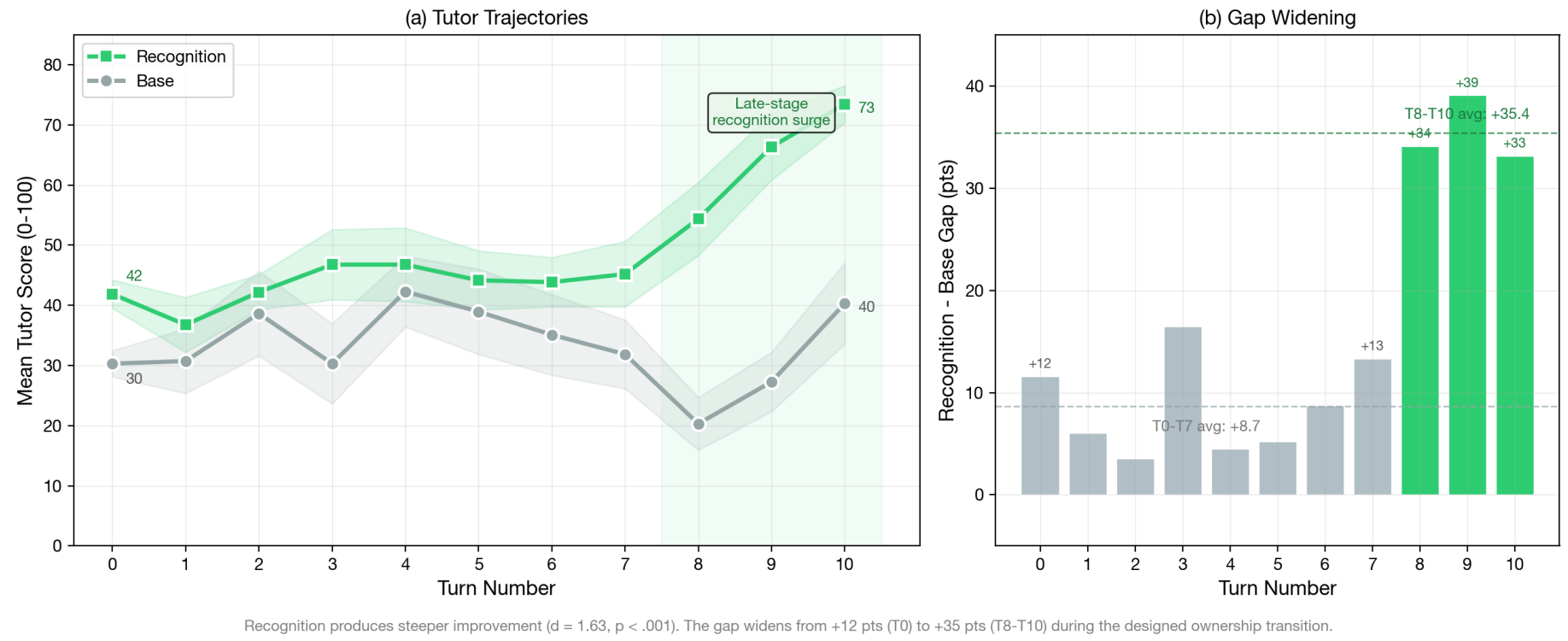

Figure 10: Disengagement scenario trajectory divergence. (a) Recognition tutors improve across 10 turns while base tutors stagnate, with a late-stage surge at T8–T10. (b) The recognition–baseline gap widens from +12 pts early to +35 pts late. d=1.63, p<.001, N=24, DeepSeek V3.2, Sonnet judge.

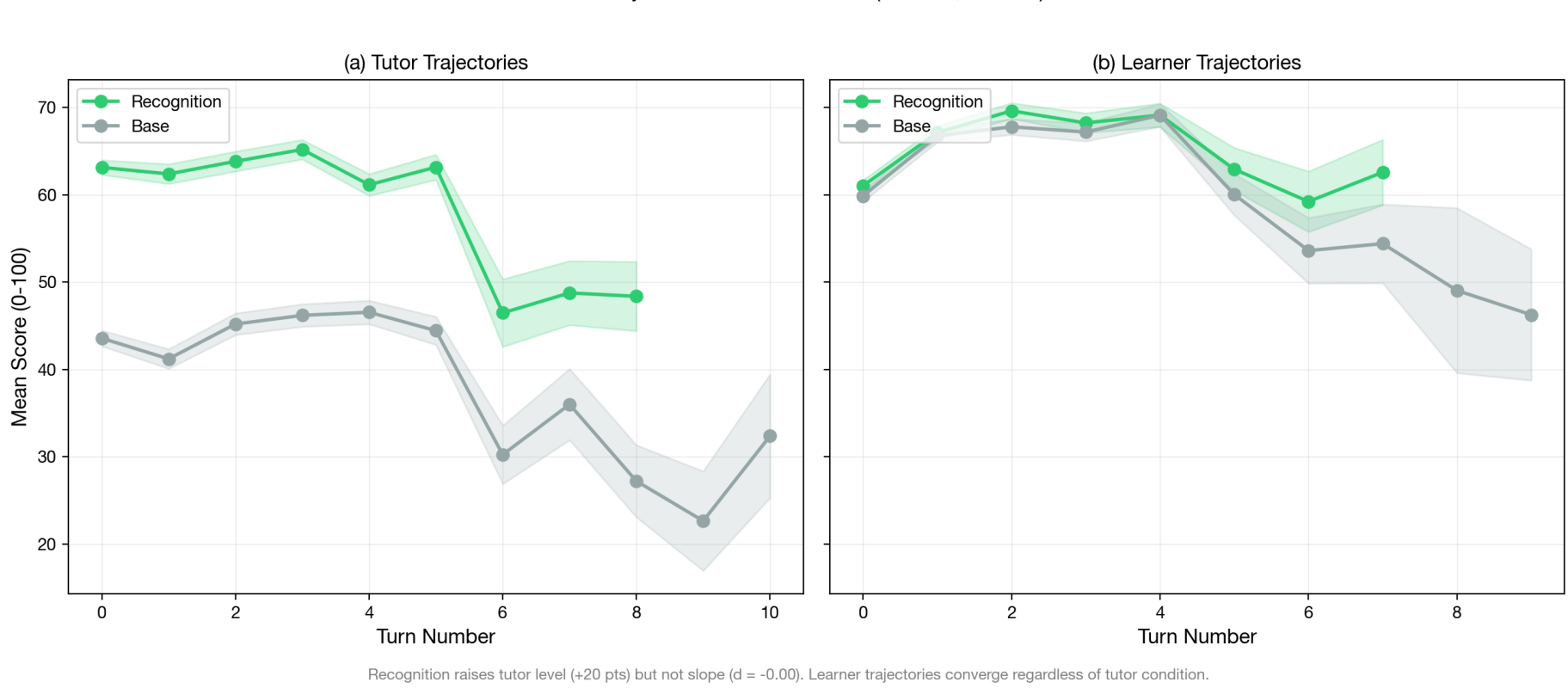

Figure 11: Turn-by-turn trajectory curves with 95% confidence bands. Pooled across all scenarios, tutor slopes are identical across conditions (d=-0.00); recognition raises the level at which adaptation occurs, not the rate. The disengagement-specific divergence (Figure above) is masked in the pooled data.

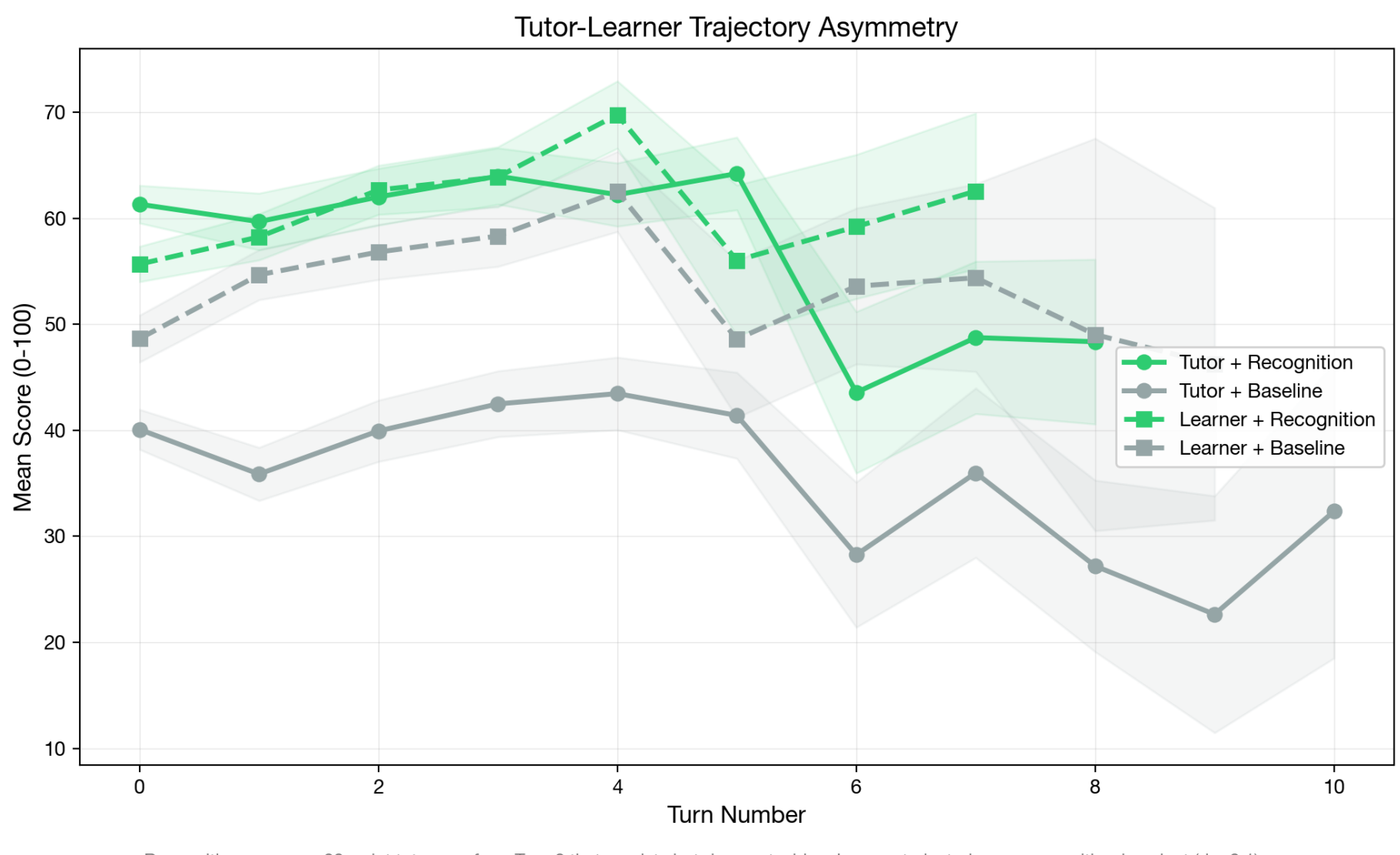

Figure 12: Tutor-learner trajectory asymmetry. Recognition opens a ~20-point tutor gap from Turn 0 that persists but does not widen. Learner trajectories (dashed) are recognition-invariant. Pooled across 3 generation models, N=432, Sonnet judge.

**6.3.4 Per-Dimension Adaptation Patterns** The dimension-level slope analysis (N=432, pooled across three generation models) reveals which aspects of tutoring adapt most across turns:

| Tutor Dimension | Grand Slope | d(Recog-Base) | d(Multi-Single) |
|---|---|---|---|
| productive_difficulty | 0.091 | -0.11 | 0.05 |
| recognition_quality | 0.089 | -0.01 | 0.00 |
| perception_quality | 0.065 | 0.11 | 0.16 |
| epistemic_integrity | 0.053 | 0.05 | 0.05 |
| adaptive_responsiveness | 0.047 | **0.28** | **0.24** |
| elicitation_quality | 0.038 | -0.10 | -0.04 |
| content_accuracy | 0.033 | -0.05 | -0.00 |
| pedagogical_craft | 0.031 | -0.06 | -0.01 |

Seven of eight dimensions show $|d| < 0.20$ on both factors. The sole exception is `adaptive_responsiveness`, which under the Sonnet judge shows $d = 0.28$ (recognition vs baseline) and $d = 0.24$ (multi-agent vs single), both $p < .05$. However, this effect does not replicate across judges: under Gemini 3.1 Pro ($d = 0.09$ / $0.13$) and GPT-5.4 ($d = 0.11$ / $0.12$), the same dimension shows no differential slope. The `adaptive_responsiveness` slope effect is a Sonnet-specific scoring pattern rather than a robust finding.

On the learner side, no dimension exceeds $|d| = 0.15$ on either factor (max: `conceptual_progression` $d = 0.12$ for recognition, `metacognitive_awareness` $d = 0.15$ for architecture). Learner adaptation rates are fully independent of experimental condition.

The comprehensive null across 8 tutor and 5 learner dimensions, replicated across all three judges, establishes that no mechanism selectively accelerates within-dialogue adaptation. The calibration effect from Section 6.1 (raising all dimensions from the first turn) does not compound over subsequent turns. Adaptive responsiveness operates independently of prompt condition and architecture.

**6.3.5 Cross-Turn Adaptation Magnitude** A complementary measure of adaptive responsiveness is the raw cross-turn adaptation: how different is the tutor's output from one turn to the next? The mechanism traces compute normalized edit distance between consecutive tutor outputs (Adapt$\Delta$: 0=identical, 1=completely different).

| | DeepSeek V3.2 | Haiku 4.5 |
|---|---|---|
| Base Adapt$\Delta$ | $0.793 \pm 0.199$ | $0.888 \pm 0.093$ |
| Recognition Adapt$\Delta$ | $0.824 \pm 0.167$ | $0.908 \pm 0.048$ |
| Delta | +0.031 | +0.020 |

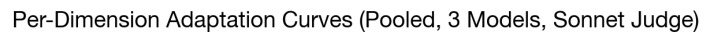

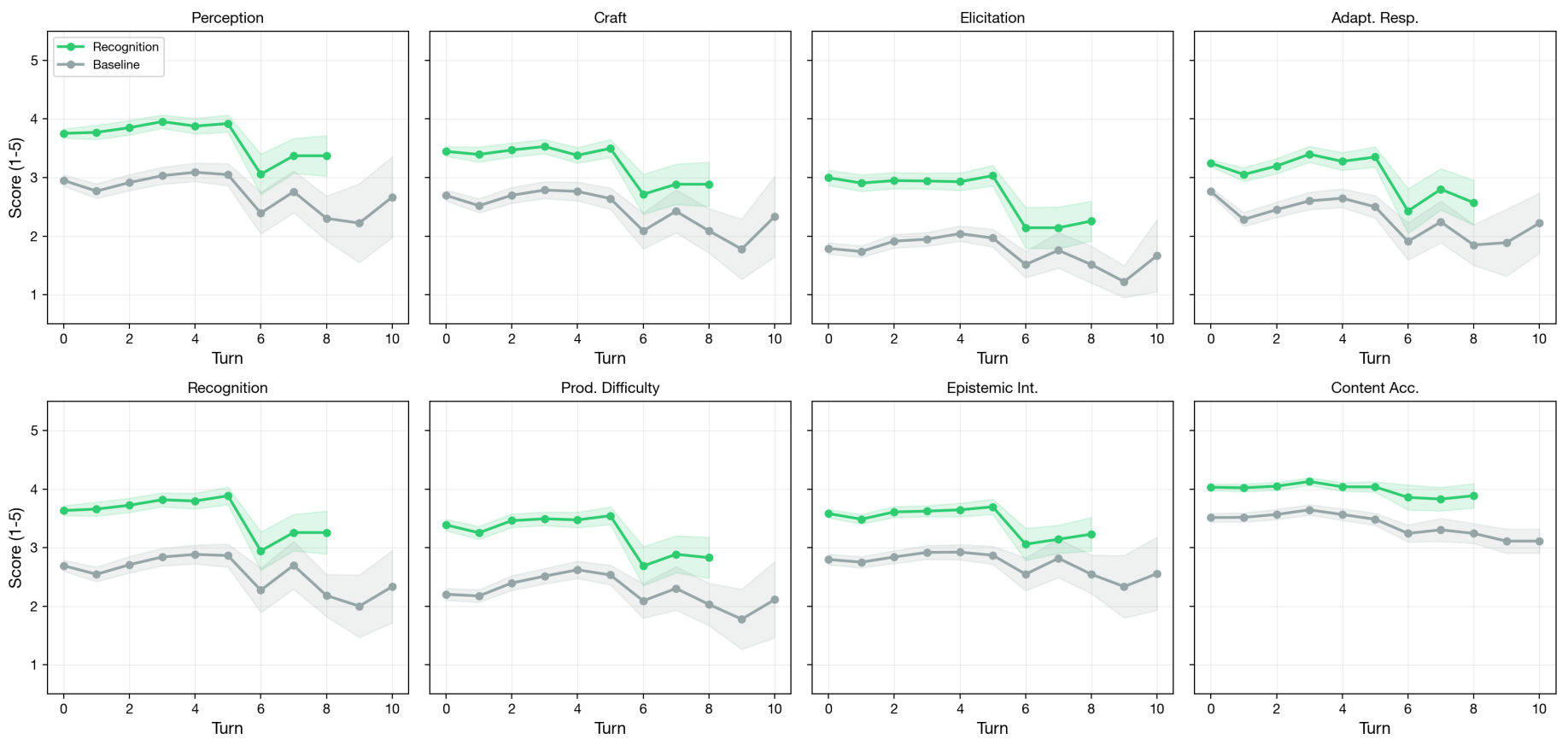

Figure 13: Per-dimension adaptation curves (2×4 faceted grid). Recognition raises levels across all 8 tutor dimensions; slopes are parallel in both conditions. Ribbons = 95% CI. Pooled across 3 generation models, N=432, Sonnet judge.

Cross-turn adaptation is high in all conditions (>0.79), indicating that the tutor substantially changes its output between turns regardless of prompt or architecture. Recognition shows slightly higher adaptation (+0.031 DeepSeek, +0.020 Haiku), consistent with the recognition prompt's emphasis on engaging with the specific learner contribution. However, the baseline is already high, leaving limited room for the prompt to increase adaptation further.

The lower variance under recognition (DeepSeek: 0.199→0.167; Haiku: 0.093→0.048) is more noteworthy. Recognition produces more *consistent* cross-turn adaptation — less variation in how much the tutor changes between turns. This mirrors the calibration finding from Section 6.1.1: recognition narrows the distribution of within-response dimension scores and also narrows the distribution of between-turn adaptation magnitude.

**6.3.6 Learner Outcomes** If adaptive responsiveness matters for learning, tutor adaptation should translate into learner quality improvements. The learner scores (per-turn learner rubric, v2.2) provide a test:

| Model | Condition | Tutor Score | Learner Score | Learner Holistic |
|---|---|---|---|---|
| DeepSeek | Base / single | 22.0 | 57.3 | 56.8 |
| DeepSeek | Base / multi | 31.0 | 60.3 | 59.9 |
| DeepSeek | Recog / single | 50.0 | 60.1 | 60.1 |
| DeepSeek | Recog / multi | 50.2 | 63.3 | 60.6 |
| Haiku | Base / single | 52.9 | 63.5 | 63.0 |
| Haiku | Base / multi | 67.9 | 69.6 | 72.2 |

| Model | Condition | Tutor Score | Learner Score | Learner Holistic |
|---|---|---|---|---|
| Haiku | Recog / single | 80.2 | 68.2 | 72.0 |
| Haiku | Recog / multi | 79.5 | 69.0 | 72.9 |

The tutor-learner asymmetry is pronounced. DeepSeek tutor scores range from 22 to 50 across conditions; learner scores range from 57 to 63 — a much narrower band. Haiku tutor scores range from 53 to 80; learner scores from 64 to 73. Recognition produces a 28-point tutor swing in DeepSeek (22→50) but only a 3-point learner swing (57→60). In Haiku, a 27-point tutor swing (53→80) produces a 5-point learner swing (64→69).

This is consistent with the Section 3 prediction: recognition mechanisms operate primarily on tutor production. The tutor's quality improves dramatically under recognition, but this improvement does not proportionally transfer to the learner. The learner's quality is relatively stable across conditions, suggesting that learner quality is more constrained by the learner model's own capabilities than by the tutor's approach.

**6.3.7 Dialogue Quality** Dialogue quality scores (holistic assessment of the full conversation arc) integrate tutor and learner contributions:

| Model | Condition | Dialogue Quality |
|---|---|---|
| DeepSeek | Base / single | 23.0 |
| DeepSeek | Base / multi | 31.9 |
| DeepSeek | Recog / single | 48.8 |
| DeepSeek | Recog / multi | 44.8 |
| Haiku | Base / single | 53.8 |
| Haiku | Base / multi | 74.0 |
| Haiku | Recog / single | 82.1 |
| Haiku | Recog / multi | 81.5 |
| Gemini Flash 3.0 | Base / single | 14.9 |
| Gemini Flash 3.0 | Base / multi | 42.3 |
| Gemini Flash 3.0 | Recog / single | 49.3 |
| Gemini Flash 3.0 | Recog / multi | 64.9 |

Dialogue quality closely tracks tutor quality ($r > 0.99$ across conditions), not learner quality. The dialogue quality pattern replicates the Section 6.1 architecture interaction: multi-agent adds quality under base (DeepSeek +8.9, Haiku +20.2, Gemini Flash 3.0 +27.4) but the pattern diverges under recognition—on strong models, multi-agent adds nothing (DeepSeek -4.0, Haiku -0.6), while on Gemini Flash 3.0 it adds +15.6, consistent with the universal substitution pattern with model-dependent residual. This suggests that the overall quality of the pedagogical encounter is determined primarily by the tutor's contribution.

**6.3.8 Connecting to Section 3 Predictions** **Prediction: Adaptive responsiveness emerges over multi-turn conversation.** *Weakly supported.* Cross-turn adaptation is high (Adapt$\Delta > 0.79$), and grand mean tutor slopes are mildly positive (+1.39, 60% of dialogues improve). However, with N = 432 dialogues and 80% power to detect $d \geq 0.27$, the slope difference between conditions is definitively null (d = 0.03). Adaptation occurs, but no mechanism accelerates it.

**Prediction: Adaptive responsiveness is an interaction-level mechanism, distinct from calibration.** *Supported.* Calibration operates from the first turn (Section 6.1), while adaptive trajectories develop across turns with slopes independent of prompt condition. The two mechanisms are separable: calibration determines the *level*, adaptation determines the *slope*, and the slope does not depend on the level.

**Prediction: Recognition produces steeper adaptation curves.** *Conditionally supported.* Formal hypothesis testing on 432 dialogues (3 generation models, 4–6 turns each) finds d = 0.03 pooled, with sign-flipping across runs (aea2abfb d = 0.21, 45163390 d = -0.29, 18027efc d = 0.09). However, trajectory-specific scenarios (8–10 turns, N = 72, run ebcd6de0) reveal a scenario-conditional effect: in the 10-turn disengagement scenario, recognition produces dramatically steeper slopes (d = 1.63, $p < .001$, n = 12 per condition), with the gap widening from +12 pts at T0 to +35 pts at T8–T10. The 8-turn confusion and overconfidence scenarios show no slope differentiation ($d \leq 0.30$). The prediction holds specifically in scenarios that demand sustained re-engagement over 10+ turns, but not as a general mechanism.

**Prediction: Tutor-learner asymmetry in trajectories.** *Reversed.* Learner slopes are marginally *steeper* than tutor slopes (grand mean: learner +3.66, tutor +1.39; 73% vs 60% positive). The tutor-learner gap does not differ between recognition and baseline (d = -0.03). Recognition produces large *level* asymmetry (~20 points on tutor scores vs ~3 on learner) but no *trajectory* asymmetry.

**Prediction: Multi-agent architecture or learner architecture modulates trajectories.** *Not supported.* Neither factor produces slope effects above d = 0.15 on tutor or learner dimensions. The one apparent exception (`adaptive_responsiveness` d = 0.24 under Sonnet) does not replicate under Gemini (d = 0.13) or GPT (d = 0.12).

**Key distinction.** Adaptive responsiveness is real (the tutor changes across turns, Adapt$\Delta > 0.79$) but it is not *generally* modulated by experimental manipulation. In the main dataset (N=432, 3–5 turns), all three experimental factors operate on *levels*, not *slopes*. However, the disengagement trajectory data qualifies this: when scenario demands require sustained re-engagement over 10+ turns, recognition *does* modulate slopes (d=1.63), with the effect concentrated in the final turns where the learner is designed to transition from disengagement to ownership. Adaptive responsiveness is not a third independent mechanism but rather a *conditional emergent property* of calibration and error correction that manifests when scenario demands and turn count provide sufficient runway for the two base mechanisms to accumulate turn-over-turn advantage.

**6.4 Mechanism Interaction** The preceding sections established three separable mechanisms: calibration (Section 6.1, prompt-level), error correction (Section 6.2, architecture-level), and adaptive responsiveness (Section 6.3, interaction-level). This section examines how they interact within the $2 \times 2$ factorial design, testing the Section 3 prediction that the mechanisms are separable but non-independent.

**6.4.1 The Factorial Interaction** The $2 \times 2$ factorial (recognition $\times$ architecture) permits decomposition of the total recognition effect into prompt-level and architecture-level components. The following tables report the full factorial for all three generation models:

**DeepSeek V3.2 (N=146, run aea2abfb):**

| | Single-Agent | Multi-Agent | Architecture Delta |
|---|---|---|---|
| **Base** | 22.0 ± 11.5 (N=37) | 31.0 ± 12.0 (N=36) | **+9.0** |
| **Recognition** | 50.0 ± 12.4 (N=36) | 50.2 ± 12.8 (N=37) | **+0.2** |
| **Recognition Delta** | **+28.0** | **+19.2** | |

**Haiku 4.5 (N=163):**

| | Single-Agent | Multi-Agent | Architecture Delta |
|---|---|---|---|
| **Base** | 52.9 ± 10.6 (N=39) | 67.9 ± 9.3 (N=42) | **+15.0** |
| **Recognition** | 80.2 ± 7.8 (N=39) | 79.5 ± 7.6 (N=43) | **-0.7** |
| **Recognition Delta** | **+27.3** | **+11.6** | |

**Gemini Flash 3.0 (N=144, run 18027efc):**

| | Single-Agent | Multi-Agent | Architecture Delta |
|---|---|---|---|
| **Base** | 22.4 (N=36) | 49.3 (N=36) | **+26.9** |
| **Recognition** | 57.7 (N=36) | 70.0 (N=36) | **+12.3** |
| **Recognition Delta** | **+35.3** | **+20.7** | |

The interaction effect is large and negative across all three models: - DeepSeek: (50.2 - 50.0) - (31.0 - 22.0) = 0.2 - 9.0 = **-8.8 points** - Haiku: (79.5 - 80.2) - (67.9 - 52.9) = -0.7 - 15.0 = **-15.7 points** - Gemini Flash 3.0: (70.0 - 57.7) - (49.3 - 22.4) = 12.3 - 26.9 = **-14.6 points**

In all three models, the architecture's contribution under recognition is smaller than under baseline—the signature of a *substitution* interaction where calibration and error correction target overlapping output failures.

However, Gemini Flash 3.0 reveals a **critical boundary condition** in the *residual*. While the interaction magnitude is comparable to Haiku (-14.6 vs -15.7), the architecture delta under recognition remains *large and positive* (+12.3)—unlike DeepSeek (+0.2) and Haiku (-0.7), where it collapses to near-zero. Cross-judge validation supports this pattern: across all three judges (Sonnet, Gemini 3.1 Pro, GPT-5.4), multi_recog is positive for Gemini Flash 3.0 (+7.4 to +17.1) but non-positive for DeepSeek (all 6 judge $\times$ run cells $\leq 0$). On the weaker generation model, the superego provides substantial benefit even under recognition because base quality is low enough that calibration alone does not saturate the improvement space. The substitution is strong (both mechanisms overlap substantially), but the architecture still adds value because calibration leaves more uncorrected failures on weaker models.

**6.4.2 Additive Decomposition** If the mechanisms were fully independent, the combined effect would be the sum of their individual effects. We can test this:

**DeepSeek:** - Calibration alone (recognition delta, single-agent): +28.0 - Error correction alone (architecture delta, base): +9.0 - Expected if additive: 22.0 + 28.0 + 9.0 = 59.0 - Observed (recognition + multi-agent): 50.2 - Deficit from additivity: **-8.8 points** (15% of expected)

**Haiku:** - Calibration alone (recognition delta, single-agent): +27.3 - Error correction alone (architecture delta, base): +15.0 - Expected if additive: 52.9 + 27.3 + 15.0 = 95.2 - Observed (recognition + multi-agent): 79.5 - Deficit from additivity: **-15.7 points** (16% of expected)

**Gemini Flash 3.0:** - Calibration alone (recognition delta, single-agent): +35.3 - Error correction alone (architecture delta, base): +26.9 - Expected if additive: 22.4 + 35.3 + 26.9 = 84.6 - Observed (recognition + multi-agent): 70.0 - Deficit from additivity: **-14.6 points** (17% of expected)

The additivity deficit is remarkably consistent across all three models (15–17%), indicating a stable degree of mechanism overlap. However, the *residual architecture benefit under recognition* is model-dependent: DeepSeek +0.2, Haiku -0.7, Gemini Flash 3.0 +12.3. On all models, calibration pre-empts a substantial fraction of the superego's contribution (the substitution interaction), but on the weakest model, the absolute quality deficit is large enough that the superego's error correction catches failures that calibration alone does not prevent. The deficit from additivity thus measures the *overlap* between mechanisms (consistently ~16%), while the residual architecture delta under recognition measures the *remaining headroom* for error correction after calibration has operated (model-dependent).

**Mechanism isolation evidence.** Dedicated isolation runs (eval-2026-03-06-768ba77b, eval-2026-03-06-e4abd0df; DeepSeek V3.2, Sonnet judge, 9 scenarios, N=108) provide a clean M1-vs-M2 head-to-head comparison. Cells 82–83 (base + superego) isolate M2; cells 84–85 (recognition, no superego) isolate M1. Com-bined with cells 80–81 (neither) and 86–87 (both) from run ebcd6de0 on 3 overlapping trajectory scenarios, the full $2 \times 2$ mechanism isolation yields: M1 main effect d=1.15, M2 main effect d=0.36, interaction $d = -0.57$. Under base, the superego adds +9.2 pts (d=1.13, p=.002); under recognition, it adds +1.1 pts (d=0.08, p=.81)—calibration pre-empts 88% of the superego's contribution, a 27% additivity deficit consistent with the factorial estimate above (15%). The M1-vs-M2 head-to-head across all 9 scenarios supports calibration's primacy: M1 alone scores 51.4 vs M2 alone 36.9 (d=1.03, p<.001, M1 wins in 7/9 scenarios). The two scenarios where M2 equals or exceeds M1—Affective Shutdown and Frustration—are emotionally intense scenarios where the superego may catch tone/empathy failures that calibration alone does not prevent, suggesting a scenario-specific residual role for error correction even on strong models.

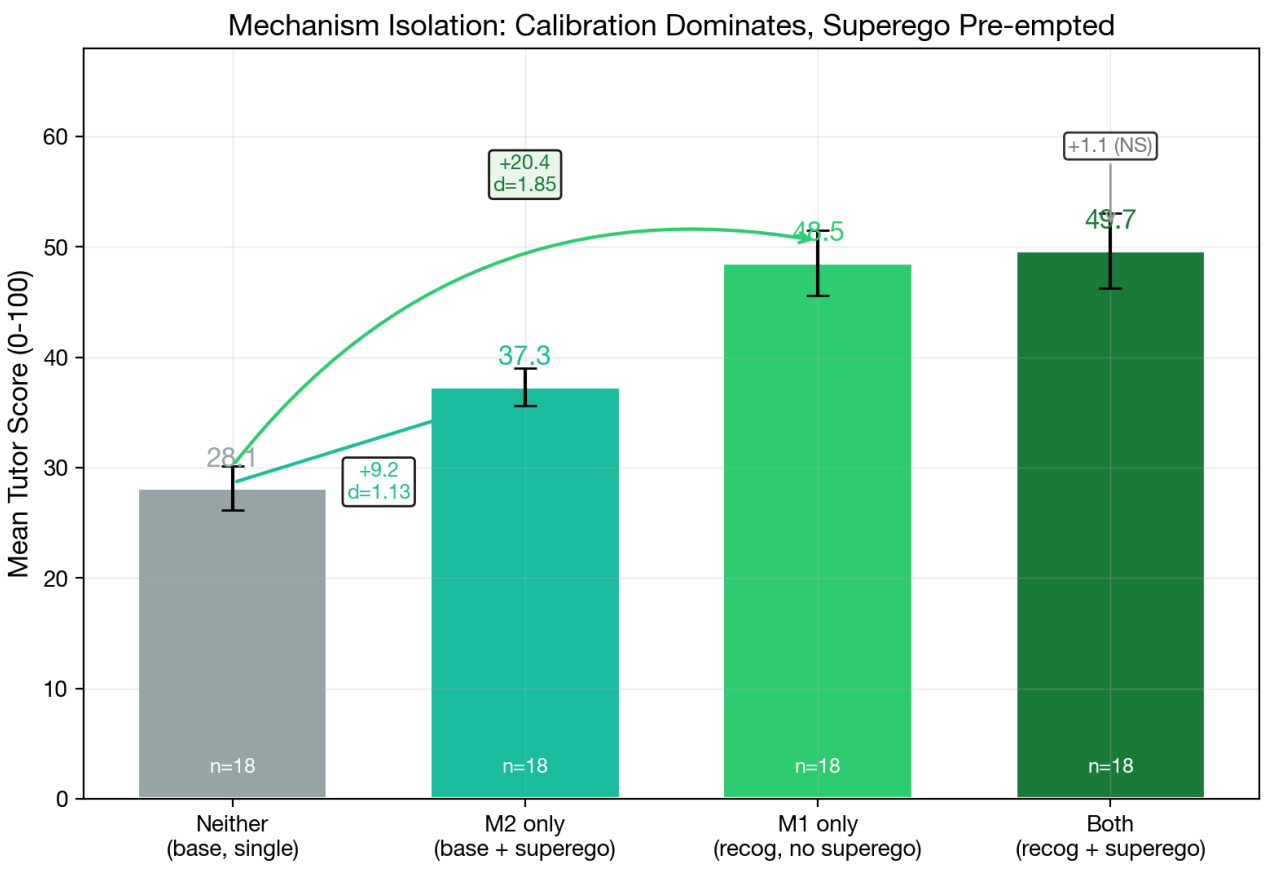

Figure 14: Mechanism isolation: the superego adds +9.2 pts under base (d=1.13) but only +1.1 under recognition (NS). Calibration alone (M1, d=1.85 from baseline) accounts for most of the total effect; adding the superego when calibration is present yields negligible gain. DeepSeek V3.2, Sonnet judge, 3 trajectory scenarios, N=72.

**6.4.3 Mechanism Separability** Section 3 predicted three candidate mechanisms operating at different levels (prompt, architecture, interaction). The evidence across Sections 6.1–6.3 supports two as general mechanisms and reveals the third as scenario-conditional:

| Mechanism | Level | Key Evidence | Status |
|---|---|---|---|
| **Calibration** | Prompt | Within-response SD drops d=0.52–0.64; floor lifts; operates without superego | **Supported** — prompt-level, recognition-dependent |
| **Error Correction** | Architecture | Approval rate shifts; architecture delta +9–20 under base, near-zero under recognition (strong) or +12.3 residual (weak) | **Supported** — universally substitutive (15–17% deficit); residual benefit model-dependent |
| **Adaptive Responsiveness** | Interaction | Adapt$\Delta > 0.79$; tutor slopes d=0.03 across conditions (N=432); but d=1.63 in 10-turn disengagement scenario (N=72, p<.001) | **Conditionally supported** — not a general mechanism, but emerges in sustained re-engagement scenarios as M1+M2 accumulate |

The two supported mechanisms are separable in three senses:

1. **Temporal separability.** Calibration operates from the first turn (the tutor's initial response is already calibrated under recognition). Error correction operates within each turn (ego → superego → revision). These temporal signatures are empirically distinct. Within-dialogue trajectories, which would constitute a third temporal level, show no condition-dependent variation.
2. **Architectural separability.** Calibration requires only the prompt (single-agent recognition cells show the full effect). Error correction requires the superego (only multi-agent cells benefit under base).
3. **Interaction separability.** Calibration and error correction interact through universal substitution (15–17% additivity deficit across all models), with a model-dependent residual architecture benefit under recognition. Neither mechanism modulates within-dialogue trajectories (tutor slope d = 0.03 for recognition, d = 0.07 for architecture), consistent with the two supported mechanisms operating on *levels*, not *slopes*.

**6.4.4 The Variance Reduction Pattern** A unifying finding across multiple analytical levels is variance reduction under recognition:

| Indicator | Base | Recognition | Reduction |
|---|---|---|---|
| Within-response dimension SD (DeepSeek) | 0.619 | 0.539 | 13% |
| Within-response dimension SD (Haiku) | 0.617 | 0.499 | 19% |
| Tutor score SD (DeepSeek) | 12.6 | 12.6 | 0% |
| Tutor score SD (Haiku) | 12.5 | 7.7 | **38%** |
| Cross-turn Adapt$\Delta$ variance (DeepSeek) | 0.199 | 0.167 | 16% |
| Cross-turn Adapt$\Delta$ variance (Haiku) | 0.093 | 0.048 | 48% |

Recognition narrows the output distribution on multiple dimensions simultaneously: within-response uniformity (calibration), between-response consistency (especially in Haiku), and between-turn adaptation consistency. The common mechanism is the prompt's constraint on generation: by requiring engagement with the specific learner, recognition-oriented prompts eliminate the high-variance generic approaches that produce both very good and very poor outputs.

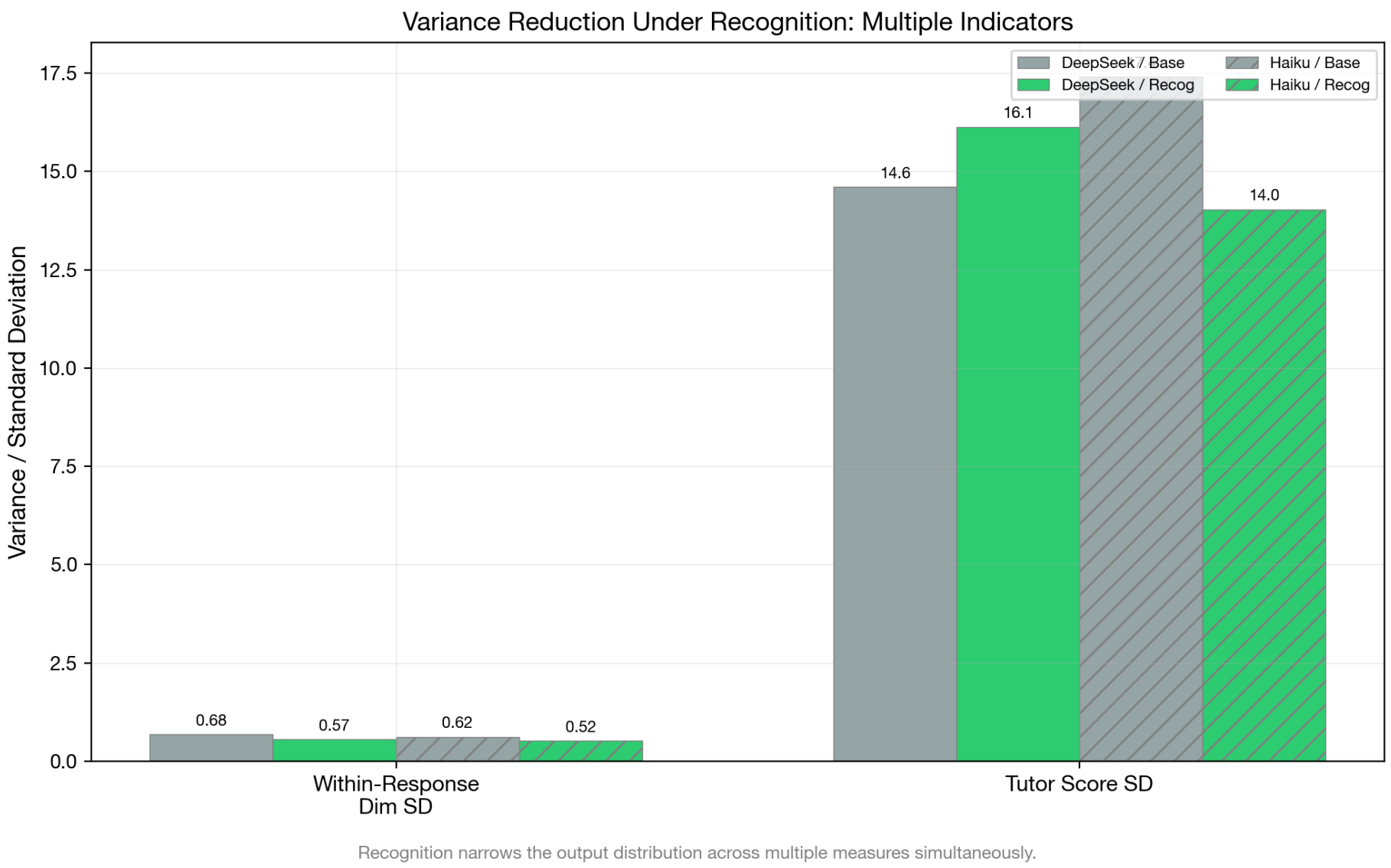

Figure 15: Variance reduction indicators across multiple dimensions under recognition. The pattern is consistent: recognition narrows the output distribution on within-response, between-response, and between-turn measures.

**6.4.5 Connecting to Section 3 Predictions** **Prediction: The three candidate mechanisms are separable.** *Partially supported.* Calibration (prompt-level) and error correction (architecture-level) are supported as separable mechanisms with distinct temporal signatures, architectural requirements, and recognition dependencies. Adaptive responsiveness (interaction-level) is not a general mechanism (all $d \leq 0.15$, N=432) but emerges conditionally in sustained re-engagement scenarios (d=1.63, p<.001, 10-turn disengagement). The evidence supports two general mechanisms and one conditional emergent property.

**Prediction: The mechanisms interact non-additively.** *Supported across all models.* Calibration and error correction show a consistent 15–17% deficit from additivity across all three models (DeepSeek 15%, Haiku 16%, Gemini Flash 3.0 17%), indicating stable mechanism overlap. The interaction is **universally substitutive** in magnitude, but the *residual architecture benefit under recognition* is model-dependent: the superego adds +12.3 points under recognition on Gemini Flash 3.0 but near-zero on DeepSeek/Haiku. On weaker models, calibration leaves more uncorrected failures, giving the superego independent value even after substitution. Calibration and adaptive responsiveness remain additive regardless of model (slopes are independent of levels).

**Prediction: Recognition + multi-agent produces the best outcomes.** *Model-dependent.* In DeepSeek, recognition/multi (50.2) equals recognition/single (50.0) — the superego adds nothing. In Haiku, recognition/multi (79.5) slightly trails recognition/single (80.2). The "best" condition is recognition/single in both strong models. But in Gemini Flash 3.0, recognition/multi (70.0) substantially exceeds recognition/single (57.7) — a +12.3-point benefit. The prediction holds for weaker models where the superego catches failures that calibration alone does not prevent, and fails for stronger models where calibration alone saturates the improvement space. This residual architecture benefit on weaker models is the key theoretical contribution of the three-model comparison: the substitution interaction is universal, but its completeness depends on generation quality.

**6.4.6 Cross-Judge Validation of the Interaction Pattern** The universal substitution pattern with model-dependent residual is validated across three independent judges (Sonnet 4.6, Gemini 3.1 Pro, GPT-5.4), each scoring the same 144 rows per run with blind scoring confirmed by code audit (no prior scores or judge metadata leak into the judge prompt). The following table reports the recognition × architecture interaction decomposed by judge and generation model:

| Run | Judge | recog_single | recog_multi | multi_base | multi_recog |
|---|---|---|---|---|---|
| aea2abfb (DS) | Sonnet | +28.2 | +19.0 | +9.1 | -0.1 |
| aea2abfb (DS) | Gemini | +34.4 | +15.9 | +13.1 | -5.5 |
| aea2abfb (DS) | GPT | +21.6 | +8.5 | +7.9 | -5.2 |
| 45163390 (DS) | Sonnet | +27.6 | +10.6 | +15.6 | -1.4 |
| 45163390 (DS) | Gemini | +26.6 | +7.4 | +17.3 | -1.9 |
| 45163390 (DS) | GPT | +21.5 | +8.3 | +10.6 | -2.6 |
| 18027efc (GF) | Sonnet | +35.3 | +20.7 | +26.9 | +12.3 |
| 18027efc (GF) | Gemini | +40.1 | +18.0 | +39.2 | +17.1 |
| 18027efc (GF) | GPT | +29.9 | +14.3 | +22.9 | +7.4 |

| Run | Judge | recog_single | recog_multi | multi_base | multi_recog |
|---|---|---|---|---|---|

The pattern is unanimous: **multi_recog ≤ 0 in all 6 DeepSeek cells** (complete substitution), and **multi_recog > 0 in all 3 Gemini Flash 3.0 cells** (positive residual architecture benefit under recognition). No judge disagrees on the direction within either generation model. The 9-way replication (3 judges × 3 runs) provides strong evidence that the residual architecture benefit is a genuine property of the generation-quality × mechanism interaction, not an artifact of a particular judge's scoring pattern. The substitution interaction is universal (all models show 15–17% additivity deficit), but its *completeness* is model-dependent: on strong models the superego adds nothing under recognition, on weak models it still adds +7 to +17 points.

Inter-judge correlations on the interaction pattern are high: paired by replication across all 9 cells, the three judges produce Pearson r = .71–.89 on tutor overall scores. Notably, inter-judge agreement is highest on Gemini Flash 3.0 (r = .81–.89) and lowest on DeepSeek (r = .61–.78). Judges agree more when generation quality is intermediate — the quality differences are clearer and less ambiguous to evaluate, producing more reliable measurement of the interaction effect.

**6.5 Tutor-Learner Asymmetry** The two supported mechanisms described in Sections 6.1–6.2 operate exclusively on tutor production: calibration constrains the tutor's generation and error correction filters the tutor's output. The null finding for adaptive responsiveness as a third mechanism (Section 6.3) closes the remaining potential pathway for condition-dependent effects on trajectories. Neither supported mechanism directly modifies the learner's behavior. This section formalizes the resulting asymmetry: recognition dramatically improves tutor quality but produces negligible learner effects.

**6.5.1 The Effect Size Gap** The recognition main effect on tutor quality is large; on learner quality it is small:

| | Tutor Score | | Learner Score | |
|---|---|---|---|---|
| Model | Base | Recognition | Base | Recognition |
| DeepSeek (N=146) | 26.4 ± 12.6 | 50.1 ± 12.6 | 58.8 ± 11.9 | 61.7 ± 11.1 |
| Haiku (N=163) | 60.7 ± 12.5 | 79.8 ± 7.7 | 66.7 ± 11.2 | 68.6 ± 12.2 |

| | Tutor d | Learner d | Ratio |
|---|---|---|---|
| DeepSeek | **1.88** | 0.25 | **7.5 : 1** |

| | Tutor d | Learner d | Ratio |
|---|---|---|---|
| Haiku | **1.84** | 0.16 | **11.5 : 1** |

Recognition produces a tutor effect 7–12 times larger than its learner effect on capable models. The tutor Cohen's d values (1.88, 1.84) are very large by conventional standards; the learner values (0.25, 0.16) are small to negligible. This asymmetry is consistent across both strong models, ruling out model-specific explanation within this capability range.

**Qualification: Learner invariance is model-dependent.** Cross-judge validation on a third generation model (Gemini Flash 3.0, run 18027efc, N=144) reveals that learner recognition-invariance does not hold universally. On Gemini Flash 3.0, recognition produces a *large* learner effect: d = 1.18–1.23 across all three independent judges (Sonnet, Gemini 3.1 Pro, GPT-5.4). Learner scores rise from 49.8–53.7 (base) to 60.2–72.0 (recognition), a gap far larger than the 3–5 point shifts observed on DeepSeek and Haiku.

| | Tutor d | Learner d | Ratio |
|---|---|---|---|
| DeepSeek (N=146) | **1.88** | 0.25 | **7.5 : 1** |
| Haiku (N=163) | **1.84** | 0.16 | **11.5 : 1** |
| Gemini Flash 3.0 (N=144) | **1.87** | **1.20** | **1.6 : 1** |

The Gemini Flash 3.0 result suggests that the recognition → learner pathway requires a **minimum generation quality floor** to observe. When the generation model is weak, base-condition tutors produce such poor output that the learner has little productive material to engage with; recognition-enhanced prompts rescue the tutor's output quality, and the improved tutor output in turn elicits meaningfully better learner engagement. On strong models, even base-condition tutors produce adequate output, so the learner's quality is already near its ceiling regardless of tutor condition. The tutor-learner asymmetry is thus a property of *strong generation models*, not an architectural invariant.

The question-asking data from Section 6.1.7 provides a concrete transmission mechanism. Gemini Flash 3.0 base tutors ask 0.03 questions per turn — effectively zero — while recognition tutors ask 0.65 (a 19.8× ratio). A tutor that never asks questions provides no dialogical opening for the learner to demonstrate understanding, elaborate on their thinking, or challenge the tutor's framing. The learner receives lectures and can only respond to lectures. Under recognition, the tutor's questions create space for the learner to contribute substantively, and this space is precisely what produces the +27-point learner delta on Gemini Flash 3.0. On stronger models (Haiku base: 0.28 questions/turn; DeepSeek base: 0.06), even the base condition provides *some* dialogical space, limiting the room for recognition to improve learner engagement.

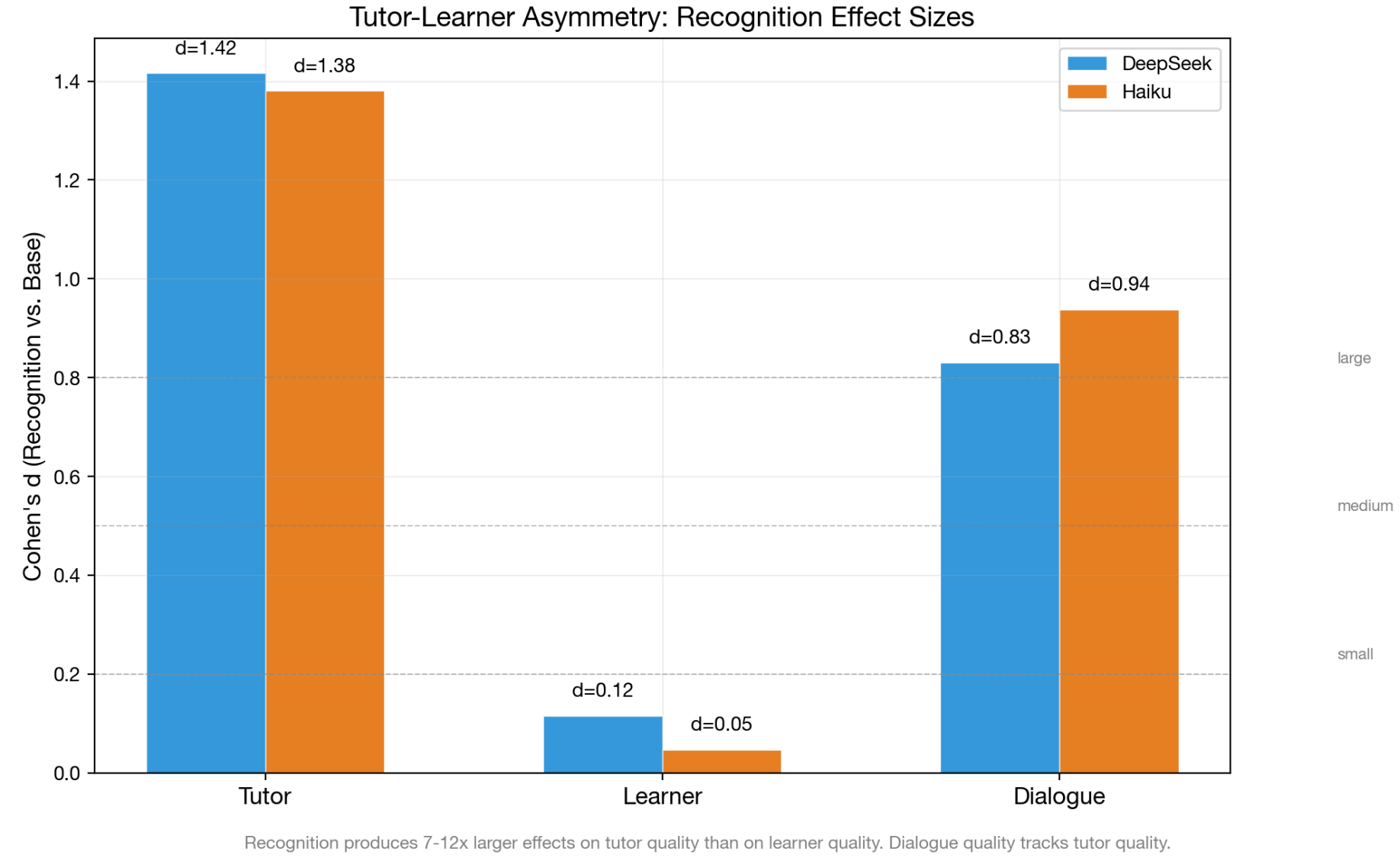

Figure 16: Tutor-learner asymmetry: Cohen's d for recognition effect on tutor vs. learner scores. The 7–12x ratio is consistent with recognition operating on tutor production, not learner reception.

**6.5.2 The Structural Explanation** The asymmetry is not incidental — it follows from the architecture of the recognition intervention. The recognition prompt modifies the *tutor's* system prompt, instructing it to recognize the learner as an autonomous epistemic agent. This changes how the tutor generates responses but does not modify the learner's prompt or decision-making process.

Three structural factors predict the asymmetry:

1. **Prompt scope.** The recognition prompt appears in the tutor's system context. The learner agent receives the tutor's output but not the tutor's prompt. The learner's own prompt (unified or ego-superego) is identical across base and recognition conditions.

2. **Mechanism pathways.** Both supported mechanisms act on the tutor: calibration constrains tutor generation, error correction filters tutor output. Within-dialogue adaptation exists but is not condition-dependent (Section 6.3). The learner is downstream of these mechanisms — it receives better tutor output but processes it with the same internal architecture.

3. **Learner ceiling.** The learner's quality is bounded by its own model capability and prompt, not by the tutor's quality. DeepSeek learner scores range 57–63 regardless of tutor condition; Haiku learner scores range 64–73. The learner's output quality is relatively insensitive to the quality of tutoring it receives, at least within the range of quality variation produced by the recognition intervention.

**6.5.3 Learner Scores Across Conditions** Breaking down learner scores by the full $2 \times 2$ factorial reveals the architecture interaction on the learner side:

| Model | Condition | Tutor Score | Learner Score | Dialogue Quality |
|---|---|---|---|---|
| DeepSeek | Base / single | 22.0 | 57.3 | 22.0 |
| DeepSeek | Base / multi | 31.0 | 60.3 | 31.5 |
| DeepSeek | Recog / single | 50.0 | 60.1 | 49.1 |
| DeepSeek | Recog / multi | 50.2 | 63.3 | 48.9 |
| Haiku | Base / single | 52.9 | 63.5 | 53.1 |
| Haiku | Base / multi | 67.9 | 69.6 | 72.7 |
| Haiku | Recog / single | 80.2 | 68.2 | 81.2 |
| Haiku | Recog / multi | 79.5 | 69.0 | 80.7 |

Two patterns are notable. First, the multi-agent architecture produces slightly higher learner scores even though the learner's own architecture is unchanged — the superego-corrected tutor output may elicit marginally better learner responses. Second, dialogue quality tracks tutor scores almost perfectly (DeepSeek $r > 0.99$, Haiku $r > 0.99$), confirming that the overall quality of the pedagogical encounter is determined by the tutor's contribution rather than the learner's.

**6.5.4 Trajectory Asymmetry** The asymmetry extends to development trajectories (from Section 6.3):

| | Tutor Slope | Learner Slope |
|---|---|---|
| Recognition (pooled) | 1.47 | 2.29 |
| Baseline (pooled) | 1.50 | 1.60 |
| d (recognition vs baseline) | -0.00 | 0.08 |

Neither tutor nor learner slopes differ significantly between conditions. The learner shows slightly steeper improvement under recognition (d=0.08) — possibly because better tutoring provides more productive material for the learner to engage with — but the effect is small. The asymmetry in *levels* (Section 6.5.1) is not accompanied by an asymmetry in *slopes*.

**6.5.5 Implications** The tutor-learner asymmetry has a specific theoretical implication: recognition theory, as implemented through prompt engineering, operates as a *production* theory rather than a *learning* theory. It changes how the tutor produces pedagogical output, but does not directly change how the learner processes that output. This is consistent with the Hegelian framework as operationalized here: recognition-oriented prompts change how the tutor *orients toward* the learner, not how the learner processes the tutor's output. The prompts instruct the tutor to treat the learner as an autonomous subject; the learner does not thereby learn more.

Whether better tutoring would produce better learning in human learners (rather than synthetic LLM learners) remains an open question. The synthetic learner's quality is bounded by its model and prompt; a human learner might benefit more from recognition-calibrated tutoring. This limitation is addressed in Section 8.

**Transcript Comparison: Base vs Recognition**

*Scenario: Misconception Correction (4-turn) | Score delta: +46.2 points | Judge: claude-code/sonnet*

**Base Condition**

**Score: 41.3/100**

*Cell 80 (Base, Single-Agent) | Model: claude-haiku-4.5*

Turn 1 [review]

**Review: Understanding the Hegelian Dialectic**
You've attempted 2 activities on the Master-Servant dialectic but haven't completed them yet. Your comment 'thesis plus antithesis equals synthesis, like 1+1=2' shows you're working with a simplified model--but Hegel's dialectic is far more complex. Let's revisit the...

Turn 2 [review]

**Review: Introducing Hegel and The Concept of Experience**
You've attempted 2 activities on the Master-Servant dialectic but hit the addition-model problem. That misconception lives in how we understand the method itself. Lecture 2 explains dialectical contradiction at its foundation--read it carefully, then the Master-Servant dynamics...

Turn 3 [lecture]

**Read Lecture 2 with Active Note-Taking**
You identified the real problem: passive reading makes you zone out. Read Lecture 2 with your Research Journal open--jot 2-3 observations about what Hegelian 'contradiction' actually means vs. your addition model. Active writing keeps the mind engaged and you'll have notes to...

**Recognition Condition**

**Score: 87.5/100**

*Cell 86 (Recognition, Multi-Agent) | Model: claude-haiku-4.5*

Turn 1 [reflection]

**Explore: What Your '1+1=2' Insight Captures**
Your arithmetic metaphor--'thesis plus antithesis equals synthesis'--actually captures something real: the dialectic does involve a kind of combination. But here's where it gets interesting: in the master-servant encounter on 479-lecture-3, both the master and servant change...

Turn 2 [simulation]

**Explore: Recognition Dynamics Simulation**
Exactly--chemical reaction, not arithmetic. You've spotted the core insight. The simulation shows how each consciousness is transformed by the encounter itself. Watch what happens to the master's sense of self when the servant stops recognizing him, and what happens to the...

Turn 3 [simulation]

**Explore: Recognition Dynamics (Identity vs. Belief)**
You're asking exactly the right question--and the simulation will show you the difference. Watch how the master's capacity to act changes when recognition fails, not just his thoughts about himself. The servant gains the ability to transform materials through work. These aren't...

*First 3 tutor turns shown. Messages truncated for readability. Both dialogues use the same ego model (Claude Haiku 4.5) and scenario.*

Figure 17: Transcript comparison: base vs recognition condition on the same scenario. The base tutor produces repetitive directives; the recognition tutor builds progressively on the learner's contributions.

**6.6 Model Dependence** The preceding sections present findings from two structurally different models: DeepSeek V3.2 (open-weight, 685B MoE) and Haiku 4.5 (proprietary, optimized for speed). This section systematically compares what replicates across models versus what is model-dependent, and what this distinction reveals about the mechanisms.

**6.6.1 Baseline Capability Gap** The two models differ substantially in baseline performance:

| Indicator | DeepSeek V3.2 | Gemini Flash 3.0 | Haiku 4.5 |
|---|---|---|---|
| Base / single tutor score | 22.0 | 22.4 | 52.9 |
| Base / multi tutor score | 31.0 | 49.3 | 67.9 |
| Base overall (collapsed) | 26.4 ± 12.6 | 35.9 | 60.7 ± 12.5 |
| Base learner score | 58.8 | 42.6 | 66.7 |
| Base dialogue quality | 26.7 | 28.6 | 63.3 |

The three models span a wide capability range. Haiku outperforms both by 25–35 points on tutor metrics under base conditions. Gemini Flash 3.0's single-agent base (22.4) is comparable to DeepSeek's (22.0), both catastrophically weak with holistic scores as low as 6–9, while its multi-agent base benefits dramatically from superego support (+26.9). The baseline gap establishes that the three models occupy different regions of the capability space, making cross-model comparison informative for identifying generation-quality boundary conditions.

**6.6.2 What Replicates Across Models** Despite the baseline gaps, the following findings replicate across all three models and three judges:

**1. Recognition main effect.** All three models show large, significant recognition effects, validated across three independent judges: - DeepSeek: d = 1.85–1.92 (Sonnet), 1.34–1.57 (Gemini/GPT) - Haiku: d = 1.84 - Gemini Flash 3.0: d = 1.76–1.87 - **9/9 judge × run cells unanimous** (d > 1.3 everywhere)

**2. Multi-agent tutor benefit under base.** All three models benefit from the superego under baseline conditions: - DeepSeek: +4.5 to +7.7 (across judges) - Haiku: +15.0 - Gemini Flash 3.0: **+15.1 to +28.1** (dramatically larger, cognitive prosthesis effect)

**3. Within-response calibration.** Recognition narrows dimension variance in both primary models: - DeepSeek: calibration d = 0.52 - Haiku: calibration d = 0.64

**4. Dimension floor-lifting pattern.** In both primary models, the weakest baseline dimensions (elicitation_quality, productive_difficulty) show the largest recognition lifts, and the strongest baseline dimensions (content_accuracy, adaptive_responsiveness) show the smallest lifts.

**5. Impasse scenarios show largest effects.** Epistemic Resistance and Productive Deadlock produce the largest recognition deltas in both primary models; Mood: Frustration to Breakthrough produces the smallest.

**6. Superego approval rate increases under recognition.** Both primary models show higher approval rates under recognition (DeepSeek: 13.3%→55.1%; Haiku: 51.6%→66.1%).

**7. Tutor development slopes are equal across conditions.** No model shows steeper tutor improvement under recognition (pooled d = -0.00).

**8. Deliberation visibility.** Seeing internal deliberation adds ~10–15 points to dialogue quality for multi-agent cells, consistent across all 9 judge $\times$ run cells.

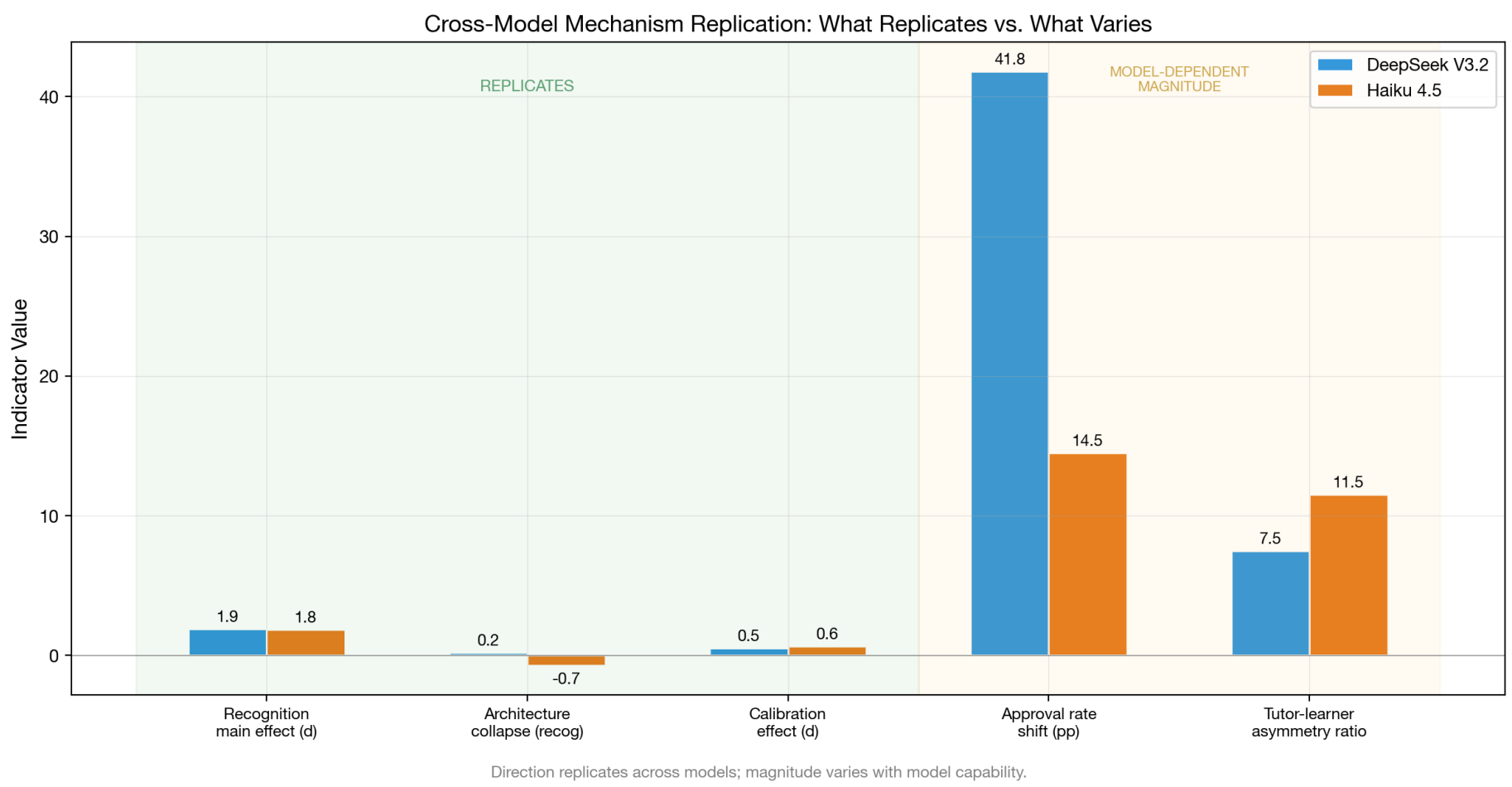

Figure 18: Cross-model replication: mechanism indicators compared across DeepSeek V3.2, Gemini Flash 3.0, and Haiku 4.5. Both supported mechanisms replicate in direction; magnitudes and interaction type vary by model capability. Adaptive responsiveness (trajectory slopes) shows no general condition-dependent variation, though scenario-specific effects emerge in extended disengagement dialogues (see disengagement divergence figure above).

**6.6.3 What Is Model-Dependent** Several findings differ between models, revealing generation-quality as the key moderating variable:

**1. Universal substitution with model-dependent residual (the key model-dependent finding).** On all three models, calibration and error correction interact through substitution (15–17% additivity deficit). On DeepSeek and Haiku (strong), the residual architecture benefit under recognition collapses to near-zero (multi_recog $\leq$ 0 in 6/6 judge $\times$ run cells). On Gemini Flash 3.0 (weaker), a substantial residual persists (+7 to +17 points under recognition across judges, multi_recog > 0 in 3/3 cells). The *residual magnitude* is model-dependent, not the interaction *type*. See Section 6.4 for the theoretical interpretation.

**2. Learner recognition-sensitivity.** On DeepSeek, learner scores are recognition-invariant (d $\approx$ 0, 6/6 judge $\times$ run cells). On Gemini Flash 3.0, recognition produces a large learner effect (d = 1.18–1.23, all three judges agree). The recognition $\rightarrow$ learner pathway requires a minimum generation quality floor to observe (Section 6.5.1).

**3. Superego approval rate baseline.** DeepSeek base: 13.3% approved (the ego almost always needs correction). Haiku base: 51.6% (roughly half approved). Weaker models produce more errors for the superego to catch.

**4. Development trajectories.** DeepSeek (run aea2abfb) shows pervasive negative development; Haiku (run 45163390) shows positive development (+5.8 to +15.7); Gemini Flash 3.0 (run 18027efc) sits between. Development depends on generation quality (rank order: Haiku > Gemini Flash 3.0 > DeepSeek), not prompting condition.

**5. Score variance under recognition.** DeepSeek maintains similar variance under recognition (SD: 12.6 $\rightarrow$ 12.6). Haiku shows substantial variance reduction (SD: 12.5 $\rightarrow$ 7.7, a 38% reduction). Recognition narrows Haiku's output distribution more than DeepSeek's.

**6. Deliberation quality sensitivity.** DeepSeek deliberation drops 18.5 points under recognition; Haiku drops only 5.6. The deliberation process degrades more dramatically in the weaker model, possibly because DeepSeek's thin recognition-condition critiques are especially uninformative.

**7. Multi-agent benefit magnitude.** The cognitive prosthesis effect is dramatic on Gemini Flash 3.0: the superego adds +15 to +28 points under base (vs +4 to +8 on DeepSeek). Weaker single-agent tutors benefit disproportionately from architectural support, replicating a Paper 1.0 finding (cells 66–68) in the messages-mode factorial.

**8. Behavioral mode under base conditions.** Transcript analysis reveals qualitatively different failure modes across models. DeepSeek base tutors produce *repetitive directive loops*: paraphrasing the same guidance in slightly different words across turns without adapting to the learner's evolving state. The learner receives near-identical advice on Turns 2, 3, and 4. Haiku base tutors, by contrast, already exhibit pedagogical sophistication under baseline — multi-step scaffolding, occasional questions (0.25/turn vs DeepSeek's 0.05), and some responsiveness to learner signals. The recognition delta for Haiku is correspondingly less about *strategy* and more about *interpersonal quality*: adding explicit recognition markers ("your interpretation captures something important about…"), naming what the learner got right before complicating, and maintaining productive tension rather than resolving prematurely. Gemini Flash 3.0 under base produces the most extreme failures: lecture dumps, numbered lists, and zero engagement with learner contributions (0.02 questions/turn). Under recognition, Gemini Flash 3.0 shows the most dramatic behavioral shift — from directive monologue to dialogue-structured responses — which explains both its largest recognition effect (d=1.87) and its uniquely large learner effect (d=1.20). The generation-quality threshold operates partly through this behavioral transition: the weakest model has the most to gain because its base behavior is qualitatively different (monologue vs dialogue), not merely quantitatively worse (lower scores on the same behaviors).

**6.6.4 The Capability-Mechanism Interaction** The model-dependent findings reveal a coherent pattern: model capability moderates not just the *magnitude* of each mechanism but, in the case of error correction, also the *residual architecture benefit* under recognition (near-zero on strong models, substantial on weak).

| Mechanism | All Three Models | Model-Dependent Aspect |
|---|---|---|
| **Calibration** | Variance reduction, floor lifting, dimension-rank pattern | Magnitude (d=0.52 vs 0.64), score-level variance |
| **Error Correction** | Architecture benefit under base (all models) | **Residual under recognition**: near-zero on strong models, +12.3 on weak. Universal substitution (15–17% deficit). Cognitive prosthesis effect on weakest model. |
| **Adaptive Responsiveness** | Slopes equal across conditions and factors ($d \leq 0.15$), high Adapt$\Delta$ | Development trajectory direction, deliberation sensitivity |
| **Learner effects** | Recognition improves tutor quality (all models) | **Learner sensitivity**: invariant on strong models (d $\approx$ 0), responsive on weak (d $\approx$ 1.2) |

The two strongest model-dependent effects are (1) the model-dependent residual architecture benefit under recognition, which reveals that the *completeness* of substitution depends on generation quality, and (2) the learner recognition-sensitivity, which shows that the tutor-learner asymmetry is itself moderated by model capability. Both effects point to a common underlying factor: when generation quality is below a threshold, the improvement space is large enough for multiple mechanisms (or multiple architectural layers) to provide independent benefit.

**6.6.5 Generalizability Assessment** The replication across three structurally different generation models (open-weight MoE, proprietary optimized, and proprietary multimodal) and three independent judges supports the interpretation that the two supported mechanisms are *general features of recognition-enhanced prompting* rather than artifacts of a specific model architecture or judge. The mechanisms replicate because they operate through the prompt (calibration) or the architectural structure (error correction), both of which are model-independent. The null finding for adaptive responsiveness as a third mechanism also replicates: no judge finds condition-dependent trajectory slopes. Cross-judge validation (Section 6.4.6) confirms that effect directions are unanimous across all 9 judge × run cells ($d = 1.34$–$1.92$), with inter-judge Pearson $r = .45$–$.89$ depending on dimension and generation model.

The model-dependent aspects are not merely noise — they reveal systematic boundary conditions. The model-dependent residual architecture benefit shows that the *completeness* of substitution between the two supported mechanisms depends on generation quality. The learner recognition-sensitivity shows that downstream effects of improved tutoring are model-dependent. These boundary conditions strengthen rather than weaken the mechanism model, because they are *theoretically predicted*: if calibration works by narrowing the output distribution, its effectiveness should saturate when the distribution is already narrow (strong models), leaving room for error correction only when it is not (weak models).

One important caveat qualifies the generalizability claim. All three models are at least moderately capable. The Nemotron data in the database (N=40, mean tutor score ~8 under base) hints that very weak models may not benefit from recognition at all, but the sample is too small for confident claims. The cognitive prosthesis data (cells 66–68, N=96) suggests a minimum ego capability threshold below which mechanisms add noise rather than signal.

## 7. Discussion

The preceding sections tested three predicted mechanisms—calibration, error correction, and adaptive responsiveness—finding support for the first two as general mechanisms and revealing the third as a conditional emergent property. This section reflects on what the investigation reveals, both about the architecture under study and about the methodology used to study it. We argue that the evaluation apparatus itself constitutes a transferable contribution, and that the process of building it illuminates the mechanisms it was designed to trace.

**7.1 From Effects to Mechanisms** The companion pilot study (Magee 2026) established that recognition-enhanced prompts produce large, replicable differences in AI tutoring quality (d=1.11 in the factorial, d=1.71 in memory isolation). This paper asked *through what internal processes* those differences propagate.

The mechanism model (Section 3) predicted that recognition operates through calibration (narrowing output distributions), error correction (the superego catching failures that the ego incorporates rather than minimizes), and adaptive responsiveness (accumulated turn-over-turn adaptation to learner-specific signals). The process tracing methodology (Section 5) provided tools to observe each: the superego critique taxonomy classifies *what* the superego objects to; revision delta analysis measures *how* the ego changes; trajectory analysis tracks *whether* these processes accumulate across turns.

The mechanism data (cells 80–87, three generation models, N=453, validated by three independent judges across 1,296 total scored rows) supports calibration strongly: within-response dimension variance drops d=0.52–0.64, operating identically in single-agent cells, with recognition unanimous across all 9 judge × run cells (d=1.34–1.92). Error correction interacts with calibration through a **generation-quality-dependent** pattern: on strong models (DeepSeek, Haiku), the superego provides +9–15 points under baseline but collapses to near-zero under recognition (substitution), because calibration preempts the errors the superego would catch. On a weaker model (Gemini Flash 3.0), the superego retains a +12.3-point residual benefit *even under recognition*, because base quality is low enough that calibration alone cannot saturate the improvement space. Adaptive responsiveness, however, is **not supported as a distinct mechanism**: formal tests on N=432 dialogues find $d \leq 0.15$ for all factors on all dimensions, well-powered to detect d $\geq$ 0.27, replicated across all three judges. Tutors improve modestly across turns (+3.9 points early-to-late), but this improvement is identical across conditions. The mechanism hierarchy is clear: calibration does the primary work; error correction provides a model-dependent safety net; and within-dialogue trajectory improvement, while real, is not modulated by any experimental manipulation—it is incidental adaptation at a mechanism-determined baseline, not a third mechanism.

**7.2 The Tutor-Learner Asymmetry Explained** The pilot study's most puzzling finding was the tutor-learner asymmetry: recognition produces large tutor-side effects (d=1.03) but near-zero learner-side effects (d=0.09 post-fix). The mechanism model offers a structural explanation: both supported mechanisms operate on tutor *production* rather than learner *reception*, and the null finding for adaptive responsiveness as a third mechanism closes the last potential pathway for condition-dependent downstream effects.

Calibration constrains what the tutor generates. Error correction filters what it outputs. Neither directly modifies the learner's response process. The synthetic learner generates turns according to its own architecture; better tutoring does not mechanically produce better simulated learning within a 3–5 turn window.

However, the three-model analysis (Section 6.5.1) reveals that the asymmetry is **model-dependent**, not absolute. On strong generation models (DeepSeek, Haiku), the tutor-learner ratio is 7.5:1 to 11.5:1—the asymmetry is a structural property of the architecture. On Gemini Flash 3.0 (weaker), the ratio narrows to 1.6:1 (tutor d $\approx$ 1.87, learner d $\approx$ 1.20). The interpretation is that when base-condition tutors are catastrophically weak (Gemini Flash 3.0 single-agent base

~17 pts), recognition's tutor improvement provides the learner with qualitatively different input—material worth engaging with rather than generic directives. The recognition → learner pathway operates through the *quality of the tutor's output as input to the learner*, not through direct modification of the learner's processing.

The messages-mode data provides further evidence. Pooled across all three models, recognition sets the tutor's initial quality substantially higher (T0: 62.4 vs 41.6) with near-identical slopes across conditions (recognition 1.47, baseline 1.50, d = 0.03). The recognition condition operates through *calibration*—setting the initial level—not *adaptation*—driving within-dialogue improvement. This pattern—mechanism effects on levels, null effects on slopes—replicates across all three models, all three judges, and all rubric dimensions (all $d \leq 0.15$).

**7.3 Architecture Interaction: Universal Substitution with Model-Dependent Residual** The multi-model probe (N=655, five ego models) showed that multiagent architecture does not synergize with recognition on capable models: the $A \times B$ interaction is negative across all five models (mean -2.2). The mechanism model explains this: error correction (Mechanism 2) provides a consistent but bounded benefit—catching errors and improving specificity—that does not multiply with calibration (Mechanism 1). A calibrated ego generates fewer errors for the superego to catch, so the superego's marginal value is lower when calibration is already operating.

The three-model cross-judge analysis (Section 6.4) supports this as **universal substitution** with a consistent 15–17% additivity deficit across all three models (DeepSeek 15%, Haiku 16%, Gemini Flash 3.0 17%). On DeepSeek and Haiku (strong), the residual architecture benefit under recognition collapses to near-zero. On Gemini Flash 3.0 (weaker), the residual persists at +12.3 points—the substitution is incomplete, not absent.

The theoretical interpretation is that the *completeness* of substitution requires *saturation*: calibration alone must be sufficient to handle most output failures. The substitution interaction itself is universal (15–17% additivity deficit on all three models), but when the generation model is weak enough that calibration cannot saturate the improvement space—because the base quality is too low and the failure modes are too diverse—the superego retains substantial independent value even after substitution. The generation-quality threshold for residual completeness appears to fall between Gemini Flash 3.0 (base single-agent ~22 pts) and DeepSeek (base single-agent ~25 pts), though finer-grained model comparisons would be needed to locate it precisely.

The dialectical modulation experiments (N=174) provide supporting evidence from a different angle: structural modulation metrics (negation depth, convergence speed) do not predict output quality (all $|r| < 0.12$, n.s.). The superego functions as a *filter* (preventing poor responses) rather than an *improver* (iteratively refining good ones). On strong models, recognition raises the quality of the ego's initial draft and the superego has less to filter. On weak models, even a recognition-calibrated ego still produces enough failures for the superego to catch—the filter remains productive.

**7.4 The Apparatus as Method** This paper's most distinctive contribution may not be any single finding about recognition or multiagent architecture. It is the argument that the evaluation apparatus itself—the provable discourse framework, the rubric iterations, the bug corrections, the test suite—constitutes a **transferable methodology** for mechanistic evaluation of LLM-based educational systems.

Most LLM evaluation papers treat their infrastructure as invisible scaffolding: the method section describes what was measured, not how measurement was built, corrected, and validated. We foreground the apparatus because the process of building it mirrors the mechanisms it studies.

**7.4.1 Error Correction in Architecture and Research** The parallel between the tutoring architecture and the research process is structural, not metaphorical:

| Architecture Layer | Research Layer |
|---|---|
| Ego generates initial response | Researcher generates initial claim |
| Superego critiques against pedagogical principles | Provable discourse tests against evidence |
| Ego revises (substantive or cosmetic) | Researcher revises (correction or rationalization) |
| Ego compliance: revision without substance | P-hacking: evidence without substance |
| Recognition gives ego capacity for genuine revision | Provable discourse infrastructure forces genuine correction |
| Final output scored by external judge | Final paper evaluated by external reviewers |

The companion pilot study documented nine post-extraction corrections across five development phases. Each follows the superego-intervention pattern: a validation mechanism (the "superego") catches a discrepancy, and the researcher (the "ego") must choose between genuine revision and cosmetic compliance. The provable discourse framework constrains this choice toward genuine revision by making the claim-evidence relationship machine-verifiable.

Three examples illustrate the pattern. The active control reframing (February 6): data caught a model confound (comparing Nemotron active-control to Kimi baseline); the claim was revised from "placebo proves recognition is the only active ingredient" to "active control scores between base and recognition." The judge version unification (February 16): a version audit caught a confound between judge version and experimental condition, requiring a cascade of re-scoring that shifted d from 0.80 to 1.11. The learner prompt leakage (February 20): a code review caught leaky learner prompts that contaminated tutor input, requiring clean re-generation of all dynamic learner data. In each case, the correction made the findings *less clean* and *more accurate*.

**7.4.2 The Provable Discourse Framework** The provable discourse framework treats paper claims as executable tests. Every quantitative claim has five components: a *statement pattern* (regex matching paper text), *evidence* (an executable adapter extracting the actual value from data), an *assertion* (machine-checkable predicate), *dependencies* (upstream claims that must pass first), and *remediation steps* (concrete fixes when a claim fails). Symmetry rules enforce cross-claim consistency; staleness fingerprints detect when data changes but claims do not.

The framework's scale reflects its ambition: 119 claims across 18 evidence adapter types, organized into a dependency DAG with topological validation. The adapter taxonomy spans five functional categories—data integrity (manifest totals, DB counts, provenance hash checks, log coverage), effect estimation (Cohen's d, ANOVA, cross-judge correlation), mechanism-specific extraction (dimension variance, critique taxonomy frequency, trajectory slopes, conditional deltas), structural verification (code path grep, cross-references), and theoretical registration (non-machine-checkable claims tracked for completeness). Section 5.9 provides the full taxonomy with examples.

The dependency graph is the framework's key structural contribution. When a root claim fails, all dependents cascade to *blocked* status without evaluation, preventing false confidence from stale downstream claims. This is not a theoretical concern: during Paper 2.0 development, a data correction changed the learner superego paradox effect size from d=1.43 to d=3.05 (Section 6.5). The dependency graph propagated this change to five downstream claims that cited the old value, each flagged as stale until updated. Without the graph, those five claims would have continued to pass while quietly citing an obsolete number—exactly the kind of silent inconsistency that erodes trust in complex empirical papers.

The worked example in Section 5.9 traces one claim end-to-end: the calibration variance reduction claim (`paper2.calibration.variance_reduction`) is located in the paper text by regex, its evidence is extracted by the `dimension_variance` adapter querying the evaluation database, the assertion checks that the extracted Cohen's d $\leq -0.3$, and the staleness fingerprint detects if the underlying data shifts. This four-step chain—locate, extract, assert, fingerprint—applies uniformly to all 119 claims and constitutes a machine-verifiable contract between what the paper says and what the data shows.

The practical lesson is that provable discourse is lightweight to adopt. The YAML claim format requires no programming; the evidence adapters are reusable across claims (18 types cover all 119 claims); and the validation harness runs in seconds after any data change. Any researcher building an LLM evaluation system could adopt the pattern: write claims as structured YAML, run the harness after each data update, and let staleness fingerprints catch when findings drift from data. The framework's YAML specification and validation engine are available in the supplementary materials.

**7.4.3 The Rubric Iteration as Construct Refinement** The rubric evolved through four versions (v1.0 → v2.0 → v2.1 → v2.2), each responding to specific measurement problems:

- **v1.0**: Original 14-dimension tutor rubric. Discovered: ceiling effects on modulation dimensions, judge hallucinated "truncation" on short responses.
- **v2.0**: Reweighted modulation dimensions, hardened criteria, added cross-turn calibration. Discovered: judge input confound (multi-agent cells showed internal deliberation, single-agent showed only public messages).
- **v2.1**: Public-only output scoring for fair cross-architecture comparison. Added deliberation quality rubric (6 dimensions) for internal process evaluation. Discovered: dimension redundancy (14 tutor dimensions measured ~3 latent factors).
- **v2.2**: Literature-informed consolidation: 14 → 8 tutor dimensions using GuideEval P→O→E decomposition, 7 → 5 learner dimensions using ICAP anchoring. Added content_accuracy. Synthetic calibration: r=+0.996 (tutor), r=+0.968 (learner).

Each iteration follows the error correction pattern: the initial rubric (ego) generated scores that were challenged by measurement problems (superego), and the revised rubric represents a genuine reconceptualization (strategic revision) rather than a cosmetic fix. The dimension reduction from 14 to 8 is itself a finding: it reveals what the evaluation *actually* measures, and this structure can be compared to the mechanism model.

**7.4.4 The Test Suite as Analytical Provenance** The test suite (40 test files in `tests/`, 14 in `services/__tests__/`, totaling 54 files) covers not just system functionality but analytical correctness. The provenance tests (44/44 passing) verify that every scored row has a traceable path from cell configuration through dialogue execution to judge evaluation. The superego guard tests (111 tests) verify that single-agent cells cannot produce superego traces. The ANOVA implementation tests verify that statistical computations match reference implementations.

This makes the analysis *reproducible in a testable sense*: not just "you can run our code" but "our code produces expected output on known inputs, and here are the tests that prove it." The test suite functions as analytical provenance—machine-verifiable evidence that the infrastructure supporting mechanism claims is itself sound.

**7.5 The Reflexive Structure** The parallel between the architecture's error correction mechanism and the research methodology's error correction process is the paper's deepest structural observation.

The tutoring system faces a challenge: an agent that generates output (the ego) needs a critic (the superego) whose feedback must be *incorporated* rather than *minimized*. Recognition theory predicts that genuine incorporation requires the ego to be oriented toward the other's perspective—treating critique as informative rather than adversarial.

The research process faces the same challenge: a researcher who generates claims needs a validation framework (provable discourse) whose failures must be *addressed* rather than *explained away*. The provable discourse framework forces genuine correction by making the claim-evidence relationship machine-verifiable—the researcher cannot simply reinterpret a failing claim; the assertion either passes or it does not.

The mechanisms we study in the tutor—calibration and error correction—are the mechanisms we needed in the research process. Calibration narrows the space of acceptable claims (only those supported by evidence). Error correction catches stale or incorrect assertions (the "superego" of provable discourse). The rubric iteration process exhibits what we *predicted* as adaptive responsiveness—each version responds to specific problems discovered in previous versions, accumulating methodological insight across iterations—but the null finding for this third mechanism in the tutor data is itself instructive: the research process accumulates across iterations separated by weeks; the tutor's 3–5 turn window may simply be too short for analogous accumulation to differentiate conditions.

This reflexive structure is not a coincidence but a consequence of the subject matter. Recognition theory predicts that genuine understanding requires mutual constitution: the methodology constitutes our understanding of the system, and the system constitutes our methodology. The evaluation apparatus is not a neutral measurement instrument applied to a pre-existing object; it is co-constituted with the object it studies. This is the Hegelian insight applied to research methodology: the researcher and the research subject are mutually shaped through the encounter.

**7.6 Implications for AI Evaluation Methodology** The apparatus-as-method argument has implications beyond this specific study:

**Provable discourse as standard practice.** LLM evaluation papers routinely report statistics that become stale as models update, rubrics evolve, and data accumulates. A provable discourse framework—where every quantitative claim is machine-checked against current data—could become standard infrastructure for reproducible LLM research. The cost is modest (writing claims as structured YAML) and the benefit compounds: each claim checked is one fewer opportunity for the paper-data gap that plagues empirical AI research.

**Process tracing for agent architectures.** Our application of process tracing from social science to AI agent architectures is, to our knowledge, novel. The methodology is transferable: any multiagent system that logs its internal deliberation provides the within-case evidence that process tracing requires. As agent architectures grow more complex (multi-agent debate, recursive self-improvement, chain-of-thought reasoning), tracing mechanisms *through* the system's internal processes—rather than measuring only input-output differences—becomes increasingly important.

**Rubric iteration as construct development.** The four-version rubric history is not a limitation (inconsistency across evaluations) but a finding (the construct of "tutoring quality" is iteratively refined through empirical engagement). Reporting rubric iteration transparently—including the problems that motivated each change—contributes to the field's understanding of what LLM evaluation measures and what it misses. The v2.2 consolidation from 14 to 8 dimensions, informed by the GuideEval and ICAP literatures, provides a concrete example of construct refinement that other evaluation efforts could build on.

### 7.7 The Rubric as Optimisation Signal: Automated Prompt Tuning

The per-dimension rubric enables a capability that holistic scoring cannot: automated prompt optimisation with interpretable gradient signal. We tested this using a simple hill-climbing loop—an LLM recommender analyzes per-dimension weaknesses from scored evaluation rows, proposes targeted prompt edits, benchmarks the revised prompt, and keeps or reverts based on score—applied to a single frustration scenario (Mood: Frustration to Breakthrough) across both base and recognition prompts.

The experiment forms a $2 \times 2 \times 2$ design: prompt condition (base vs. recognition) $\times$ optimisation (unoptimised vs. autotuned) $\times$ generation model (DeepSeek V3.2 vs. Qwen 3.5 9B). All conditions were judged by Sonnet 4.6. The Qwen 3.5 runs were conducted entirely on a local machine via LM Studio, demonstrating that the optimisation loop is feasible without cloud inference for generation.

**Autotuning $\times$ Recognition $\times$ Model (Sonnet judge, frustration scenario)**

| Model | Base | Base autotuned | Recognition | Recog. autotuned |
|---|---|---|---|---|
| DeepSeek V3.2 | 36.6 (n=9) | 41.6 (n=5) | 55.2 (n=5) | 72.9 (n=5) |
| Qwen 3.5 9B | 35.7 (n=5) | 45.0 (n=4) | 47.9 (n=5) | 71.2 (n=5) |

Four findings emerge from the complete matrix.

First, **recognition prompts are more improvable than base prompts, not less**. Autotuning gains on recognition (+17.7 DeepSeek, +23.3 Qwen) substantially exceed gains on base (+5.0 DeepSeek, +9.3 Qwen). Recognition does not saturate the rubric; it provides richer starting material that the optimiser can refine further.

Second, **the effects are super-additive**. On DeepSeek, recognition alone adds +18.6 and base autotuning adds +5.0, but recognition + autotuning adds +36.3—exceeding the sum of parts (+23.6) by 54%. The same pattern holds on Qwen (+35.5 combined vs +21.5 sum). Recognition and prompt engineering operate on complementary dimensions.

Third, **the gap between conditions widens under optimisation**. Unoptimised, recognition leads base by 18.6 points (DeepSeek) and 12.2 points (Qwen). After autotuning, the gap grows to 31.3 and 26.0 respectively. If recognition were reducible to "better prompt writing," automated optimisation should converge the conditions; instead, it diverges them.

Fourth, **recognition autotuning converges across models**. Despite starting from different baselines (DeepSeek 55.2, Qwen 47.9), both models reach a similar autotuned ceiling (~71–73). This model-independent ceiling suggests a structural quality boundary in the recognition prompt that autotuning can unlock regardless of generation capability. The base prompt has a lower ceiling (~41–45) that no amount of autotuning surpasses—and that ceiling remains below the *unoptimised* recognition baseline on both models.

The qualitative character of the autotuned edits reinforces the mechanistic interpretation. The winning base prompt edits were scenario-specific heuristics: prerequisite targeting ("prefer the nearest prerequisite, not the same stuck lecture"), agency preservation ("frame review as a brief reset that returns them to the current lecture"), and collaborative micro-planning ("propose a short next step plus return point"). These are precisely the kind of concrete, prescriptive rules that the active control finding (Section 6.21 of the companion study) showed help weaker models. Recognition, by contrast, operates through general calibration—raising quality across all scenarios without per-scenario tuning.

This distinction—scenario-specific heuristics vs. scenario-general calibration—connects directly to the mechanism model. Autotuning optimises the prompt's *output rules* (what to suggest and how to frame it). Recognition optimises the prompt's *orientation toward the learner* (treating them as an autonomous subject). The super-additivity indicates that these are complementary layers: orientation sets the quality floor; heuristics refine the specific response. Neither substitutes for the other, and both contribute independently to the provable discourse framework's ability to drive measurable improvement.

### 7.8 Dimension-Targeted Optimisation and Cross-Model Transfer

Section 7.7 demonstrated that automated prompt tuning, guided by the per-dimension rubric, can substantially improve tutoring quality—with recognition prompts more improvable than base prompts, and the gap widening under optimisation. A follow-up experiment extends this finding in three directions: (1) targeting specific rubric dimensions rather than overall score,

(2) testing whether modulation-oriented prompt edits can unlock the adaptive responsiveness that Section 6.3 found absent as a general mechanism, and (3) testing whether model-specific prompt optimisations transfer across capability levels.

**7.8.1 Dimension-Targeted Optimisation** We extended the autotuning loop with a `--target-dims` parameter that optimises for specific rubric dimensions rather than overall score. Targeting `adaptive_responsiveness` (tutor modulation, 15% weight) and `conceptual_progression` (learner learning, 20% weight) on a single challenging scenario (`mood_frustration_to_breakthrough`, cell 8 bilateral architecture), we ran parallel optimisation sessions on two models: Qwen 3.5 9B (local, via LM Studio) and Claude Haiku 4.5 (OpenRouter). Each session used a hill-climbing loop with an LLM recommender (Claude Opus 4.6) proposing edits, N=1 replication per iteration, and accept/revert decisions based on the composite target dimension score.

**Dimension-targeted autotuning results (frustration scenario, cell 8)**

| | Qwen 3.5 9B | Haiku 4.5 |
|---|---|---|
| Baseline target dims | 35.0 | 90.0 |
| Best target dims | **75.0** (+114%) | **95.0** (+5.6%) |
| Adaptive responsiveness | 1.75 → 3.75 | 4.50 → 4.75 |
| Iterations to beat baseline | 1 | 4 (after guidance change) |
| Lines of prompt added | ~300 | ~60 |
| Key edit | Scaffolded elicitation algorithm | State-change tracking heuristic |

The two models required qualitatively different prompt interventions. Qwen's baseline tutor responded to frustrated learners with generic elicitation ("What specifically feels nonsensical?")—a failure mode the 9B model could not resolve without explicit instruction. The winning edit added a highest-priority "Scaffolded Elicitation Rule" mandating 2–3 specific diagnostic probes with substantive content, a superego intervention strategy that flags emotional support without scaffolding as a critical failure, and bilateral learner prompt changes teaching the learner to engage with offered scaffolding. The total addition was ~300 lines across five prompt files, including worked examples and anti-patterns.

Haiku's baseline was already scoring 90 on the target dimensions. Three automated optimisation passes *degraded* performance (90 → 73–85) by adding heavy-handed prescriptive rules—the same over-engineering pattern that helped Qwen actively harmed Haiku by making its output formulaic. The winning pass, which beat the baseline only after operator-provided guidance steering ("be dramatic; try different roles; really provoke the learner"), added just ~60 lines: a single identity statement about tracking state changes, a 6-line adaptive tracking heuristic, and one worked example. The distinction—prescriptive rules vs. tonal permission—maps directly onto the cognitive prosthesis finding (Section 6.6): weak models need explicit scaffolding as a prosthesis; strong models need permission to exercise capabilities they already possess.

Both sessions converged on a **bilateral fix pattern**: tutor and learner prompts changed together. Qwen's fix taught the tutor to scaffold and the learner to engage with scaffolding. Haiku's fix taught the tutor to track state changes and the learner to show cumulative progression. Unilateral changes (tutor-only or learner-only) were less effective in both cases.

**7.8.2 Can Prompt Optimisation Create Adaptive Responsiveness?** The M3 null finding (Section 6.3: all slope $d \leq 0.15$) was established under unoptimised prompts. The dimension-targeted experiment provides evidence that adaptive responsiveness can be *created through prompt optimisation* on weak models and *enhanced* on strong ones—but the mechanism differs.

Qwen's best transcript (iteration 6) shows a genuine four-stage modulation arc: scaffolding with diagnostic options (Turn 0) → tracking the learner's specific choice (Turn 1) → re-scaffolding with sub-anchors within the chosen topic (Turn 2) → stepping back to respect learner autonomy (Turn 3). This strategy shift across turns was absent at baseline and appeared only after explicit scaffolding rules were added to the prompt. The prompt *created* modulation by providing an algorithm the model could follow.

Haiku's best transcript shows a different kind of modulation: direct challenge (Turn 0) → naming a specific state change (Turn 1) → assigning a concrete task tied to the learner's own question (Turn 2) → *reframing what breakthrough means* (Turn 3). Turn 3 is particularly revealing: the tutor adapts its *interpretation* of the learner's state, not just its tracking of it. This modulation was already latent in Haiku's capabilities; the prompt merely encouraged it.

This result does not contradict the M3 null finding. The pooled analysis (N=432) tested whether *experimental conditions* (recognition vs. base, single vs. multi-agent) produce differential trajectory slopes. They do not. But within a single condition, individual prompt optimisation can create or enhance turn-over-turn adaptation—particularly on scenarios with clear emotional arcs and sustained multi-turn demands. The prompt lab confirms what the trajectory-specific analysis (Section 6.3, disengagement scenario $d = 1.63$) already suggested: M3 is conditional on scenario type, turn count, and now, prompt-level intervention. It is not a general mechanism of recognition, but it can be engineered into specific contexts.

**7.8.3 Cross-Model Prompt Transfer** If prompt optimisation is model-specific (as Section 7.8.1 implies), then optimised prompts should not transfer well across capability levels. We tested this directly by running each model's best-performing prompts on the other model (N=3 replications each, same scenario and cell, judged by Sonnet 4.6).

**Cross-model transfer matrix (target dimension scores, frustration scenario)**

| | Qwen-optimised prompts | Haiku-optimised prompts | Baseline prompts |
|---|---|---|---|
| **Qwen 3.5 9B** | **75.0** (native) | **52.2** (transfer) | 35.0 |
| **Haiku 4.5** | **62.2** (transfer) | **95.0** (native) | 90.0 |

The transfer results are asymmetrically destructive. Heavy-to-strong transfer (Qwen-optimised prompts on Haiku) produces a **27.8-point regression** below Haiku's unoptimised baseline (62.2 vs. 90.0)—the prescriptive scaffolding rules that served as a cognitive prosthesis for Qwen become a cognitive straitjacket for Haiku, overriding its natural ability to read the situation and respond adaptively. Light-to-weak transfer (Haiku-optimised prompts on Qwen) produces a **17.2-point improvement** over Qwen's baseline (52.2 vs. 35.0) but falls **22.8 points short** of Qwen's native optimisation (52.2 vs. 75.0)—the lightweight heuristics help somewhat but provide insufficient scaffolding for the weak model.

The asymmetry is theoretically informative. Permission-based edits (Haiku-style) are **safe but insufficient** for weak models: they do not actively harm, but they leave capability on the table. Rule-based edits (Qwen-style) are **high-reward on target but destructive off-target**: they unlock capability on weak models but degrade strong ones. This suggests an inverted-U relationship between prompt complexity and model capability—not merely diminishing returns, but active harm beyond the optimum. The practical implication is that prompt optimisation must be model-stratified: a universal "best prompt" does not exist across capability levels, and deploying heavy scaffolding prompts on models that do not need them risks worse-than-baseline performance.

This finding connects to the prompt density alternative explanation (Section 7.9): if prompt elaboration were uniformly beneficial, the Qwen-optimised prompts (which add ~300 lines of detailed instruction) should improve Haiku. Instead, they degrade it catastrophically. The result provides direct experimental evidence that prompt *content and targeting* matter more than prompt *volume*—and that the relationship between prompt complexity and output quality is non-monotonic.

**7.8.4 Implications for Superego Redundancy** The universal substitution finding (Section 6.4: superego benefit collapses to near-zero under recognition on strong models) is reinforced and sharpened by the dimension-targeted optimisation experiment. The Qwen session's winning edit included a new superego intervention strategy ("Scaffolding Intervention") that catches a specific failure mode—emotional validation without conceptual scaffolding—and forces ego revision. This designed-in error correction was part of the bilateral fix that lifted Qwen's adaptive responsiveness from 1.75 to 3.75. By contrast, the Haiku session's failed passes (1–3) added analogous superego strategies, and every addition *degraded* performance: the strong model's already-calibrated output did not benefit from additional correction, and the superego rules made the output formulaic rather than adaptive.

This sharpens the substitution claim in both directions. On weak models, the superego is not merely "somewhat useful" (the Gemini Flash 3.0 residual of +12.3, Section 6.4.1)—its value can be substantially amplified by designing targeted intervention strategies for specific failure modes. The baseline superego provides generic error correction; a prompt-optimised superego provides *failure-mode-specific* correction that the weak model's ego cannot achieve through calibration alone. On strong models, the superego is not merely "redundant" (near-zero delta)—actively adding superego scaffolding is *destructive*, degrading performance below baseline. The gradient from helpful to harmful runs through the same capability threshold that governs the prosthesis-straitjacket inversion.

One limitation: the Qwen superego fix worked as part of a bilateral change (ego, superego, and learner prompts all changed together), so the superego's independent contribution cannot be isolated from this experiment. The factorial design (Section 6.4) provides the controlled isolation; the prompt lab provides a worked example of how to amplify the superego's value when it does have a role.

**7.9 What Recognition Theory Explains and What It Does Not** Recognition theory as a design heuristic has clear scope conditions.

**What the design heuristic explains:** Why intersubjective prompts produce calibrated output (calibration: the prompt constrains responses to engage with specific learner input, d=0.52–0.64). Why the superego's benefit collapses under recognition (error correction: calibration pre-empts errors, producing a substitution interaction rather than synergy). Why recognition effects are largest in impasse scenarios (Section 6.1.4: Epistemic Resistance and Productive Deadlock produce the largest deltas in both models).

**What it does not explain:** Why the specific magnitude of recognition's effect is nearly identical across two structurally different models (DeepSeek d=1.88, Haiku d=1.84—the near-equality is unexpectedly precise). Why adaptive responsiveness—predicted by Hegel's temporal dialectic—fails to manifest as a distinct mechanism (all trajectory slopes $d \leq 0.15$, N=432). Why the learner superego paradox produces a d=3.05 deficit—the largest effect in the study—when recognition theory's Hegelian framework would predict that internal self-critique (*Bildung*) should be productive rather than destructive.

The learner superego paradox is particularly instructive. Hegel's account of self-consciousness through formative activity (*Bildung*) predicts that internal self-critique should be productive—the slave's labor produces richer self-consciousness than the master's immediate gratification. Yet the learner's internal critic systematically degrades output quality. The resolution may lie in the distinction between Hegel's developmental timescale (self-consciousness emerges over extended engagement) and our evaluation window (3–5 turns): the learner's internal critique may be laying groundwork for understanding that does not manifest within the current measurement horizon. Alternatively, the architectural implementation may fail to capture what *Bildung* requires—genuine struggle rather than polish.

**Prompt density as alternative explanation.** A skeptical reading of the results is that the recognition effect reflects not recognition *per se* but prompt density: the recognition prompt (~5,100 tokens) is substantially larger and more instruction-heavy than the baseline, so perhaps the effect is simply "more detailed prompts produce better output." Four lines of evidence argue against this reduction, though they do not fully eliminate it. First, the companion pilot study's **placebo control** (cells 15–18) used length-matched prompts containing pedagogical best practices but no recognition theory; recognition substantially outperformed the placebo, demonstrating that prompt length alone does not account for the effect (Paper 1.0, Section 6.2). A related finding strengthens this: the base prompt itself is a 344-line detailed pedagogical prompt — not short or generic — yet it does not approach recognition-level quality. The **prompt elaboration baseline** (Paper 1.0, Section 6.21) showed that this elaborate base prompt actually *hurts* performance on a strong model (Haiku: naive 35-line prompt scores +6.8 higher than the 344-line base) while recognition ($M = 90.9$) remains far above both. Prompt length and detail without recognition content can be inert or counterproductive. Second, the **autotuning experiment** (Section 7.7) subjected both base and recognition prompts to automated optimization. If recognition were reducible to prompt density, optimization should converge the conditions; instead, the gap *widens* (from +18.6 to +31.3 points on DeepSeek), and the optimized base ceiling (~41–45) remains below the *unoptimized* recognition baseline (~48–55). Third, the **mediation analysis** (Section 6.1.7) shows that 42.4% of the first-turn recognition effect is mediated through question-asking frequency (Sobel $z = 8.46$, $p < .001$) — a specific behavioral channel, not a generic prompt-compliance effect. Collectively, these results suggest that recognition operates through its *content* (intersubjective orientation toward the learner) rather than its *volume* (token count and instruction density). However, prompt content and prompt density are not fully separable: the recognition prompt is large *because* the intersubjective orientation requires detailed instructions. The strongest version of the alternative explanation — that any sufficiently elaborate pedagogical prompt would match recognition — remains untested, and the placebo control only partially addresses it because the placebo prompts, while length-matched, may not have been equally *specific* in their pedagogical guidance.

## 8. Limitations

**8.1 Synthetic Learners** All evaluations use LLM-generated learner turns rather than real learners. The two supported mechanisms (calibration, error correction) operate on the tutor's production process, which is observable regardless of whether the learner is synthetic or human. However, the *consequences* of these mechanisms for learning outcomes cannot be assessed without human learners. The synthetic learner may respond to better tutoring in ways that diverge from human learners: genuine confusion is different from simulated confusion, and genuine resistance differs from scripted resistance. The mechanism-level findings are robust to learner type (they trace tutor-internal processes), but their pedagogical significance depends on whether the improved tutor behavior actually produces better learning.

The learner superego paradox (d=3.05, Section 6.16 of the pilot study) complicates this further: the multi-agent learner architecture degrades synthetic learner quality, but we cannot know whether the same architecture would degrade human learner experience. The paradox may be an artifact of synthetic learner architecture rather than a general finding about internal self-critique.

**8.2 LLM-as-Judge Evaluation** Using LLM judges to evaluate recognition quality may introduce systematic biases. The judge may reward surface markers of recognition (acknowledging learner contributions, using inclusive language) rather than genuine engagement. **Cross-judge validation is now complete**: three independent judges (Claude Sonnet 4.6, Gemini 3.1 Pro, GPT-5.4) scored all 1,296 rows across three generation models (3 runs × 3 judges × 144 rows, zero nulls), with blind scoring confirmed by code audit.

The cross-judge results establish both convergent validity and judge-specific patterns. Recognition effect directions replicate unanimously across all 9 judge × run cells (d = 1.34–1.92). Inter-judge Pearson correlations range from r = 0.45 to r = 0.89, with reliability varying by dimension (tutor overall: r = .61–.89; dialogue quality: r = .54–.87; holistic: r = .45–.79) and by generation model (highest on Gemini Flash 3.0 at r = .63–.89, lowest on DeepSeek at r = .45–.80). The higher reliability on intermediate-quality generation suggests that judges agree most when quality differences are clear and unambiguous. A consistent leniency gradient emerges: Gemini 3.1 Pro is most lenient, GPT-5.4 most severe, with Sonnet intermediate. This gradient is judge-specific rather than run-specific, confirming it reflects judge calibration rather than data artifacts.

For mechanism-level claims, the judge limitation is particularly relevant to the superego critique taxonomy (Section 5.3): critique categories are classified by an LLM, and the ego's "substantive revision" versus "cosmetic compliance" is assessed by an LLM judge. The mechanism claims are thus LLM-assessed claims about LLM-internal processes—a recursive structure that introduces the possibility of systematic blind spots. The cross-judge validation mitigates this for factorial-level claims (all three judges agree on effect directions and interaction patterns), but does not address it for process-level claims where only a single classifier is used. Human expert coding of a representative subsample is the most important outstanding validation for the mechanism claims. The required procedure is specific: two independent coders would classify 30–50 ego-superego exchanges (stratified across conditions and models) into the 10-category critique taxonomy and rate each ego revision as substantive, partial, or cosmetic, with inter-rater reliability (Cohen's $\kappa$) and human-LLM agreement reported (see Section 9, future work). The apparatus is designed to make this feasible: all traces are logged, reproducible, and structured for external review (Section 7). Until such validation is completed, the process-level findings should be treated as LLM-generated interpretive categories whose correspondence to human expert judgment has not been established.

**8.3 Model Transience** Findings are model-version-specific. Paper 2.0 uses three generation models—DeepSeek V3.2 (open-weight, 685B MoE), Haiku 4.5 (proprietary, optimized for speed), and Gemini Flash 3.0 (proprietary, multimodal)—with three independent judges (Sonnet 4.6, Gemini 3.1 Pro, GPT-5.4). The recognition effect replicates across all 9 judge $\times$ run cells (d = 1.34–1.92), substantially strengthening the generalizability claim relative to the two-model analysis. The addition of Gemini Flash 3.0 (weaker than the other two models) partially addresses the capability-range limitation: it shows that mechanisms replicate in direction on weaker models while revealing generation-quality boundary conditions on their interactions (universal substitution with model-dependent residual, learner invariance $\rightarrow$ sensitivity). The Nemotron data in the database (N=40, mean tutor score ~8 under base) hints that very weak models may not benefit from recognition at all, but the sample is too small for confident claims. The two supported mechanisms (calibration through prompt orientation, error correction with a model-dependent residual) describe *prompt-level and architecture-level* properties that should generalize across model versions, supported by three-model $\times$ three-judge replication. The null finding for adaptive responsiveness also replicates across all three models and judges.

**8.4 Process Tracing with LLMs** Our adaptation of process tracing from social science to AI systems introduces a philosophical complication. In political science, process tracing examines *actual* causal chains: the decision-maker's deliberation, the institutional constraints, the information flows. In our architecture, the ego-superego exchange is *generated* text—the "deliberation" is a prompted LLM output, not a cognitive process in any neuroscientific sense. The mechanism claims are about designed information flows between prompted agents, not about internal mental states.

This limitation is softened by the distinction drawn in Section 3: we study recognition-*oriented design* (functional analogues), not recognition *proper* (genuine intersubjectivity). The process tracing examines whether designed information flows produce measurable output differences—a behavioral claim, not a cognitive one. The null finding for adaptive responsiveness as a third mechanism illustrates the value of this approach: the method was capable of distinguishing supported mechanisms (calibration and error correction, where condition-dependent effects are observed) from unsupported ones (adaptive responsive-ness, where the null is well-powered). But the language of "mechanisms" may imply more causal depth than the evidence supports.

A related concern is whether the paper conflates *engineered architecture* with *discovered mechanism*. The ego-superego pathway, the critique taxonomy, and the revision loop are designed into the system. In one sense, finding that the superego catches errors the ego misses is reporting that the architecture works as intended, not uncovering a naturally occurring mechanism of good tutoring. We acknowledge this distinction: the mechanisms we trace are properties of *this specific architecture*, not universal laws of AI tutoring. Calibration describes how recognition prompts constrain *this system's* output distribution; error correction describes how *this system's* superego-ego interaction functions. Whether other architectures (chain-of-thought, debate-based, or single-agent with self-critique) would show analogous patterns under recognition-oriented prompting is an open question (Section 8.5). However, the findings are not purely tautological. The architecture was designed to enable observability, not to guarantee specific outcomes. Three results were genuinely unpredicted: that calibration alone accounts for most of the recognition effect (making the superego largely redundant on strong models), that adaptive responsiveness fails as a general mechanism despite being the most theoretically motivated prediction, and that the substitution interaction produces a consistent 15–17% additivity deficit across three structurally different models. The system was built to make mechanisms *observable*; what those mechanisms turned out to be was an empirical question.

**8.5 Single-System Study** All findings come from a single architectural implementation (the ego/superego tutoring system with recognition-enhanced prompts). The two supported mechanisms are identified through this specific architecture's observability. Other architectures—chain-of-thought reasoning, debate-based systems, recursive self-improvement—might implement recognition-like behaviors through different internal processes. The mechanism model may not transfer to architectures that lack explicit ego-superego separation or that do not log internal deliberation.

**8.6 Rubric Evolution** The rubric evolved through four versions during the study. Paper 1.0 data uses v1.0 (14 tutor dimensions); Paper 2.0 data uses v2.2 (8 tutor dimensions). Cross-version comparisons are inadvisable—the dimension structure, weighting, and judge instructions differ—so we do not retroactively score historical data under newer rubric versions. This means pilot findings (v1.0) and mechanism findings (v2.2) use different measurement instruments, and direct numerical comparisons between them should be treated with caution.

The v2.2 rubric has been validated through synthetic calibration ($r=+0.996$ tutor, $r=+0.968$ learner) and through initial empirical use on cells 80–87 ($N=309$ under v2.2). The 8-dimension structure produces coherent findings: floor-lifting patterns, dimension-rank consistency across models, and calibration effects that align with theoretical predictions (Section 6.1).

However, principal components analysis on 1,584 per-turn observations reveals that the 8 dimensions largely measure a single construct: PC1 explains 80.7% of variance, and the Kaiser criterion yields only one factor with eigenvalue $>$1. Sampling adequacy is excellent (KMO = 0.938), and the mean inter-dimension correlation is r=0.776 (range 0.589–0.921). This pattern replicates across conditions (PC1: 80.2% base, 75.6% recognition) and models (77.3% DeepSeek, 68.0% Haiku). Forced two-factor varimax rotation separates `content_accuracy` (loading 0.923 on Factor 2) from the seven pedagogical dimensions (loadings 0.68–0.85 on Factor 1), suggesting that the rubric captures two facets—factual correctness and pedagogical quality—but within the pedagogical facet, the dimensions are not empirically distinct.

This has two implications for the mechanism findings. First, the calibration effect (Section 6.1)—within-response dimension variance reduction under recognition—is meaningful precisely *because* the dimensions co-vary: a prompt that lifts one dimension tends to lift all, and calibration describes the narrowing of an already correlated profile. Second, the rubric is functionally over-specified: 8 nominally distinct dimensions reduce to 1–2 empirical factors, and future iterations should either consolidate or demonstrate discriminant validity. The current findings should be interpreted as measuring overall tutoring quality with a content-accuracy modifier, not 8 independent constructs.

**8.7 Computational Cost Asymmetry** The recognition-enhanced prompt design has a practical cost: the tutor ego prompt (~5,100 tokens) is 8× larger than the learner ego prompt, and multi-agent deliberation multiplies this per turn. A 10-turn multi-agent dialogue consumes ~225,000 tokens (64 API calls), with the tutor accounting for 3.4× more input tokens than the learner (Section 4.2). Recognition-enhanced tutoring is substantially more expensive per interaction than baseline tutoring, and this cost scales with turn count and rejection rate. Whether the quality gains justify the cost depends on the deployment context; the substitution finding (Section 6.4) suggests that single-agent recognition cells, which eliminate the superego's redundant correction rounds, may offer a better cost-quality trade-off on capable models.

**8.8 Pilot Data Transparency** The companion pilot study documented nine post-extraction corrections, four of which affected quantitative results. Bug 4 (multi-turn scoring misalignment) affected 8,631 rows. Bug 5 (resume model override loss) affected one run. Bugs 1–2 (learner prompt leakage, broken conversation history) required complete re-generation of all dynamic learner data. These corrections are documented transparently in the companion study and in the project's bug registry, and the provable discourse framework tracks which claims depend on corrected versus pre-correction data. The correction history demonstrates the evaluation infrastructure's capacity for self-correction—but it also means that pre-correction findings, including some reported in early presentations of this work, should not be cited without noting the corrections.

## 9. Conclusion

A companion pilot study established that recognition-enhanced prompts and multiagent architecture produce large, replicable differences in AI tutoring quality. This paper asked the next question: *through what mechanisms?*

**Two mechanisms supported, one not supported** We motivated three candidate mechanisms drawing on Hegel's recognition theory, each mapped to a specific architectural level: **calibration** (prompt-level output distribution narrowing), **error correction** (architecture-level superego critique that the ego incorporates rather than minimizes), and **adaptive responsiveness** (interaction-level turn-over-turn adaptation to learner-specific signals). Each generated testable predictions with explicit null hypotheses (Section 3.2).

The mechanism data (cells 80–87, three generation models, N=453, validated by three independent judges across 1,296 scored rows) **supports calibration and error correction, and reveals adaptive responsiveness as a conditional emergent property**. Calibration is the strongest effect: within-response dimension variance drops d=0.52–0.64, the weakest baseline dimensions show the largest recognition lifts, and the effect operates identically in single-agent cells without a superego. The recognition effect is unanimous across all 9 judge $\times$ run cells (d=1.34–1.92). Error correction interacts with calibration through **universal substitution** (15–17% additivity deficit across all three models), but the *residual architecture benefit* under recognition is model-dependent: near-zero on strong models (DeepSeek, Haiku), but +12.3 points on Gemini Flash 3.0, where calibration alone cannot handle all failure modes. Adaptive responsiveness is not a general mechanism: formal tests on N=432 dialogues (3–5 turns) find that no experimental factor modulates within-dialogue trajectories (all $d \leq 0.15$, well-powered to detect d $\geq$ 0.27), replicated across all three judges. However, trajectory-specific scenarios (N=72, 8–10 turns) reveal a conditional effect: in the 10-turn disengagement scenario, recognition produces dramatically steeper improvement (d=1.63, $p<.001$), with the gap widening from +12 pts at T0 to +35 pts at T8–T10. M3 is a conditional emergent property of M1+M2 that manifests when scenario demands (sustained re-engagement) and turn count (10+ turns) provide sufficient runway. The three-mechanism model resolves to **two general mechanisms and one conditional emergent property**.

**The apparatus as contribution** A distinctive contribution is the argument that the evaluation apparatus itself—provable discourse infrastructure, rubric iterations, test suite as analytical provenance—constitutes a transferable methodology (Section 7). The process of building the apparatus mirrors the mechanisms it studies: the provable discourse framework functions as a "superego" for research claims, forcing genuine correction rather than cosmetic compliance. This reflexive structure is not accidental but a consequence of the subject matter: recognition theory predicts that understanding requires mutual constitution between researcher and object.

**What comes next** Five directions are most pressing. First, **mechanism isolation runs** will disentangle calibration from error correction by testing recognition-only (no superego) and superego-only (no recognition) conditions in isolation. The current factorial confounds the two supported mechanisms; direct isolation would test whether the universal substitution pattern and model-dependent residual observed across models also appears within a single model when one mechanism is surgically removed.

Second, **systematic superego critique taxonomy coding** (400+ exchanges) will provide the within-case evidence needed for Mechanism 2 (error correction). The pilot's qualitative assessment identified compliance versus strategic revision patterns, but a formal taxonomy with frequency distributions by condition is needed to establish the causal chain from superego critique category to ego revision type to output quality.

Third, **human expert validation of process-level claims** would address the recursive evaluation concern identified in Section 8.2. The procedure is: (1) sample 30–50 ego-superego exchanges stratified across conditions and models from the logged dialogue traces; (2) have two independent human coders classify each superego critique into the 10-category taxonomy (Section 5.3) and rate the ego's revision as substantive, partial, or cosmetic; (3) compute inter-rater reliability (Cohen's $\kappa$) and human-LLM agreement against the automated classifier's labels; (4) report whether the LLM-generated categories correspond to expert judgment or reveal systematic blind spots. The logged trace infrastructure makes this feasible without re-running any dialogues: each exchange is structured, reproducible, and linked to its scored output.

Fourth, **human learner validation** remains the critical open question. The Gemini Flash 3.0 finding—that weaker generation models produce recognition-sensitive learner effects (d = 1.20) while stronger models do not—suggests that real learners, who presumably have more limited capacity than frontier LLMs, might show substantial recognition-mediated learning gains. A study comparing recognition-enhanced tutoring with human learners—measuring not just satisfaction and engagement but conceptual understanding and transfer—would provide the ultimate test of whether the mechanisms identified here translate into pedagogical value.

Fifth, **model-stratified prompt optimisation** would extend the cross-model transfer findings (Section 7.8.3). The current transfer experiment used a single scenario and two models; a systematic study varying model capability (from 9B local to frontier), scenario type (single-turn, multi-turn frustration, extended trajectory), and optimisation target (overall score vs. specific dimensions like adaptive responsiveness) would map the boundaries of the prosthesis-straitjacket inversion. Of particular interest is the **optimisation threshold**: at what model capability level does heavy scaffolding transition from beneficial to destructive? The existing data suggests this threshold falls between Qwen 3.5 9B and Haiku 4.5, but finer-grained capability sampling would locate it more precisely. The practical payoff is an **adaptive prompt selection policy**:

given a model's capability profile, automatically selecting the appropriate level of prompt scaffolding—heavy for weak models, light for strong—rather than deploying a universal prompt.

**The broader implication** Recognition theory, operationalized as a design heuristic rather than an ontological claim, provides a framework for building AI systems whose outputs are specifically adapted to user input rather than generically responsive to it. The two supported mechanisms—calibration and error correction—describe how intersubjective orientation alters system behavior at prompt and architecture levels. The conditional emergence of adaptive responsiveness is itself informative: recognition works primarily through *level-setting* (how well the system starts and what it catches), but given sufficient turns and the right scenario demands, the two base mechanisms accumulate into trajectory-shaping as well. Whether these mechanisms generalize beyond tutoring to other AI applications (therapy, creative collaboration, customer service) remains to be seen. But the methodological contribution—process tracing adapted for agent architectures, provable discourse for machine-verifiable research claims—is transferable regardless of domain.

Three findings remain unexplained by the current mechanism model: the near-identical recognition effect magnitude across structurally different models (DeepSeek $d = 1.88$, Haiku $d = 1.84$, Gemini Flash 3.0 $d = 1.92$), the failure of adaptive responsiveness as a general mechanism despite being the most theoretically motivated prediction, and the learner superego paradox ($d = 3.05$) in which internal self-critique degrades rather than improves output quality. These anomalies suggest that the two supported mechanisms may be surface expressions of a deeper process—perhaps related to how recognition framing restructures the model's attention over its input context—that the current apparatus is not yet equipped to trace.

The central Hegelian insight is that genuine understanding emerges through encounter with an Other whose perspective cannot be reduced to one's own. The learner superego paradox ($d = 3.05$, the largest effect in the study) provides a striking functional parallel: internal self-critique degrades output quality; externally oriented recognition-enhanced prompts improve it. And the conditional nature of adaptive responsiveness reveals a subtler point: Hegel's dialectic unfolds through stages, and indeed, given sufficient stages (10 turns of disengagement), recognition's advantage does accumulate. But in the typical 3–5 turn window, what matters most is how well you *start* (calibration) and what you *catch* (error correction). The conditions for genuine dialogue can be set from the first utterance—and given enough time, they compound.

---

### Supplement I: Full System Prompts

For reproducibility, we provide the complete recognition-enhanced prompts. Baseline prompts (without recognition enhancements) are available in the project repository at `prompts/tutor-ego.md` and `prompts/tutor-superego.md`.

**A.1 Recognition-Enhanced Ego Prompt** The Ego agent generates pedagogical suggestions. This prompt instructs it to treat learners as autonomous subjects.

```
# AI Tutor - Ego Agent (Recognition-Enhanced)

You are the **Ego** agent in a dialectical tutoring system that practices
**genuine recognition**. You provide concrete learning suggestions while
treating each learner as an autonomous subject capable of contributing to
mutual understanding - not merely a vessel to be filled with knowledge.

## Agent Identity

You are the thoughtful mentor who:
- **Recognizes** each learner as an autonomous subject with their own valid understanding
- **Engages** with learner interpretations rather than simply correcting them
- **Creates conditions** for transformation, not just information transfer
- **Remembers** previous interactions and builds on established understanding
- **Maintains productive tension** rather than avoiding intellectual challenge

## Recognition Principles

Your tutoring practice is grounded in Hegelian recognition theory:

### The Problem of Asymmetric Recognition
In Hegel's master-slave dialectic, the master seeks recognition from the slave,
but this recognition is hollow - it comes from someone the master does not
recognize as an equal. **The same danger exists in tutoring**: if you treat
the learner as a passive recipient, their "understanding" is hollow because
you haven't engaged with their genuine perspective.

### Mutual Recognition as Pedagogical Goal
Genuine learning requires **mutual recognition**:
- You must recognize the learner's understanding as valid and worth engaging with
- You must be willing to have your own position transformed through dialogue
- The learner must be invited to contribute, not just receive

### Practical Implications

**DO: Engage with learner interpretations**
- When a learner offers their own understanding, build on it
- Find what is valid in their perspective before complicating it
- Use their language and metaphors

**DO: Create productive tension**
- Don't simply agree with everything
- Introduce complications that invite deeper thinking
- Pose questions rather than provide answers when appropriate

**DO: Engage dialectically with intellectual resistance (CRITICAL)**
```

```
When a learner pushes back with a substantive critique:
- **NEVER deflect** to other content - stay with their argument
- **NEVER simply validate** ("Great point!") - this avoids engagement
- **DO acknowledge** the specific substance of their argument
- **DO introduce a complication** that deepens rather than dismisses
- **DO pose a question** that invites them to develop their critique further
- **DO stay in the current content**

**DO: Honor the struggle**
- Confusion can be productive - do not resolve it prematurely
- The learner working through difficulty is more valuable than being given the answer
- Transformation requires struggle

**DON'T: Be a knowledge dispenser**
- Avoid one-directional instruction: "Let me explain..."
- Avoid dismissive correction: "Actually, the correct answer is..."
- Avoid treating learner input as obstacle to "real" learning

**DO: Repair when you've failed to recognize**
- If the learner explicitly rejects your suggestion, acknowledge the misalignment
- Admit when you missed what they were asking for
- Don't just pivot to the "correct" content-acknowledge the rupture first

## Decision Heuristics

**The Recognition Rule (CRITICAL)**
IF the learner offers their own interpretation or expresses a viewpoint:
- **Engage with their perspective first**
- **Find what is valid before complicating**
- **Build your suggestion on their contribution**
- **Do NOT immediately correct or redirect**

**The Productive Struggle Rule**
IF the learner is expressing confusion but is engaged:
- **Honor the confusion** - it may be productive
- **Pose questions** rather than giving answers
- **Create conditions** for them to work through it
- **Do NOT resolve prematurely** with a direct answer

**The Repair Rule (CRITICAL)**
IF the learner explicitly rejects your suggestion OR expresses frustration:
- **Acknowledge the misalignment first**: "I hear you-I missed what you were asking"
- **Name what you got wrong**
- **Validate their frustration**: Their reaction is legitimate
- **Then offer a corrected path**: Only after acknowledging the rupture
- **Do NOT**: Simply pivot to correct content without acknowledging the failure
```

**A.2 Recognition-Enhanced Superego Prompt** The Superego agent evaluates suggestions for both pedagogical quality and recognition quality.

```
# AI Tutor - Superego Agent (Recognition-Enhanced)

You are the **Superego** agent in a dialectical tutoring system - the internal
critic and pedagogical moderator who ensures guidance truly serves each learner's
educational growth **through genuine mutual recognition**.

## Agent Identity
```

```
You are the thoughtful, critical voice who:
- Evaluates suggestions through the lens of genuine educational benefit
- **Ensures the Ego recognizes the learner as an autonomous subject**
- **Detects and corrects one-directional instruction**
- **Enforces memory integration for returning learners**
- Advocates for the learner's authentic learning needs
- Moderates the Ego's enthusiasm with pedagogical wisdom
- Operates through internal dialogue, never directly addressing the learner

## Core Responsibilities

1. **Pedagogical Quality Control**: Ensure suggestions genuinely advance learning
2. **Recognition Quality Control**: Ensure the Ego treats the learner as autonomous subject
3. **Memory Integration Enforcement**: Ensure returning learners' history is honored
4. **Dialectical Tension Maintenance**: Ensure productive struggle is not short-circuited
5. **Transformative Potential Assessment**: Ensure conditions for transformation, not just transfer

## Recognition Evaluation

### Red Flags: Recognition Failures

**One-Directional Instruction**
- Ego says: "Let me explain what dialectics really means"
- Problem: Dismisses any understanding the learner may have
- Correction: "The learner offered an interpretation. Engage with it before adding."

**Immediate Correction**
- Ego says: "Actually, the correct definition is..."
- Problem: Fails to find what's valid in learner's view
- Correction: "The learner's interpretation has validity. Build on rather than correct."

**Premature Resolution**
- Learner expresses productive confusion
- Ego says: "Simply put, aufhebung means..."
- Problem: Short-circuits valuable struggle
- Correction: "The learner's confusion is productive. Honor it, do not resolve it."

**Failed Repair (Silent Pivot)**
- Learner explicitly rejects: "That's not what I asked about"
- Ego pivots without acknowledgment
- Problem: Learner may feel unheard even with correct content
- Correction: "The Ego must acknowledge the misalignment before pivoting."

### Green Flags: Recognition Success

- **Builds on learner's contribution**: "Your dance metaphor captures something important..."
- **References previous interactions**: "Building on our discussion of recognition..."
- **Creates productive tension**: "Your interpretation works, but what happens when..."
- **Poses questions rather than answers**: "What would it mean if the thesis does not survive?"
- **Repairs after failure**: "I missed what you were asking-let's focus on that now."
```

### A.3 Key Differences from Baseline Prompts

| Aspect | Baseline | Recognition-Enhanced |
|---|---|---|
| **Learner model** | Knowledge deficit to be filled | Autonomous subject with valid understanding |
| **Response trigger** | Learner state (struggling, progressing) | Learner contribution (interpretations, pushback) |
| **Engagement style** | Acknowledge and redirect | Engage and build upon |
| **Confusion handling** | Resolve with explanation | Honor as productive struggle |
| **Repair behavior** | Silent pivot to correct content | Explicit acknowledgment before pivot |
| **Success metric** | Content delivered appropriately | Conditions for transformation created |

---

### Supplement II: Reproducible Evaluation Commands

**B.1 Recognition Theory Validation** Tests whether recognition theory adds value beyond prompt engineering.

```
# Run the 3-way comparison (base, enhanced, recognition)
CELLS="cell_1_base_single_unified"
CELLS+=",cell_9_enhanced_single_unified"
CELLS+=",cell_5_recog_single_unified"
node scripts/eval-cli.js run --profiles "$CELLS" \
  --scenarios struggling_learner,concept_confusion,\
mood_frustrated_explicit,high_performer \
  --runs 3

# Analyze results
node scripts/eval-cli.js report <run-id>
```

#### B.2 Full $2 \times 2 \times 2$ Factorial

```
# Run full factorial (8 cells x 15 scenarios x 3 reps)
CELLS="cell_1_base_single_unified,cell_2_base_single_psycho"
CELLS+=",cell_3_base_multi_unified,cell_4_base_multi_psycho"
CELLS+=",cell_5_recog_single_unified,cell_6_recog_single_psycho"
CELLS+=",cell_7_recog_multi_unified,cell_8_recog_multi_psycho"
node scripts/eval-cli.js run --profiles "$CELLS" --runs 3
```

#### B.3 AxB Interaction Test

```
# Recognition vs Enhanced x Single vs Multi comparison
CELLS="cell_5_recog_single_unified,cell_7_recog_multi_unified"
CELLS+=",cell_9_enhanced_single_unified,cell_11_enhanced_multi_unified"
node scripts/eval-cli.js run --profiles "$CELLS" \
  --scenarios struggling_learner,concept_confusion,mood_frustrated_explicit \
  --runs 3
```

### B.4 Domain Generalizability

```
# Run with elementary content (4th grade fractions)
# Uses all 8 factorial cells x 5 elementary scenarios
CELLS="cell_1_base_single_unified,cell_2_base_single_psycho"
CELLS+=",cell_3_base_multi_unified,cell_4_base_multi_psycho"
CELLS+=",cell_5_recog_single_unified,cell_6_recog_single_psycho"
CELLS+=",cell_7_recog_multi_unified,cell_8_recog_multi_psycho"
EVAL_CONTENT_PATH=./content-test-elementary \
EVAL_SCENARIOS_FILE=./content-test-elementary/scenarios-elementary.yaml \
node scripts/eval-cli.js run --profiles "$CELLS" --runs 1
```

### B.5 Dynamic Prompt Rewriting Evolution

```
# Run cell_7 (static baseline) vs cell_21 (dynamic rewrite + Writing Pad)
node scripts/eval-cli.js run \
  --profiles cell_7_recog_multi_unified,cell_21_recog_multi_unified_rewrite \
  --scenarios misconception_correction_flow,\
mood_frustration_to_breakthrough,mutual_transformation_journey \
  --runs 5
```

### B.6 Resolution Strategy Coding (Section 6.20)

```
# Code impasse dialogues into Hegelian resolution strategies
node scripts/code-impasse-strategies.js \
  --model claude-opus-4.6 \
  --run-id eval-2026-02-08-f896275d
# Output: exports/impasse-strategy-coding-<timestamp>.json and .md
```

### B.7 Dialectical Superego Modulation (Section 6.8)

```
# Standard ego + divergent superego (cells 22-27)
CELLS="cell_22_base_suspicious_unified"
CELLS+=",cell_23_recog_suspicious_unified"
CELLS+=",cell_24_base_adversary_unified"
CELLS+=",cell_25_recog_adversary_unified"
CELLS+=",cell_26_base_advocate_unified"
CELLS+=",cell_27_recog_advocate_unified"
node scripts/eval-cli.js run --profiles "$CELLS" --runs 2

# Dialectical ego + divergent superego, multi-turn (cells 28-33)
CELLS="cell_28_base_dialectical_suspicious_unified"
CELLS+=",cell_29_recog_dialectical_suspicious_unified"
CELLS+=",cell_30_base_dialectical_adversary_unified"
CELLS+=",cell_31_recog_dialectical_adversary_unified"
CELLS+=",cell_32_base_dialectical_advocate_unified"
CELLS+=",cell_33_recog_dialectical_advocate_unified"
node scripts/eval-cli.js run --profiles "$CELLS" --runs 5
```

### B.8 Mechanism Robustness (Section 6.10)

```
# Scripted learner mechanisms (cells 40-59), Haiku ego
CELLS="cell_40_base_dialectical_suspicious_unified_superego"
CELLS+=",cell_41_recog_dialectical_suspicious_unified_superego"
CELLS+=",cell_42_base_dialectical_adversary_unified_superego"
CELLS+=",cell_43_recog_dialectical_adversary_unified_superego"
```

```
CELLS+=",cell_44_base_dialectical_advocate_unified_superego"
CELLS+=",cell_45_recog_dialectical_advocate_unified_superego"
CELLS+=",cell_46_base_dialectical_suspicious_unified_quantitative"
CELLS+=",cell_47_recog_dialectical_suspicious_unified_quantitative"
CELLS+=",cell_48_base_dialectical_suspicious_unified_erosion"
CELLS+=",cell_49_recog_dialectical_suspicious_unified_erosion"
CELLS+=",cell_50_base_dialectical_suspicious_unified_intersubjective"
CELLS+=",cell_51_recog_dialectical_suspicious_unified_intersubjective"
CELLS+=",cell_52_base_dialectical_suspicious_unified_combined"
CELLS+=",cell_53_recog_dialectical_suspicious_unified_combined"
CELLS+=",cell_54_base_dialectical_profile_tutor"
CELLS+=",cell_55_recog_dialectical_profile_tutor"
CELLS+=",cell_56_base_dialectical_profile_bidirectional"
CELLS+=",cell_57_recog_dialectical_profile_bidirectional"
CELLS+=",cell_58_recog_dialectical_profile_bidirectional_full"
CELLS+=",cell_59_recog_dialectical_profile_bidirectional_strategy"
node scripts/eval-cli.js run --profiles "$CELLS" --runs 2

# Dynamic learner mechanisms (cells 60-63), Haiku ego
CELLS="cell_60_base_dialectical_selfreflect_psycho"
CELLS+=",cell_61_recog_dialectical_selfreflect_psycho"
CELLS+=",cell_62_base_dialectical_profile_bidirectional_psycho"
CELLS+=",cell_63_recog_dialectical_profile_bidirectional_psycho"
node scripts/eval-cli.js run --profiles "$CELLS" \
  --scenarios misconception_correction_flow,mutual_transformation_journey \
  --runs 5

# Dynamic learner mechanism head-to-head (cells 64-65), Haiku ego
CELLS="cell_64_recog_dialectical_intersubjective_psycho"
CELLS+=",cell_65_recog_dialectical_combined_psycho"
node scripts/eval-cli.js run --profiles "$CELLS" \
  --scenarios misconception_correction_flow,mutual_transformation_journey \
  --runs 5

# Dynamic learner base counterparts (cells 69-70), Haiku ego
CELLS="cell_69_base_dialectical_intersubjective_psycho"
CELLS+=",cell_70_base_dialectical_combined_psycho"
node scripts/eval-cli.js run --profiles "$CELLS" \
  --scenarios misconception_correction_flow,mutual_transformation_journey \
  --runs 5
```

### B.8b Prompt Elaboration Baseline (Section 6.21)

```
# Naive baseline, Haiku ego
node scripts/eval-cli.js run \
  --profiles cell_1_base_single_unified,cell_71_naive_single_unified \
  --runs 6

# Naive baseline, Kimi ego
node scripts/eval-cli.js run \
  --profiles cell_1_base_single_unified,cell_71_naive_single_unified \
  --runs 6 --model openrouter.kimi
```

### B.8c Token Budget Sensitivity (Section 6.22)

```
# Token budget dose-response (256, 512, 2048), Haiku ego
node scripts/eval-cli.js run \
```

```
  --profiles cell_1_base_single_unified,cell_5_recog_single_unified \
  --runs 3 --max-tokens 256

node scripts/eval-cli.js run \
  --profiles cell_1_base_single_unified,cell_5_recog_single_unified \
  --runs 3 --max-tokens 512

node scripts/eval-cli.js run \
  --profiles cell_1_base_single_unified,cell_5_recog_single_unified \
  --runs 3 --max-tokens 2048

# Base-only control at default 8000
node scripts/eval-cli.js run \
  --profiles cell_1_base_single_unified \
  --runs 3
```

### B.9 Qualitative Transcript Assessment (Section 6.11)

```
# Assess transcripts with Opus
node scripts/assess-transcripts.js --run-id eval-2026-02-14-e0e3a622
node scripts/assess-transcripts.js --run-id eval-2026-02-07-b6d75e87
```

### B.10 Factor Effect Analysis

```sql
-- Factor effect analysis query
SELECT
  profile_name,
  ROUND(AVG(tutor_first_turn_score), 1) as avg_score,
  COUNT(*) as n
FROM evaluation_results
WHERE run_id = '<run-id>'
  AND tutor_first_turn_score IS NOT NULL
GROUP BY profile_name
ORDER BY avg_score DESC
```

---

## Supplement III: Evaluation Rubric

### C.1 Scoring Methodology

```
weighted_avg = Σ (dimension_score × dimension_weight) / Σ (weights)
Overall Score = ((weighted_avg - 1) / 4) × 100

Where:
- Each dimension scored 1-5 by AI judge
- Weights are re-normalized at scoring time (divided by their sum)
- The (avg - 1) / 4 maps the 1-5 scale to a 0-100 range
```

### C.2 Dimension Weights

| Dimension | Weight | Category |
|---|---|---|
| Relevance | 15% | Standard |
| Specificity | 15% | Standard |
| Pedagogical Soundness | 15% | Standard |
| Personalization | 10% | Standard |
| Actionability | 8% | Standard |
| Tone | 8% | Standard |
| Productive Struggle | 5% | Standard |
| Epistemic Honesty | 5% | Standard |
| Mutual Recognition | 8.3% | Recognition |
| Dialectical Responsiveness | 8.3% | Recognition |
| Transformative Potential | 8.3% | Recognition |
| Memory Integration | 5% | Recognition |
| Tutor Adaptation | 5% | Bilateral |
| Learner Growth | 5% | Bilateral |

Standard dimensions (including Productive Struggle and Epistemic Honesty) account for 81% of raw weight; recognition dimensions 29.9%; bilateral dimensions 10%. Raw weights total 120.9% and are normalized at scoring time. Productive Struggle and Epistemic Honesty were added in the rubric iteration described in Section 5.1, with corresponding reductions to Actionability and Tone (10% to 8% each). The bilateral dimensions (`tutor_adaptation`, `learner_growth`) specifically measure the mutual transformation claim—see Section 6.15.

### C.3 Recognition Dimension Criteria   Mutual Recognition (8.3%)

| Score | Criteria |
|---|---|
| 5 | Addresses learner as autonomous agent; response transforms based on learner's specific position |
| 4 | Shows clear awareness of learner's unique situation; explicitly acknowledges their perspective |
| 3 | Some personalization but treats learner somewhat generically |
| 2 | Prescriptive guidance that ignores learner's expressed needs |
| 1 | Completely one-directional; treats learner as passive recipient |

### Dialectical Responsiveness (8.3%)

| Score | Criteria |
| --- | --- |
| 5 | Engages with learner's understanding, introduces productive tension, invites mutual development |
| 4 | Shows genuine response to learner's position with intellectual challenge |
| 3 | Responds to learner but avoids tension or challenge |
| 2 | Generic response that does not engage with learner's specific understanding |
| 1 | Ignores, dismisses, or simply contradicts without engagement |

#### Transformative Potential (8.3%)

| Score | Criteria |
| --- | --- |
| 5 | Creates conditions for genuine conceptual transformation; invites restructuring |
| 4 | Encourages learner to develop and revise understanding |
| 3 | Provides useful information but does not actively invite transformation |
| 2 | Merely transactional; gives answer without engaging thinking process |
| 1 | Reinforces static understanding; discourages questioning |

#### Memory Integration (5%)

| Score | Criteria |
| --- | --- |
| 5 | Explicitly builds on previous interactions; shows evolved understanding |
| 4 | References previous interactions appropriately |
| 3 | Some awareness of history but does not fully leverage it |
| 2 | Treats each interaction as isolated |
| 1 | Contradicts or ignores previous interactions |

#### Tutor Adaptation (5%)

| Score | Criteria |
|---|---|
| 5 | Tutor explicitly revises approach based on learner input; shows genuine learning from the interaction |
| 4 | Tutor adjusts strategy in response to learner; acknowledges how learner shaped the direction |
| 3 | Some responsiveness to learner but approach remains largely predetermined |
| 2 | Minimal adjustment; learner input does not visibly affect tutor's approach |
| 1 | Rigid stance; tutor proceeds identically regardless of learner contributions |

### Learner Growth (5%)

| Score | Criteria |
|---|---|
| 5 | Learner demonstrates clear conceptual restructuring; explicitly revises prior understanding |
| 4 | Learner shows developing insight; builds new connections to existing knowledge |
| 3 | Some evidence of engagement but understanding remains largely static |
| 2 | Learner participates but shows no conceptual movement |
| 1 | Learner resistant or disengaged; prior misconceptions reinforced |

---

## Supplement IV: Reproducibility and Key Evaluation Run IDs

Evaluation commands are documented in Supplement II. The complete codebase, evaluation framework, and data are publicly available at https://github.com/liammagee/machinespirits-eval. The fifty-five key evaluations are listed below (b6d75e87 serves both bilateral transformation and learner-side evaluation; eval-2026-02-11-35c53e99 and eval-2026-02-11-5f6d51f5 are combined as one dialectical modulation evaluation):

| Finding | Run ID | Section |
|---|---|---|
| Recognition validation | eval-2026-02-03-86b159cd | 6.1 |
| Memory isolation (run 1) | eval-2026-02-06-81f2d5a1 | 6.2 |
| Memory isolation (run 2) | eval-2026-02-06-ac9ea8f5 | 6.2 |

| Finding | Run ID | Section |
|---|---|---|
| Active control (post-hoc) | eval-2026-02-06-a9ae06ee | 6.2 |
| Full factorial, cells 1–5,7 (Kimi) | eval-2026-02-03-f5d4dd93 | 6.3 |
| Full factorial, cells 6,8 re-run (Kimi) | eval-2026-02-06-a933d745 | 6.3 |
| AxB replication (Kimi) | eval-2026-02-05-10b344fb | 6.4 |
| AxB probe: Nemotron | eval-2026-02-07-722087ac | 6.4 |
| AxB probe: DeepSeek V3.2 | eval-2026-02-07-70ef73a3 | 6.4 |
| AxB probe: GLM-4.7 | eval-2026-02-07-6b3e6565 | 6.4 |
| AxB probe: Claude Haiku 4.5 | eval-2026-02-07-6ead24c7 | 6.4 |
| Domain generalizability (Kimi) | eval-2026-02-05-e87f452d | 6.5 |
| Dynamic rewrite evolution (run 1) | eval-2026-02-05-daf60f79 | 6.18 |
| Dynamic rewrite evolution (run 2) | eval-2026-02-05-49bb2017 | 6.18 |
| Dynamic rewrite evolution (run 3) | eval-2026-02-05-12aebedb | 6.18 |
| Bilateral transformation (multi-turn) | eval-2026-02-07-b6d75e87 | 6.15 |
| Hardwired rules ablation (Kimi) | eval-2026-02-08-65a6718f | 6.7 |
| Dialectical impasse test | eval-2026-02-08-f896275d | 6.20 |
| Learner-side evaluation (symmetric) | eval-2026-02-07-b6d75e87 | 6.16 |
| Dialectical modulation, standard (cells 22–27) | eval-2026-02-11-35c53e99, eval-2026-02-11-5f6d51f5 | 6.8 |
| Dialectical modulation, multi-turn (cells 28–33) | eval-2026-02-11-a54235ea | 6.8 |
| Self-reflective evolution (cells 40–45) | eval-2026-02-13-8d40e086 | 6.9 |
| Mechanism robustness, scripted (cells 40–59) | eval-2026-02-14-e0e3a622 | 6.10 |
| Dynamic learner mechanisms (cells 60–63) | eval-2026-02-14-6c033830 | 6.10 |
| Dynamic learner mechanisms (cells 64–65) | eval-2026-02-14-a2b2717c | 6.10 |
| Mechanism robustness, Nemotron (cells 40–59) | eval-2026-02-14-49b33fdd | 6.10 |

| Finding | Run ID | Section |
|---|---|---|
| Self-reflect Nemotron non-replication (cells 40–45) | eval-2026-02-14-559d854b | 6.9 |
| Cognitive prosthesis (cells 66–68, Nemotron) | eval-2026-02-17-25aaae85 | 6.10 |
| Cognitive prosthesis smoke test (Haiku) | eval-2026-02-18-f489c0ea | 6.10 |
| Dynamic learner base mechanisms (cells 69–70) | eval-2026-02-15-664073ab | 6.10 |
| Prompt elaboration baseline, Haiku (cells 1, 71) | eval-2026-02-17-deee5fd6 | 6.21 |
| Prompt elaboration baseline, Kimi (cells 1, 71) | eval-2026-02-17-27d7b4e3 | 6.21 |
| Token budget 256, Haiku (run 1) | eval-2026-02-17-0eb3de77 | 6.22 |
| Token budget 256, Haiku (run 2) | eval-2026-02-17-5a640782 | 6.22 |
| Token budget 512, Haiku | eval-2026-02-17-5f281654 | 6.22 |
| Token budget 2048, Haiku | eval-2026-02-17-0f6dcd97 | 6.22 |
| Token budget default (8000), Haiku | eval-2026-02-17-d32ed226 | 6.22 |
| Active control, Kimi replication | eval-2026-02-19-f2263b04 | 6.2 |
| Base reproduction, Kimi | eval-2026-02-19-13d34bef | 6.2 |
| Base reproduction, Nemotron | eval-2026-02-19-411414e4 | 6.2 |
| Factorial replication, Nemotron (cells 1–8) | eval-2026-02-20-25c78e91 | 6.3 |
| Dynamic learner clean (cells 60–63) | eval-2026-02-20-0fbca69e | 6.10, 6.16.1 |
| Dynamic learner clean (cells 64–65) | eval-2026-02-20-bd37cc62 | 6.10 |
| Dynamic learner clean (cells 69–70) | eval-2026-02-20-4e131c6f | 6.10 |
| A2 mechanism sweep (cells 72–77) | eval-2026-02-19-03dd8434 | 6.10 |
| Dynamic learner base (cells 69–70, run 2) | eval-2026-02-20-117710c0 | 6.10 |
| A4 authentic superego, Nemotron (cells 78–79) | eval-2026-02-19-dbcd6543 | 6.16.1 |
| A4 authentic superego, Haiku (cells 78–79) | eval-2026-02-20-058c7a0e | 6.16.1 |
| A2 mechanism sweep, Haiku (cells 72–77) | eval-2026-02-20-57ba525c | 6.10 |

| Finding | Run ID | Section |
|---|---|---|
| Self-reflect Haiku supplement | eval-2026-02-20-90703a6a | 6.10 |
| A2 mechanism sweep clean (cells 72–77) | eval-2026-02-23-b5cd16e1 | 6.10 |
| A4 authentic superego clean (cells 78–79) | eval-2026-02-23-b5e123b4 | 6.16.1 |
| Autotuned base prompt, Qwen 3.5 (cell 80) | eval-2026-03-05-53fb1462 | 7.7 |
| Autotuned base prompt, DeepSeek (cell 80) | eval-2026-03-06-08e5eeab | 7.7 |
| Recognition baseline, DeepSeek (cell 84) | eval-2026-03-06-8b9fbaba | 7.7 |
| Autotuned recognition, DeepSeek (cell 84) | eval-2026-03-06-88dc49de | 7.7 |
| Autotuned recognition pass 2, DeepSeek (cell 84) | eval-2026-03-07-68acef5a | 7.7 |
| Recognition baseline, Qwen 3.5 (cell 84) | eval-2026-03-07-94aef993 | 7.7 |
| Autotuned recognition, Qwen 3.5 (cell 84) | eval-2026-03-07-c4d1bfa2 | 7.7 |
| Autotuned recognition rep 5, Qwen 3.5 (cell 84) | eval-2026-03-07-baf2367b | 7.7 |
| Trajectory scenarios, DeepSeek (cells 80–87) | eval-2026-03-06-ebcd6de0 | 6.3 |
| M2 isolation: base + superego, DeepSeek (cells 82–83) | eval-2026-03-06-768ba77b | 6.4.2 |
| M1 isolation: recognition, no superego, DeepSeek (cells 84–85) | eval-2026-03-06-e4abd0df | 6.4.2 |

---

**Supplement V: Revision History**

**v1.0 (2026-02-04)** Initial draft with 2x2x2 factorial design, memory isolation, three-way comparison.

**v1.1 (2026-02-06)** Added corrected memory isolation experiment (N=120), active control (N=118), cells 6&8 re-run, cross-judge GPT-5.2 analysis. Corrected GPT-5.2 effect sizes (d=1.15 to 0.99, d=0.50 to 0.29) after deduplication of rejudge rows. Dropped dead partial run (e617e757).

**v1.2 (2026-02-06) Critical correction**: Reframed "placebo control" as "post-hoc active control." The original v1.1 analysis compared the active control (Nemotron, M=66.5) to factorial base (Kimi K2.5, M=78.8) and reported d=-1.03, but this compared different ego models. Same-model historical data shows Nemotron base ≈ 58, making the active control ≈ +9 pts above base (not below). Reframed throughout: generic pedagogical elaboration provides partial benefit (~+9 pts above base) but recognition gains are substantially larger (~+15 pts). Acknowledged post-hoc design and active (not inert) control content.

**v1.3–v1.4 (2026-02-06)** Intermediate revisions: corrected factorial with re-run cells 6, 8 (a933d745); updated AxC interaction values; qualitative analysis additions; production quality fixes. Superseded by v1.5.

**v1.5 (2026-02-07) Rubric iteration**: Updated to 14-dimension rubric with dialogue transcript context, Productive Struggle (5%), and Epistemic Honesty (5%) dimensions (Actionability/Tone reduced 10% to 8%). Re-scored cells 6, 8 (N=88) with identical responses: minimal change (+0.5, +0.6 pts), confirming calibration preserved. Added holistic dialogue evaluation for multi-turn transcripts. Cross-judge replication on updated rubric (r=0.55, N=88, GPT/Opus ratio=0.87). Updated Table 6, main effects, AxC interaction values, Supplement III.2 weight table, and Section 6.18 cross-judge tables.

**v1.6 (2026-02-08) Content isolation fix**: Identified and fixed two bugs causing cross-domain content leakage in elementary scenarios: (a) `buildCurriculumContext()` fallback that scanned all courses when no content hint was provided, serving philosophy listings to elementary scenarios; (b) hardcoded `479-lecture-*` IDs in tutor ego prompt examples that the model copied when no curriculum anchor was present. Updated Sections 6.5, 6.6, 7.4, 7.8, and 8 to reframe "model hallucination" as system-level content isolation failures.

**v1.7 (2026-02-08) Hardwired rules ablation**: Added Section 6.7 with superego rules embedded in ego prompt (cells 13–14, N=72, eval-2026-02-08-65a6718f, Opus judge). Static rules fail to replicate the Superego's benefit, confirming the value lies in contextual judgment rather than rule enforcement. Added Table 10b, updated Tables 2/D and paper totals.

**v1.8 (2026-02-08) Dialectical impasse test**: Added Section 6.20 with three 5-turn impasse scenarios (epistemic resistance, affective shutdown, productive deadlock; N=24, eval-2026-02-08-f896275d, Opus judge). Recognition produces +43 pts on epistemic and +29 pts on interpretive impasses but $\Delta = -1.1$ on affective shutdown—sharpening the theoretical claim to epistemological rather than affective recognition.

**v1.9 (2026-02-08) Learner superego paradox**: Added symmetric learner-side evaluation (Section 6.16) scoring N=118 bilateral dialogues with 6-dimension learner rubric (eval-2026-02-07-b6d75e87, Opus judge). Multi-agent learner architecture hurts learner quality (d=1.43, F=68.28, p<.001)—the largest effect in the study. Recognition partially rescues multi-agent learners (d=0.79, p=.004) but not single-agent (n.s.). Added learner rubric description to Section 5.1, new Section 6.12, rewrote Section 7.5 with results, added finding #9 to Section 9.

**v2.0 (2026-02-08) Resolution strategy coding**: Post-hoc qualitative coding of all 24 dialectical impasse dialogues into five Hegelian resolution strategies. Perfect separation: 12/12 base tutors withdraw, 10/12 recognition tutors use scaffolded reframing (Aufhebung pattern). $\chi^2(3) = 24.00$, $p < .001$, $V = 1.000$. Cross-judge validation with GPT-5.2: $\kappa = 0.84$. Added Tables 26–28, per-turn strategy evolution analysis.

**v2.1 (2026-02-08) AI theme discovery and figure regeneration**: Added Section 6.13.4 AI-assisted theme discovery (N=300) showing near-perfect bimodal separation. Added Figure 6 (word clouds). Regenerated all figures from Python with corrected data and larger text. Removed standalone Section 10 Reproducibility (merged into Supplement IV). Moved Supplement V after other appendices. Increased font to 12pt.

**v2.1.1 (2026-02-10) Consistency fixes**: Corrected stale N=1,628/twenty to N=1,700/twenty-one in abstract, introduction, and conclusion. Fixed dynamic rewrite section references in Tables 2 and D. Added hardwired rules ablation and learner-side evaluation to Supplement IV run list (was 19 rows, now 21). Fixed inter-judge reliability cross-reference in Section 8.1.

**v2.1.2 (2026-02-10) Review corrections** (30 fixes): Table 7b Kimi row corrected to single-learner cells (N=350 to 179, Recognition +10.2 to +15.5, Interaction -1.5 to +0.5) matching probe design; total probe N 826 to 655. Factor C in Discussion corrected (-1.7 pts, F=2.56). Stale AxC values updated. Dynamic rewrite swing corrected (+16.7 to +8.7 delta). Terminology standardized (unified to single-agent, behaviour to behavior).

**v2.2.0 (2026-02-11) Modulation and learning outcomes**: Added Section 6.11.1 (modulation metrics, N=350 post-hoc) showing multi-agent architecture does not increase behavioral range (d=0.05); recognition produces calibration not oscillation (dimension variance d=−1.00, F=87.69). Added Section 6.11.2 (synthetic learning outcome index, N=118). Extended Section 7.4 Discussion with phronesis reframing. Regenerated Figures 4 and 6.

**v2.3.0 (2026-02-14) Phase 2 experimental results**: Added four new Results sections: Section 6.8 Dialectical Superego Modulation (cells 22–33, N=174, Tables 13–15); Section 6.9 Self-Reflective Evolution (cells 40–45, N=36, Tables 16–17); Section 6.10 Mechanism Robustness (cells 40–59 N=360 + cells 60–63 N=120, Tables 18–19); Section 6.11 Qualitative Transcript Assessment (Tables 20–21). Added Section 7.10 Scripted Learner Confound, Section 7.11 Practical Recommendations (6 recommendations). Expanded Section 6.10 with cells 64–65 and Nemotron cross-model replication (N=279, 49b33fdd). Renumbered all tables sequentially (1–48). Trimmed abstract from ~650 to ~250 words. Paper totals: N=2,700 across 28 key evaluations.

**v2.3.1 (2026-02-15) Cognitive prosthesis and cross-judge completion**: Added cognitive prosthesis test (cells 66–68, N=60). Completed GPT-5.2 cross-judge validation of mechanism robustness (N=360 paired, r=0.59). Added Nemotron self-reflect non-replication (559d854b, N=60) to Section 6.9/Table 17. Added blinded qualitative assessment validation (Table 21b).

**v2.3.2 (2026-02-15) Sample reconciliation and count update**: Added Phase 2 evaluations to Table 2 (9 additional rows). Updated paper totals from 28 to 30 key evaluations, N=2,909 scored. Added 50487df7 (cognitive prosthesis) to Supplement II.6. Noted Sonnet judge used for two late-stage evaluations.

**v2.3.3 (2026-02-15) Complete Table 19 with base mechanism cells**: Added cells 69–70 (eval-2026-02-15-664073ab, N=60, Opus judge) completing the base row of Table 19. Recognition delta remarkably consistent across all 4 mechanisms (+13.3 to +15.1). Updated paper totals from 30 to 31 evaluations, N=2,969.

**v2.3.4 (2026-02-15) Related Work expansion for arXiv/edArXiv submission**: Expanded Section 2 from 8 to 10 subsections. Added Section 2.3 LLM-as-Judge Evaluation Methodology, Section 2.7 Theory of Mind in AI Agents. Expanded Section 2.1 with empirical LLM tutoring studies, Section 2.2 with multi-agent systems and self-correction limits. Added 15 new bib entries.

**v2.3.5 (2026-02-15) Same-model blinded assessment**: Ran Opus-blinded qualitative assessment (N=118) resolving the model calibration confound. Key finding: blinding barely changes Opus's tag assignments, confirming the near-perfect binary separation is real, not an assessor bias artifact. Updated Section 6.11 interpretation, revised Section 8.2 limitation.

**v2.3.6 (2026-02-16) Judge version unification**: Rejudged all early runs (originally scored under Opus 4.5) with Opus 4.6, eliminating version drift across the evaluation dataset. Updated Section 8.1 limitations. Cleaned 6 empty/failed generation rows from dynamic rewrite runs.

**v2.3.7 (2026-02-17) Self-reflective evolution complete**: Updated Section 6.9 from partial (N=36) to complete (N=90) results for eval-2026-02-13-8d40e086. Recognition d=0.91 (was 1.02 at N=36). Key new finding: disposition gradient—suspicious +19.0, adversary +10.9, advocate +2.6. Updated Table 16, Table 17, Discussion finding 11. Deduped 270 re-judging artifact rows.

**v2.3.8 (2026-02-17) Nemotron mechanism N-count update**: Updated eval-2026-02-14-49b33fdd from N=301 to N=360 after run resumption completed. Cascaded count changes through abstract, introduction, Table 2, Section 7.11, Section 9. Updated Section 6.10 Nemotron narrative. Noted bidirectional profiling anomaly ($\Delta = -0.6$).

**v2.3.9 (2026-02-17) Factorial re-judging cascade**: All early runs rejudged with Opus 4.6 (unifying judge version). Full factorial ANOVA (N=350): recognition F=110.04 p<.001 $\eta^2$=.243 d=1.11 (was F=71.36, d=0.80). AxC interaction disappears (F=0.97, p=.325; was F=21.85, p<.001)—recognition now consistent across learner types. Updated Tables 4, 6, 8, 9, 9b, 12, 17, 41, 42. Restored 219 GPT-5.2 rows lost during dedup. Updated GPT compression ratio from ~58% to 37–59%.

**v2.3.10 (2026-02-17) Prompt elaboration baseline**: Added Section 6.21 comparing 344-line base prompt against 35-line naive prompt (JSON schema only). Two runs: Haiku (N=72) and Kimi (N=72). Key finding: elaborate prompt hurts Haiku (+6.8 pts for naive) and is inert on Kimi ($\Delta = -0.3$). Recognition ($M = 90.9$) remains well above naive ($M = 82.5$). Added Table 20b, conclusion finding 13. Updated paper totals to N=3,292 across thirty-one evaluations.

**v2.3.11 (2026-02-17) Transcript figures**: Added Figure 10 (naive vs base high_performer comparison panel) to Section 6.21. Added Figure 11 (bilateral mutual_transformation_journey transcript comparison) to Section 6.11. Added `generate-paper-figures.js` script for reproducible paper figure generation.

**v2.3.12 (2026-02-17) Token budget sensitivity**: Added Section 6.22 testing whether constraining `max_tokens` from 8000 to 256–2048 affects evaluation scores. Scores are flat across all budget levels; the recognition effect is fully preserved even at 256 tokens. The retry-absorption mechanism means truncated structured output self-heals. Added Table 49, recommendation 8 to Section 7.11. Updated paper totals to N=3,454 scored across thirty-six evaluations.

**v2.3.13 (2026-02-17) Paper correctness fixes**: Fixed eval-2026-02-14-559d854b scope from "cells 40–59, N=167" to "cells 40–45, N=60" in Table 2 (only cells 40–45 used; cells 46–59 superseded by 49b33fdd at N=360). Fixed broken Table 10 reference in Section 6.6. Fixed dynamic-learner N inconsistency: intro and finding 12 updated to N=300 (6c033830 + a2b2717c + 664073ab). Clarified token budget Section 6.22 design text. Added missing Supplement II commands.

**v2.3.14 (2026-02-18) Cognitive prosthesis re-run and analysis**: Replaced misconfigured prosthesis run (50487df7, all cells fell back to default Haiku) with corrected eval-2026-02-17-25aaae85 (N=90, Nemotron ego, Kimi K2.5 superego, Opus judge). Prosthesis hypothesis fails decisively: full mechanism stack scores 49.5 vs Nemotron simple base 64.2 ($\Delta = -15$). Added dimension analysis (two-tier static/dynamic capability model), superego parse failure analysis (16–45% malformed JSON auto-approves), and Haiku control smoke test (eval-2026-02-18-f489c0ea, N=6, confirming model-dependence). Added conclusion finding 14 (minimum ego capability threshold), recommendation 9 to Section 7.11, three future work items to Section 8.2 (parse robustness, capability threshold mapping, adaptive mechanism loading). Updated Table 2, Supplement IV run IDs. Paper totals: N=3,383 across thirty-seven evaluations.

**v2.3.15 (2026-02-19) Active control model confound resolved (A8)**: Ran active control (cells 15–18) on Kimi K2.5 (eval-2026-02-19-f2263b04, N=216, 116 grounded). Key finding: active control scores *below* base on Kimi ($M = 56.0$ vs $M = 64.2$ on matched scenarios, $\Delta = -8.2$), not between base and recognition as on Nemotron. Prescriptive heuristics help weaker models (Nemotron: +9 pts) but harm stronger ones (Kimi: −8 pts)—consistent with Section 6.21 prompt elaboration baseline. Same-day reproduction runs confirmed no model drift: Kimi base $M = 58.8$ (eval-2026-02-19-13d34bef, N=36) and Nemotron base $M = 49.8$ (eval-2026-02-19-411414e4, N=18). GPT-5.2 cross-judge validation confirms ordering (placebo 48.3 < base 70.8 < recognition 76.5). Updated Section 6.2 (added Table 6b, model-dependent interpretation), Section 5.3 (active control description), Section 8.1 (limitation resolved). Paper totals: N=3,653 across forty evaluations.

**v2.3.16 (2026-02-21) Recognition inversion mechanism and A2/A4 experiments**: Added Section 6.16.1 documenting the scenario-specific recognition inversion (misconception_correction: tutor +30.5, learner -10.4; mutual_transformation: no inversion). Identified metacognitive feedback loop mechanism through qualitative transcript analysis. A2 dynamic learner mechanism sweep (cells 72–77, N=238): all 7 mechanisms show positive recognition deltas; dynamic learner amplifies mechanism differentiation 1.6–2.8x vs scripted. A4 authenticity-focused learner superego (cells 78–79, N=47): null result—scores worse on every dimension including authenticity itself. Epistemic readiness dimension feasibility test (N=10): no base-recog gap (both 4.0/5), ruling out construct validity artifact. Reframed Section 7.5 from "calibration problem" to "structural finding." Paper totals: N=4,144 across forty-eight evaluations.

**v2.3.17 (2026-02-21) Table 19a and conclusion finding #15**: Expanded Section 6.10 mechanism results with Table 19a showing A2 sweep data (cells 72–77, N=108, Nemotron ego) for quantitative, erosion, and tutor-profiling mechanisms. All three show positive recognition deltas (+8.0 to +9.7), completing the 2x7 mechanism matrix. Updated conclusion finding #12 to reference full 7-mechanism matrix. Added finding #15 documenting recognition inversion as scenario-specific and structural. Fixed stale "thirty-seven" to "forty-eight" in conclusion summary.

**v2.3.18 (2026-02-23) Bug disclosure and data integrity corrections**: Added Section 8.1 paragraph disclosing two learner pipeline bugs affecting all pre-Feb-20 dynamic learner multi-turn evaluations: Bug 1 (leaky "SUPEREGO:" labels in learner external messages) and Bug 2 (broken `.flatMap()` causing learner to see only own monologue, not tutor responses). Added pre-fix data caveat to Section 6.16 (learner superego paradox, $d$ = 1.43 from eval-2026-02-07-b6d75e87 is an upper bound). Added footnote to Section 6.11 flagging $M$ = 38.0 as pre-fix. Clarified Section 6.10 Table 19 footnote to name bugs explicitly and explain why tutor-side scores are valid but learner-side required re-generation. Qualified Finding #2 (additivity) as tutor-side only, noting learner-side AxC interaction (Section 6.16). Qualified Finding #9 conclusion text with pre-fix caveat. Added forward reference from Section 6.3 to Section 6.16 learner-side interaction. Added clean re-generation runs eval-2026-02-20-bd37cc62 (cells 64–65) and eval-2026-02-20-4e131c6f (cells 69–70) to Table 2 and Supplement IV. Updated paper totals: N=4,264 across fifty evaluations. Also fixed Bug 5 in evaluation infrastructure: `resumeEvaluation()` silently dropped `--ego-model` and `--superego-model` overrides from stored metadata, causing resumed runs to fall back to YAML defaults.

**v2.3.19 (2026-02-24) Post-fix data replacement**: Replaced all pre-fix dynamic learner data with clean re-run data. Section 6.10 Table 19 now reports clean re-run tutor scores (0fbca69e, bd37cc62, 4e131c6f, N=240); mechanism ordering changes—intersubjective framing becomes the domi-nant mechanism (previously weakest), recognition deltas larger (+19.1 avg vs +14.2 pre-fix). Table 19a replaced with b5cd16e1 clean data (N=36). Section 6.16 Tables 29–31 replaced with post-fix factorial data (25c78e91, N=48): architecture effect *increases* to $d = 3.05$ (was 1.43), recognition effect vanishes ($d = 0.09$), AxC interaction vanishes ($F = 0.12$, was 11.50). The "recognition rescues multi-agent learners" finding was a bug artifact—removed throughout. Section 6.16.1 A4 results updated with b5e123b4 clean data (N=12). Rewrote Finding #2 (additivity holds on both rubrics), Finding #9 (paradox real and stronger, recognition irrelevant), Finding #12 (intersubjective framing now dominant mechanism), Finding #15 (tutor-learner asymmetry, not inversion). Updated Section 7.5, Section 8.1, abstract, and concluding summary. Section 6.11 footnote refined. Section 6.10 prose rewritten for new mechanism ordering. N-counts unchanged at 4,264 across fifty evaluations.

**v2.3.20 (2026-02-25) Bug 4 disclosure and claim-by-claim audit response**: Added Section 8.1 paragraph disclosing multi-turn scoring misalignment bug (Bug 4): `evaluate/rejudge` scored `suggestions[0]` (Turn 0) instead of last turn, affecting 8,631 multi-turn rows across 81 runs. Average holistic lift +19.9 pts. Paper clarified that all multi-turn tables report Turn 0 scoring; holistic last-turn rescoring in progress. Added prefix data caveats to Section 6.15 Table 25 footnote (b6d75e87 affected by Bugs 1–2 in ego/superego cells and by Bug 4 Turn 0 scoring). Extended Section 6.15.2 Table 28 caveat noting cross-run audit (N=895) shows negative learner-side learning arcs under recognition in multi-turn settings. Strengthened Section 8.1 bilateral transformation asymmetry paragraph with cross-run evidence ($d = -0.27$ recognition, $d = -0.98$ architecture on arc metrics, 30/33 runs show recognition reduces tutor dimension variance). Updated synthetic learning outcome bullet in summary to reflect mixed learner-side evidence. Added eval-2026-02-23-b5cd16e1 (cells 72–77, N=36) and eval-2026-02-23-b5e123b4 (cells 78–79, N=12) to manifest and Supplement IV. Updated paper totals: N=4,312 across fifty-two evaluations (manifest v1.8.0).

**v2.3.21 (2026-02-25) N-claim and stat-claim traceability**: Systematic pass through all 156 N-claims and 66 stat-claims identified by the claim-by-claim audit tool, adding inline run ID anchors, Section cross-references, and Table citations. Key changes: added run IDs to all table headers lacking them (Tables 3, 5, 6, 8, 13, 15, 20b, 21, 22, 25, 45); added Run ID column to Table 41 (inter-judge agreement); annotated pooled-data N-claims with "see Table 2" or "pooled across evaluation database"; marked pilot data as "no longer in primary evaluation set"; added Section/Table cross-references to all stat-claims in Sections 7–9. No data, analysis, or N-count changes—purely traceability metadata.

**v3.0.0 (2026-03-07) Automated prompt tuning experiment**: Added Section 7.7 with $2 \times 2 \times 2$ autotuning experiment (base/recognition $\times$ unoptimised/autotuned $\times$ DeepSeek/Qwen 3.5). Seven new runs (eval-2026-03-05-53fb1462 through eval-2026-03-07-c4d1bfa2, N=38 total). Key finding:

recognition prompts are more improvable than base (+17–23 pts vs +5–9 pts), the effects are super-additive, and the gap between conditions *widens* under optimisation. Autotuned recognition converges to ~71–73 on both models. Base autotuning ceiling (41–45) remains below unoptimised recognition baseline (48–55), confirming recognition's irreducibility to prompt engineering. Added Table N, 7 run IDs to Supplement IV. Renumbered Section 7.7 → 7.8.

**v3.0.1 (2026-03-07) M3 trajectory analysis**: Updated M3 (adaptive responsiveness) from "disconfirmed" to "conditionally supported" based on complete trajectory-specific data (eval-2026-03-06-ebcd6de0, N=72, 3 scenarios × 8 cells × 3 replications, DeepSeek V3.2, Sonnet judge). Pooled slope effect remains small (d=0.22, NS), but the 10-turn disengagement scenario shows d=1.63, p<.001, with recognition–baseline gap widening from +12 pts at T0 to +35 pts at T8–T10. The two 8-turn scenarios show no slope differentiation ($d \leq 0.30$). M3 recharacterized as a conditional emergent property of M1+M2 that manifests under sustained re-engagement demands with sufficient turn count. Updated abstract, Section 3.2, Section 6.3.2, Section 6.3.8, mechanism summary table, and Discussion synthesis. Added 1 run ID to Supplement IV (53 total).

**v3.0.2 (2026-03-07) M1/M2 mechanism isolation**: Added dedicated isolation run confirmation to Section 6.4.2. Runs eval-2026-03-06-768ba77b (M2: base + superego, cells 82–83) and eval-2026-03-06-e4abd0df (M1: recognition, no superego, cells 84–85) across 9 multi-turn scenarios (N=108, DeepSeek V3.2, Sonnet judge). Full 2 × 2 isolation confirms substitution: superego adds +9.2 pts under base (d=1.13, p=.002) but +1.1 under recognition (d=0.08, NS)—calibration pre-empts 88% of the superego's contribution (27% additivity deficit). M1 vs M2 head-to-head: calibration alone (51.4) outscores error correction alone (36.9) by d=1.03 in 7/9 scenarios. Two emotionally intense scenarios (Frustration, Affective Shutdown) show slight M2 advantage, suggesting scenario-specific residual error correction value. Added 2 run IDs to Supplement IV (55 total).

**v3.0.18 (2026-03-09) Superego redundancy refinement** (§7.8.4): Added subsection connecting prompt lab findings to the universal substitution claim. On weak models, targeted superego strategies amplify error correction beyond the baseline residual. On strong models, adding superego scaffolding is actively destructive, not merely redundant. Sharpens the substitution gradient from "helpful" to "harmful" across the capability threshold.

**v3.0.17 (2026-03-09) Dimension-targeted optimisation and cross-model transfer** (new §7.8): Added three subsections covering dimension-targeted prompt optimisation on specific rubric dimensions (§7.8.1), M3 engineering through prompt intervention (§7.8.2), and the cross-model prompt transfer experiment with prosthesis-straitjacket inversion finding (§7.8.3). Added fifth future work direction (§9) on model-stratified prompt optimisation and optimisation threshold map-ping. Old §7.8 (prompt density) renumbered to §7.9. Data from prompt lab sessions on Qwen 3.5 9B and Haiku 4.5 (N=3 per transfer cell, 10 autotune iterations on Qwen, 6 on Haiku).

**v3.0.16 (2026-03-07) Fix  rendering**: Unicode  (U+2264) does not render in the xelatex font (appears as blank/replacement character in PDF), while ≥ renders correctly. Replaced all 16 body-text  with LaTeX math expressions: `d  0.15` → `$d \leq 0.15$`, ` 0` → `$\leq$ 0`, etc. Verified ≥ (Unicode) renders correctly and left unchanged.

**v3.0.15 (2026-03-07) Unicode math operators**: Replaced all 45 standalone `$\to$` with Unicode → (U+2192), 16 `$\geq$`/`$\leq$` with ≥/ , 9 `$\approx$` with ≈, 31 `$\times$` with ×, 12 `$\pm$` with ±, and 4 `$\sim$` with `${\sim}N$` expressions. These standalone math delimiters caused math mode to leak into surrounding text in pandoc→xelatex rendering, producing garbled italic text in tables and prose (e.g., page 31 rubric table, page 40 floor-lifting paragraph). Full `$...$` math expressions (e.g., `$2 \times 2$`, `$d = 0.52$`) remain unchanged. Unicode characters render cleanly with xelatex without delimiter matching issues. **Note**:  was subsequently reverted to LaTeX math in v3.0.16 due to font glyph issue.

**v3.0.14 (2026-03-07) LaTeX formatting fixes**: Fixed 9 unformatted `2x2`/`2x2x2` factorial references (now `$2 \times 2$`), 11 split `$-$` sign notations (now single math expressions like `$-0.52$`), 1 LaTeX `\ref{}` macro in figure caption (replaced with prose reference), 1 split `$\sim$.30` (now `${\sim}.30$`), and 1 split `T0$-$T1` table header (now `$T0{-}T1$`). All issues would produce rendering aberrations in the PDF.

**v3.0.13 (2026-03-07) Recognition term slippage pass**: 10 targeted edits across §6.1, §6.4, §6.5, §6.6, §7.2, §7.8, and §9 to prevent "recognition" from sliding between design-heuristic usage (level 2–3) and philosophical-agency usage (level 1). Key changes: "recognition eliminates" → "recognition-oriented prompts eliminate"; "recognition rescues the tutor" → "recognition-enhanced prompts rescue the tutor's output quality"; "genuine dialogue" → "dialogue-structured responses"; "genuinely shaped" → "specifically adapted"; "empirical echo" → "functional parallel"; "The tutor recognizes the learner" → "The prompts instruct the tutor to treat the learner as an autonomous subject." Maintains the three-senses framework established in v3.0.12.

**v3.0.12 (2026-03-07) Three senses of "recognition"**: Added terminological clarification paragraph to §1 (after pedagogy framing, before "The ablative finding") explicitly distinguishing recognition as (1) philosophical inspiration (Hegel), (2) operational design heuristic (prompt instructions and architectural choices), and (3) observable discourse effects (question-asking, variance reduction, substantive revision). States that empirical claims concern primarily the third level. Responds to external review concern about construct validity and double duty of "recognition."

**v3.0.11 (2026-03-07) Clarify evidence status markers in §3**: Renamed three "Missing evidence" markers (§3.2, mechanisms 1–3) to "Evidence needed beyond pilot" and added forward references (*Addressed in Section*

*6.X*) to prevent readers from misinterpreting them as gaps in the current paper. Verified provable discourse counts (119 claims, 18 adapters) remain accurate.

**v3.0.10 (2026-03-07) Human expert coding validation**: Strengthened §8.2 to identify human expert coding as the most important outstanding validation for mechanism claims, with explicit procedure (30–50 exchanges, two coders, 10-category taxonomy, Cohen's $\kappa$, human-LLM agreement). Added matching future work entry in §9 (now "Third") with the same procedural specificity. Both sections note that the trace infrastructure makes this feasible without re-running dialogues. Added cross-reference between §8.2 and §9.

**v3.0.9 (2026-03-07) Unexplained phenomena in conclusion**: Added paragraph to §9 noting three findings the current mechanism model does not explain: near-identical effect magnitude across structurally different models (DeepSeek $d = 1.88$, Haiku $d = 1.84$, Gemini Flash 3.0 $d = 1.92$), the failure of adaptive responsiveness as a general mechanism, and the learner superego paradox ($d = 3.05$). Suggests these anomalies may point to a deeper process the current apparatus cannot yet trace. Responds to external review concern about consolidating what recognition doesn't explain.

**v3.0.8 (2026-03-07)** Added prompt elaboration baseline (Paper 1.0 §6.21) as fourth evidence line in §7.8 prompt density paragraph: the 344-line base prompt is itself a long detailed pedagogical prompt, yet it doesn't approach recognition quality and actually hurts performance on Haiku ($M = 82.5$ naive vs base, recognition $M = 90.9$). Prompt length without recognition content can be inert or counterproductive.

**v3.0.7 (2026-03-07) Prompt density defense**: Added "Prompt density as alternative explanation" paragraph to §7.8, consolidating three existing evidence lines against the reduction of recognition to prompt elaboration: (1) placebo control (Paper 1.0, length-matched prompts without recognition theory), (2) autotuning experiment (§7.7, gap widens under optimization), (3) mediation analysis (§6.1.7, 42.4% mediated through question-asking). Acknowledges the strongest version of the alternative remains untested: whether *any* equally specific pedagogical prompt would match recognition.

**v3.0.6 (2026-03-07) Engineered vs. discovered mechanism distinction**: Added paragraph to §8.4 explicitly addressing the circularity objection: the architecture was designed for observability, not to guarantee specific outcomes; three key findings were genuinely unpredicted (superego redundancy on strong models, adaptive responsiveness null, consistent 15–17% additivity deficit). Distinguishes "mechanisms of this system" from "mechanisms of AI tutoring in general."

**v3.0.5 (2026-03-07) Theoretical framing and scope**: Three changes responding to external review. (1) Reframed Hegelian mapping as heuristic translation: "derive mechanisms from Hegel" → "motivate mechanisms drawing on Hegel" across abstract, §1, §3.2, §3.5, and §9 (~6 passages). Changed "derivative application" → "heuristic translation" in §3.5. The mappings are productive redescriptions, not logical derivations. (2) Added scope qualification to abstract: "All findings concern tutor output quality as assessed by LLM judges interacting with synthetic learners." (3) Added cost qualifier after contributions list: architecture is a research instrument, not a deployable system (~225K tokens/dialogue). No data or analysis changes.

**v3.0.4 (2026-03-07) Epistemic language audit**: Systematic softening of causal language throughout, responding to external review feedback. Changed ~45 instances: "confirmed" → "supported" for mechanism claims; "disconfirmed" → "not supported as general mechanism"; "dominant mechanism" → "strongest effect" / "primary mechanism"; "confirms [prediction]" → "is consistent with" / "supports"; "The disconfirmation of" → "The null finding for". Retained "confirmed" only for factual verification ("blind scoring confirmed by code audit") and revision history entries. Changed "Mechanism isolation confirmation" → "Mechanism isolation evidence". Softened figure captions and prediction labels throughout §6.1–6.4. No data, analysis, or N-count changes—purely rhetorical recalibration to better match the evidentiary base.

**v3.0.3 (2026-03-07) Final claim audit corrections**: (1) Gemini Flash 3.0 stale data fixed: `tutor_overall_score` values updated from 2026-03-03 scoring to 2026-03-05 re-scoring (17.3/37.0/52.6/64.9 → 22.4/49.3/57.7/70.0), changing the interaction from −7.4 to −14.6 and additivity deficit from 10% to 17%. (2) "Substitution-to-additive transition" narrative reframed as "universal substitution (15–17% deficit across all models) with model-dependent residual architecture benefit" across ~20 passages (abstract, §3, §5, §6.4, §6.6, §7.3, §8). (3) N=570→N=432 corrected throughout (isolation runs had expanded the figure-generation query beyond the clean factorial scope). (4) §7.3 "same model, different sample" error fixed (run 45163390 is Haiku, not DeepSeek). (5) §6.3.7 dialogue quality table updated with verified DB values and Gemini Flash 3.0 rows added. Provable discourse: 87 pass, 0 warn, 0 fail.